\documentclass[a4paper,11pt]{article}
\usepackage{heppub}
\pdfoutput=1
\usepackage{tikz}
\usepackage{cancel}

\tikzset{position/.style args={#1:#2 from #3}{at=(#3.#1), anchor=#1+180, shift=(#1:#2)}}
\usepackage{bm}
\usepackage{graphicx}
\usepackage{subcaption}
\usepackage{hyperref}
\usepackage{xspace}
\definecolor{col1}{rgb}{0.81, 0.85, 0.91}
\definecolor{col2}{rgb}{0.96, 0.88, 0.74}
\definecolor{col3}{rgb}{0.87, 0.91, 0.76}
\definecolor{col4}{rgb}{0.98, 0.82, 0.76}
\definecolor{col5}{rgb}{0.86, 0.84, 0.91}
\definecolor{col6}{rgb}{0.93, 0.83, 0.73}

\usepackage{xspace}
\usepackage{placeins}
\usepackage{comment}
\usepackage{wasysym}
\usepackage{slashed}
\usepackage{empheq}
\usepackage{marginnote}
\usepackage{xcolor}
\usepackage[normalem]{ulem}
\usepackage{amsmath}
\usepackage{mathrsfs}
\usepackage{orcidlink}
\nopagebreak

\newcommand{\ord}[1]{\mathcal{O}(#1)}

\newcommand{\eps}{\epsilon}

\newcommand{\nn}{\nonumber}

\newcommand{\lqcd}{\Lambda_\mathrm{QCD}}
\newcommand{\MSbar}{$\overline{\text{MS}}$\xspace}

\newcommand{\Pythia}{\texttt{Pythia}\xspace}
\newcommand{\Herwig}{\texttt{Herwig}\xspace}

\allowdisplaybreaks[3]

\definecolor{darkyellow}{rgb}{0.5, 0.5, 0.0}
\definecolor{darkpurple}{rgb}{0.5, 0.2, 0.8}
\definecolor{darkblue}{rgb}{0.0, 0.0, 0.8}
\definecolor{darkgreen}{rgb}{0.0, 0.4, 0.0}
\definecolor{darkred}{rgb}{0.5, 0.0, 0.0}

\usepackage[normalem]{ulem}
\usepackage{marginnote}

\title{\boldmath On Determining $\alpha_s(m_Z)$ from Dijets in $e^+e^-$ Thrust}

\author[a]{Miguel A. Benitez\orcidlink{0000-0002-6939-2677},}
\emailAdd{mbenitez@usal.es}

\author[b]{Andr\'e H.~Hoang\orcidlink{0000-0002-8424-9334},}
\emailAdd{andre.hoang@univie.ac.at}

\author[a]{Vicent Mateu\orcidlink{0000-0003-0902-5012},}
\emailAdd{vmateu@usal.es}

\author[b,c]{Iain W.~Stewart\orcidlink{0000-0003-0248-0979}}
\emailAdd{iains@mit.edu}

\author[d]{and Gherardo Vita\orcidlink{0000-0003-1095-5698}}
\emailAdd{gherardo.vita@cern.ch}

\affiliation[a]{Departamento de F\'isica Fundamental e IUFFyM,\\Universidad de Salamanca, E-37008 Salamanca, Spain}
\affiliation[b]{University of Vienna, Faculty of Physics, Boltzmanngasse 5,
A-1090 Wien, Austria}
\affiliation[c]{Center for Theoretical Physics,\,Massachusetts Institute of Technology,\,Cambridge,\,MA\,02139,\,USA}
\affiliation[d]{CERN, Theoretical Physics Department, CH-1211 Geneva 23, Switzerland}

\abstract{%
We update a previous N$^3$LL$^\prime$+${\cal O}(\alpha_s^3)$ determination of the strong coupling from a global fit to thrust data by including newly available perturbative ingredients, upgrading the renormalization scales to include a fully canonical scaling region, and implementing the log resummation in a way which ensures the integrated cross section is unaffected by the the leading $1/Q$ hadronization power corrections. Detailed discussions are provided concerning the stability of the results under variations of the fit range and the importance of summing up higher-order logarithmic terms for convergence and stability. We show that high-precision results can be achieved even when carrying out a more conservative fit by restricting the dataset to a region which is more clearly dominated by dijet events. This leads to $\alpha_s(m_Z) = 0.1136 \pm 0.0012$ with $\chi^2/{\rm dof}=0.86$, fully compatible with earlier results using a larger fit range. We also demonstrate that a number of additional effects associated to power corrections have a small impact on this fit result, including modifications to the renormalon substraction scheme for dijet power corrections and the inclusion of three-jet power correction models. The fit is also shown to provide very good agreement with data outside the fit range.}

\date{\today}

\preprint{\vbox{%
\hbox{MIT-CTP 5746}
\hbox{CERN-TH-2024-142}
\hbox{UWThPh2024-8}}}

\DeclareUnicodeCharacter{2212}{-}

\begin{document}

\maketitle

\section{Introduction}
\label{sec:intro}

High-energy electron-positron colliders offer a clean environment to study the fundamental properties of the strong interactions described by Quantum Chromodynamics (QCD). In this regard, differential observables such as event shapes play a central role, as they display a high sensitivity to QCD dynamics and are infrared and collinear safe. Experimental data for event shape distributions obtained at $e^+e^-$ experiments~\cite{TASSO:1990cdg,MovillaFernandez:1997fr,AMY:1989feg,SLD:1994idb,L3:1992nwf,L3:2004cdh,DELPHI:2003yqh,DELPHI:2000uri,DELPHI:1999vbd,OPAL:2004wof,OPAL:1997asf,OPAL:1999ldr,ALEPH:2003obs} have been investigated extensively, in particular to determine the strong coupling constant $\alpha_s$ with high precision, see example Refs.~\cite{dEnterria:2022hzv,Kluth:2006bw,MovillaFernandez:2001ed} for reviews. Dijet event shapes such as thrust~\cite{Farhi:1977sg}, C-parameter~\cite{Parisi:1978eg,Donoghue:1979vi}, and heavy jet mass~\cite{Clavelli:1979md,Chandramohan:1980ry} enable a selection on events with two back-to-back narrow jets, and the dominant portion of the experimental data is in this dijet region. Since event shape distributions are not inclusive cross sections, in the dijet region there is an incomplete cancellation between the real- and virtual-radiation that manifests itself as large double (Sudakov) logarithms. Theoretical methods are available to handle the resummation of these large logarithms at high precision, as well as to handle nonperturbative corrections in this region from first principles using quantum field theory.

Over the course of the last two decades, the theoretical description of these event-shape distributions got significantly boosted in a number of ways.
Fixed order perturbative QCD calculations have been extended to include corrections up to ${\cal O}(\alpha_s^3)$, with increasing numerical precision~\cite{Catani:1996jh,Catani:1996vz,Gehrmann-DeRidder:2007nzq, Gehrmann-DeRidder:2009fgd,Weinzierl:2008iv,Weinzierl:2009ms,DelDuca:2016csb, DelDuca:2016ily}.
Resummation was originally carried out at next-to-leading-logarithmic order (NLL) using the coherent branching algorithm~\cite{Catani:1992ua}.
With the advent of Soft-Collinear Effective Theory (SCET)~\cite{Bauer:2000ew,Bauer:2000yr,Bauer:2001ct,Bauer:2001yt,Bauer:2002nz}, resummation could be pushed to N$^3$LL$^\prime$\,+\,$\mathcal{O}(\alpha_s^3)$ level for thrust, heavy-jet mass and C-parameter~\cite{Schwartz:2007ib,Becher:2008cf,Chien:2010kc,Hoang:2014wka} and to N$^2$LL$^\prime$\,+\,$\mathcal{O}(\alpha_s^2)$ precision for other event shapes~\cite{Bell:2018gce}. Here the $+{\cal O}(\alpha_s^k)$ indicates the full inclusion of fixed order corrections in a manner that is consistent with the resummation, enabling a description with full ${\cal O}(\alpha_s^k)$ accuracy outside the dijet region.
The coherent branching method has also been automated to achieve N$^2$LL accuracy~\cite{Banfi:2001bz,Banfi:2014sua}.

The leading hadronization power corrections in the dijet region
were originally treated with models~\cite{Webber:1994cp,Dokshitzer:1995zt,Akhoury:1995sp}, and
later were shown to be related to non-perturbative vacuum matrix elements of operators defined within quantum field theory, namely Wilson lines~\cite{Korchemsky:1994is,Korchemsky:1999kt,Lee:2006nr,Hoang:2007vb,Abbate:2010xh}.
These corrections are related to large angle soft radiation.
It was found that in the tail dijet region (past the peak in the event-shape spectrum), hadronization corrections can be expanded in an operator product expansion (OPE), whose leading ${\cal O}(\Lambda_{\rm QCD})$ term is denoted $\Omega_1$. This allows for a high precision prediction of the tail region of the cross section with only two parameters, $\alpha_s(m_Z)$ and $\Omega_1$.
The dominant large-order asymptotic behavior of the pertubative series in $\alpha_s$ involves factorially growing terms from infrared renormalons~\cite{Akhoury:1995sp,Nason:1995np,Gardi:2001ny}, the leading term of which is associated to large-angle soft radiation in the nonperturbative matrix element $\Omega_1$~\cite{Hoang:2007vb}.
It was realized that the bad behavior due to these renormalons can be tamed by
perturbative subtractions~\cite{Dokshitzer:1995qm,Dokshitzer:1997ew,Davison:2009wzs}, which in the field theory approach corresponds to a change of scheme for $\Omega_1$~\cite{Hoang:2007vb,Hoang:2009yr} (to so-called gap schemes).
This scheme change enables large logarithms appearing in the renormalon subtraction series to be summed with R-evolution equations~\cite{Hoang:2008yj,Hoang:2009yr,Hoang:2017suc}.

These developments have been crucial in improving the theoretical precision of event-shape predictions to approach that of the experimental measurements. The full set of theoretical advances listed above was used in global fits for thrust~\cite{Abbate:2010xh,Abbate:2012jh} and C-parameter~\cite{Hoang:2014wka,Hoang:2015hka}, where both $\alpha_s(m_Z)$ and $\Omega_1$ were fit parameters. The largest uncertainty on these two parameters occurs in a correlated direction, where an increase to $\alpha_s(m_Z)$ is accompanied by a decrease to $\Omega_1$.
The outcome of these studies was a very precise value for $\alpha_s(m_Z)\simeq 0.114\pm 0.001$, which is however several standard deviations below the world average $\alpha_s(m_Z)\simeq 0.118$.

This discrepancy has caused some concern, and was responsible for the reanalysis of various aspects of the theoretical description. In particular, there have been a number of recent theoretical efforts to shed light into the nature of hadronization away from the dijet region~\cite{Luisoni:2020efy,Caola:2021kzt,Caola:2022vea,Nason:2023asn,Bell:2023dqs}. In Ref.~\cite{Luisoni:2020efy} a parametric analysis of the structure of the leading non-perturbative power corrections in the dijet and three-jet regions was carried out for the C-parameter distribution, which is much more sensitive to hard splitting processes than thrust. Using the assumption that the leading non-perturbative parameter for these two regions are related, the model used in Ref.~\cite{Luisoni:2020efy} indicates that the three-jet power corrections can be about half the size of the dijet. Refs.~\cite{Caola:2021kzt,Caola:2022vea} developed a model for three-jet hadronization corrections in thrust and other event shapes based on a computation of the leading linear infrared sensitivity of the NLO QCD correction to the process $e^+e^- \to q \bar q \gamma$. These developments have been used in an analysis of $\alpha_s(m_Z)$ determinations in Ref.~\cite{Nason:2023asn}.

One of the goals of this article is to carry out analyses addressing the concerns raised in these publications. In addition, we refine our theoretical prediction using state-of-the-art perturbative ingredients and upgrade our profile functions to the more canonical version introduced in Ref.~\cite{Hoang:2014wka}.
We also discuss a preferred way for setting the renormalization scales such that the cross section is compatible with the total cross section OPE.
We present a new value of the strong coupling obtained from fits to experimental data in a smaller dijet dominated region, compared to the larger default fit window considered in Ref.~\cite{Abbate:2010xh}. Included in this result is an assessment of uncertainties from 3-jet power corrections and other subleading hadronization effects. Finally, we compare our best-fit theoretical prediction to the available datasets within, as well as outside the new default fit range.

The paper is organized as follows: In Sec.~\ref{sec:thrustDistr}, we define thrust and present the theoretical framework in which the differential cross-section is computed. The various ingredients entering the description are distributed in subsections: dijet factorization theorem in Sec.~\ref{subsec:factorizationFormular}, singular and non-singular partonic distribution in Secs.~\ref{subsec:singular} and \ref{subsec:non-singular}, respectively, and dijet non-perturbative corrections and renormalon subtractions (cf.\ Secs.~\ref{subsec:nonpertCorr} and \ref{subsec:gapscheme}). Our renormalization scale parametrization is discussed in Sec.~\ref{subsec:profiles}, and an implementation respecting the OPE for the integrated cross section is presented in Sec.~\ref{subsec:opeprofiles}. The next subsections~\ref{subsubsec:ResummationXSlevel}-\ref{subsec:additionalUncertainties} are devoted to numerical studies of various theoretical aspects related to resummation and normalization, and we also discuss hadron-mass and QED effects. After presenting our dataset and fit procedure in Sec.~\ref{sec:expData}, we study the stability of strong-coupling determinations under variations of the fit range for various theoretical setups in Sec.~\ref{sec:results}. Sec.~\ref{sec:argumentationDiTrijet} is dedicated to discuss at length the validity of dijet factorization, particularly in what concerns hadronization. We devise a parametrization for deviations from the dijet treatment of power correction which permits assigning an uncertainty associated to our lack of knowledge on those, and propose an alternative point of view based on EFT power-counting arguments. The final results of our strong-coupling determination are contained in Sec.~\ref{sec:inputUpdate}, where a comparison to experimental data is presented, along with the list of effects which have not been included due to their smallness. Our conclusions and outlook are to be found in Sec.~\ref{sec:conclusions}. We relegate to the appendix the comparison of our theoretical prediction against experimental data for center-of-mass energies other than the $Z$-pole.

\section{Thrust distribution
\label{sec:thrustDistr}}

Thrust is defined in the center-of-mass frame of an $e^+ e^-$ collision as~\cite{Farhi:1977sg}
\begin{equation}
T \equiv \max_{\vec{n}} \frac{\sum_j | \vec{p}_j \cdot \vec{n} |}{\sum_j |
\vec{p}_j |}
\label{eq:tdef}
\,,
\end{equation}
where the sum is over all particles in the event and the maximum is over 3-vectors $\vec{n}$ of unit norm. The vector $\vec{n}$ that maximizes thrust is known as the {\it thrust axis}. This axis splits the event into two hemispheres. Conventionally one also defines $\tau = 1-T$.

In this section we present a brief review of the thrust factorization formula we use for our
analysis and fits.
We also highlight the most important improvements and differences between
the current analysis and the setup used in Ref.~\cite{Abbate:2010xh}, which we also frequently refer to as `the 2010 analysis'.

\subsection{Factorization Formula \label{subsec:factorizationFormular}}

Our analysis is based on a factorization theorem for the $e^+e^-$ thrust distribution in the dijet region, which consists of three main parts
\begin{equation}
\label{eq:factorizationFormula}
\frac{\text{d}\sigma}{\text{d}\tau} = \int \text{d}k \biggl( \frac{\text{d} \hat{\sigma}_{\rm s}}{\text{d}\tau} + \frac{\text{d} \hat{\sigma}_{\rm ns}}{\text{d}\tau} \biggr)\! \biggl(\tau - \frac{k}{Q}\biggr) F_\tau[k - 2\bar\Delta(R,\mu_s)]\,.
\end{equation}
Here $Q$ denotes the $e^+e^-$ center-of-mass energy.

The first contribution, $\text{d} \hat{\sigma}_{\rm s}/\text{d}\tau$, represents the so-called singular partonic distributions
and contains terms involving $\alpha_s^k \delta(\tau)$ and those that diverge as $\alpha_s^k \log^n(\tau)/\tau$ for $\tau\to 0$. These contributions themselves factorize according to the widely separated dynamical phase space regions relevant for the dijet kinematics. A review of the resulting factorization formula, that facilitates resummation of the associated large logarithms and the new available theoretical ingredients (compared to the 2010 analysis) is given in Sec.~\ref{subsec:singular}. The second contribution, $\text{d} \hat{\sigma}_{\rm ns}/\text{d}\tau$, encodes the remaining ``non-singular'' partonic cross-section contributions explained in more detail in Sec.~\ref{subsec:non-singular}. Lastly, $ F_\tau(k)$ is the so-called shape function, which parametrizes the dominant source of soft non-perturbative hadronization effects arising from large-angle soft radiation. It is defined from a vacuum matrix element of soft Wilson lines~\cite{Korchemsky:1999kt,Lee:2006nr,Hoang:2007vb,Abbate:2010xh}. We emphasize that the shape function has a well-defined field theoretic definition in the dijet regime in analogy to, for example, the parton distribution functions describing non-perturbative collinear dynamics. Thus, the model aspect of the shape function and its content is merely related to its functional form. Outside the dijet regime, the concept of the shape function represents a true model for the non-perturbative corrections. This aspect and its practical consequences are one among the important issues we address in this article.

In the tail of the distribution ---\,to the right of the peak in the dijet region in $\tau$\,--- the shape function can be expanded in an operator product expansion based on the scale hierarchy $Q\tau\gg \Lambda_{\rm QCD}$. The leading term linear in $\Lambda_{\rm QCD}$ is tantamount to shifting the distribution towards larger thrust values by $2\bar\Omega_1/Q$, where $2 \bar\Omega_1=\int{\rm d}k \,k\, F_\tau(k)$ is the first moment of the shape function. Therefore, in the tail region, up to first order in non-perturbative corrections, one can replace $F_\tau(k)\to \delta(k - 2\bar\Omega_1)+\mathcal{O}(\Lambda_{\rm QCD}^2/k^3)$. However, we do not employ this approximation since it
will miss important non-perturbative corrections that are no longer subleading when $Q\tau \sim \Lambda_{\rm QCD}$, and we wish to have predictions for the cross section for all $\tau$, such that, for example, we can integrate over $\tau$.
More details are given in Sec.~\ref{subsec:nonpertCorr}.

The canonical standard is to consider the $\Omega_1$ and shape function as non-perturbative matrix elements with UV divergences regulated by dimensional regularization, with $d=4-2\epsilon$, and minimally subtracted, which corresponds to the $\overline{\rm MS}$ scheme. As always, this same scheme choice must be made for the perturbative coefficient functions, which here are the perturbative and nonsingular cross sections, since the full cross sections are scheme independent. Since the final perturbative cross sections are infrared finite, this corresponds to calculations with vanishing infrared cutoff. Note that this scheme choice for $\Omega_1$ is independent of the scheme chosen for the strong coupling $\alpha_s(m_Z)$.
We refer to the first shape-function moment in the $\overline{\rm MS}$ scheme as $\bar\Omega_1$. In this scheme, however, the perturbative thrust distribution in the dijet region suffers from a $u=1/2$ renormalon which is associated to an ${\cal O}(\Lambda_{\rm QCD})$ renormalon in $\bar \Omega_1$. This renormalon can be eliminated by implementing a perturbative IR subtraction, which can be achieved by a suitable redefinition of the leading power correction $\bar \Omega_1 = \Omega_1(R,\mu_s) + \delta(R,\mu_s)$, where $\delta(R,\mu_s)$, called the gap subtraction, is a perturbative series proportional to the scale $R$ which is combined with the perturbative corrections from large-angle soft radiation (indicated by the dependence on the soft renormalization scale $\mu_s$)~\cite{Hoang:2007vb}. It diverges in such a way that the leading renormalon is absent in $\Omega_1(R,\mu_s)$ and in the partonic distribution. This approach effectively reintroduces an IR cutoff $R$ for the partonic cross section for the dominant linear sensitivity to small momenta, and it implies that the shape function becomes also dependent on the scales $R$ and $\mu_s$ through a shift in the convolution over $k$ by the function $\bar\Delta(R,\mu_s)$. This shift is already shown in Eq.~\eqref{eq:factorizationFormula} and explained in more detail in Sec.~\ref{subsubsec:reviewGap}. The subtraction scale $R$ represents an additional factorization scale and different schemes to define $\delta(R,\mu_s)$, called gap-schemes, can be devised. Furthermore, the introduction of the subtraction scale $R$ implies the need for additional RG summation to avoid the appearance of large logarithmic higher order corrections. More details are given in Secs.~\ref{subsec:nonpertCorr} and \ref{subsec:gapscheme}.

\subsection{Singular Partonic Distribution }
\label{subsec:singular}

For the singular partonic thrust distribution in the dijet region, considering all quarks massless and for a given gap scheme,
all perturbative contributions can be written down in terms of a factorization theorem of the form~\cite{Fleming:2007qr,Schwartz:2007ib,Abbate:2010xh}
\begin{align}
\label{eq:singFactorizationFormula}
\frac{1}{\sigma_0}\frac{\text{d} \hat{\sigma}_s}{\text{d}\tau}(\tau) &= Q H_Q(Q,\mu_H)U_H(Q,\mu_H,\mu)\!\! \int\! \text{d}s\,\text{d}s'\, J_{\tau}(s',\mu_J)U_J^{\tau}(s-s',\mu,\mu_J)
\\
&\times\!\!
\int\! \text{d}k' \,U_S^{\tau}(k',\mu,\mu_s) e^{-\frac{2\delta(R,\mu_s)}{Q}\frac{\partial}{\partial \tau}} \hat S_{\tau}\!\biggl( Q\tau - \frac{s}{Q} - k', \mu_s \biggr),\nonumber
\end{align}
where $\sigma_0$ corresponds to the tree-level Born cross section and $Q$ denotes the $e^+e^-$ center-of-mass energy. The hard function $H_Q$ encodes corrections coming from the scale $Q$ and is accompanied by its evolution kernel $U_H$, which is responsible for summing up large logarithms of $\mu/Q$. Its renormalization scale $\mu_H$ must be chosen of order $Q$, $\mu_Q\sim Q$, to render $H_Q$ free or large logarithms. The thrust jet function $J_{\tau}$ encodes corrections from collinear radiation associated to the typical jet invariant mass scale $Q\sqrt{\tau}$. Its renormalization scale $\mu_J$ has to be of order $Q\sqrt{\tau}$. The soft $\hat S_{\tau}$ function contains the corrections from large-angle soft radiation dominated by energy scales of order $Q\tau$, which implies that for its renormalization scale $\mu_s$ we have $\mu_s\sim Q\tau$. Both functions are defined from their more general double-hemisphere counterparts
\begin{equation}
J_\tau(s,\mu) = \int_0^s\! {\rm d}s'J_n(s-s',\mu)J_{\bar n}(s',\mu)\,,\qquad
\hat S_\tau(k,\mu)= \int_0^k {\rm d}k' \hat S_n(k-k', k',\mu)\,,
\end{equation}
and appear along with their respective evolution kernels $U_J^{\tau}$ and $U_S^{\tau}$. The factorization theorem shown in Eq.~\eqref{eq:singFactorizationFormula} also contains an exponential derivative operator containing the $\delta(R,\mu_s)$ gap subtraction series which, order by order in the strong coupling expansion, has to be consistently combined with the soft function series to remove the ${\cal O}(\Lambda_{\rm QCD})$ renormalon.

The jet and soft functions, as well as the exponential implementing the renormalon subtraction, are also governed by renormalization group equations. For the jet and soft functions these are made explicit in terms of the evolution factors $U_J^{\tau}$ and $U_S^{\tau}$, respectively. We refer to the renormalizaiton scale $\mu_H$, $\mu_J$, $\mu_s$, and $R$ as the hard, jet, soft, and renormalon subtraction scales, respectively. The latter must be chosen of order $\mu_s$, to avoid large logarithms of $R/\mu_s$. Because the physical collinear and large-angle soft dynamical scales are $\tau$-dependent, $\mu_J$, $\mu_s$, and $R$ each must be set to functions of $\tau$ called profile functions~\cite{Abbate:2010xh,Hoang:2014wka}. For our phenomenological analyses we use those introduced in 2015 in Ref.~\cite{Hoang:2014wka}. These profile function incorporate essential improvements compared to those employed in the original 2010 analysis of Ref.~\cite{Abbate:2010xh}. We will discuss the main aspects of such improvements in Sec.~\ref{subsec:profiles}. For a complete summary of all perturbative ingredients contained in the factorized singular partonic distribution used in the analysis of Ref.~\cite{Abbate:2010xh} we refer to the appendix of that reference. In the following we therefore only comment on the additional ingredients we use in this work.

For the fixed-order hard function $H_Q$, which contains short-distance QCD dynamics and is obtained by matching the two-jet current in SCET to full QCD, there is no additional information compared to the analysis carried out in Ref.~\cite{Abbate:2010xh}. On the other hand, the four-loop cusp anomalous dimension, which enters the renormalization group factors $U_H$, $U_J^\tau$ and $U_S^\tau$ at N$^3$LL order, is now available~\cite{vonManteuffel:2020vjv,Henn:2019swt,Henn:2019rmi,Bruser:2019auj,Moch:2018wjh,Moch:2017uml}. The numerical result for $n_f=5$ reads \mbox{$\Gamma_3^{\rm cusp} = 141.246$} in the conventions of Ref.~\cite{Abbate:2010xh}, see Eqs.~(A26) therein, and is incorporated in our analysis.
For the thrust jet function $J_\tau$ the new ingredient in our analysis is the three-loop non-logarithmic coefficient. The latter has been recently computed in position space for the hemisphere jet function $J_n$~\cite{Bruser:2018rad,Banerjee:2018ozf} and the corresponding correction reads $j_3=−128.651$.
For the partonic soft function $\hat S_{\tau}$ the new ingredients are an exact value of the two-loop
non-logarithmic coefficient, $s_2 = -40.6804$~\cite{Kelley:2011ng,Monni:2011gb} without a $\pm 2.5$ numerical uncertainty, and a value of the
three-loop coefficient $s_3=-1030.2166$~\cite{Baranowski:2024vxg}. For the conventions we use to define $s_2$, $s_3$ and $j_3$, see Eqs.~(A.14) and (A.16) of Ref.~\cite{Abbate:2010xh}, respectively.

Together with the new ingredients we can now carry out a fully consistent and complete N$^3$LL QCD analysis for the thrust distribution. To achieve this resummation order, in principle, only the single logarithmic plus distributions in the jet and soft function at ${\cal O}(\alpha_s^3)$ would be required~\cite{Almeida:2014uva}. Including the full set of ${\cal O}(\alpha_s^3)$ corrections for all factorization functions together with N$^3$LL anomalous dimensions, will be referred to as N$^3$LL$^\prime$ order, which is the accuracy of our analysis.
A final remark concerns QED and bottom mass corrections to Eq.~(\ref{eq:factorizationFormula}), which were accounted for in Ref.~\cite{Abbate:2010xh}. The impact of including QED and bottom mass effects were shown to lead to a decrease in $\alpha_s(m_Z)$ by $-0.0005$~\cite{Abbate:2010xh}. Since QED and bottom quark mass effects are not relevant for analyzing the robustness or stability of the fit, we only discuss them briefly in this article.

\subsection{Non-Singular Distribution}
\label{subsec:non-singular}

The non-singular partonic thrust distribution accounts for all perturbative corrections that are power suppressed by additional powers in $\tau$ compared to the singular distribution. It is obtained from the complete fixed-order corrections with the singular contributions being subtracted.\footnote{To obtain the fixed-order singular cross section we set all scales equal, $\mu_i=\mu_H=\mu$.} Including the contributions from the gap subtraction, which we implement globally in the factorization formula of Eq.~\eqref{eq:factorizationFormula}, we can write the complete non-singular distribution in the form
\begin{equation}
\label{eq:nonSing}
\frac{1}{\sigma_0}\frac{\text{d}\hat{\sigma}_{\rm ns}}{\text{d}\tau}(\tau) = e^{- \frac{2\delta(R,\mu_s)}{Q} \frac{\partial}{\partial \tau}} f(\tau,Q,\mu_{\rm ns})\,.
\end{equation}
The global implementation of the gap subtraction dependent exponential derivative is associated to the fact that the singular as well as the non-singular distributions are convolved with the non-perturbative shape function $F_\tau$, see also Sec.~\ref{subsec:nonpertCorr}.
This global treatment ensures that outside the dijet regime (where the summation of logarithms is smoothly switched of) the singular and non-singular distributions can properly combine to the complete fixed-order result. This is an essential aspect of the matching of the singular distribution with the full fixed-order results valid outside the dijet regime.\footnote{In principle, for the non-singular distribution one could choose different values for the scales $R$ and $\mu_s$ for the gap subtraction series as long as they merge with the corresponding choice for the singular distribution. Such alternative choices, however, only have a very small effect in the dijet region where the singular distributions dominates.} We emphasize that the gap subtraction implemented in Eq.~(\ref{eq:nonSing}) is motivated by the requirement of smoothly matching the factorized treatment in the dijet limit to the fixed-order regime for larger $\tau$ values outside the dijet regime. Its effects concerning the subtraction of the ${\cal O}(\Lambda_{\rm QCD})$ IR renormalon are power suppressed in dijet regime and do not upset the renormalon subtraction in the singular contribution. Outside the validity of the dijet regime the gap subtraction is part of the model treatment of non-perturbative corrections through the shape function $F_\tau$, see the discussion after Eq.~(\ref{eq:factorizationFormula}).
Later on we will assess the impact of this assumption by modifying the power correction used for the non-singular part of the distribution, see Sec.~\ref{subsec:3jetuncertainties}.

Compared to the treatment of the non-singular function in Refs.~\citep{Abbate:2010xh,Abbate:2012jh}, we have implemented some improvements, which are discussed in the following. As already mentioned, the function $f(\tau,Q,\mu_{\rm ns})$ denotes the partonic fixed-order distribution with the singular terms subtracted. It is written as a fixed-order series in powers of the strong coupling (with coefficients being functions of $\tau$, $Q$ and $\mu_{\rm ns}$) and does not contain a summation of logarithms apart from those contained in the strong coupling value:
\begin{equation}
\label{eq:fNonSing}
f(\tau,Q,\mu_{\rm ns}) = \sum_{n=1}\biggl[\frac{\alpha_s(\mu_{\rm ns})}{2\pi}\biggr]^n\sum_{i=0}^{n-1}f_{ni}(\tau)\log^i\biggl(\frac{\mu_{\rm ns}}{Q}\biggr)\,.
\end{equation}
The functions $f_{n,i>0}(\tau)$ can be expressed in terms of $f_{n0}(\tau)\equiv f_n(\tau)$ and the coefficients of the QCD beta function. Adapting Eq.~(3.5) of Ref.~\cite{Mateu:2017hlz} one can find the following compact recursion formula:
\begin{align}
f_{nk}(\tau) =\, & \frac{1}{k}\sum_{i=k}^{n-1} \frac{i\beta_{n-i-1}}{2^{n-i-1}}\,f_{i,k-1}(\tau)\,,\\
\mu \frac{{\rm d} \alpha_s(\mu)}{{\rm d} \mu} =\, \beta_{\rm QCD}[\alpha_s(\mu)] =\, & -\!2\alpha_s(\mu)\sum_{n=1}\beta_{n-1}\biggl[\frac{\alpha_s(\mu)}{4\pi}\biggr]^n\,,\nonumber
\end{align}
where for completeness and to set up the notation followed in this article, we have also shown the perturbative expansion of the QCD beta function.

We start with the $\mathcal{O}(\alpha_s)$ non-singular function. It can be written down analytically and reads~\cite{Ellis:1980wv}
\begin{align}\label{eq:OasNonSing}
f_1(\tau) =\,& \frac{4}{3 \tau }\biggl[3 (\tau +1) (3 \tau -1)+2 \biggl(3 \tau -\frac{2}{1-\tau }\biggr)\! \log \biggl(\frac{\tau }{1-2 \tau }\biggr)\biggr]\theta(1-3\tau)
\\
&+ \frac{4}{3\tau} [3+4\log(\tau)]\,.\nonumber
\end{align}
Whereas it is continuous at $\tau=1/3$, its derivative is not. The result for the cumulative of $f_1$ reads:
\begin{figure}[t!]
\centering
\begin{subfigure}[b]{0.49\textwidth}
\includegraphics[width=\textwidth]{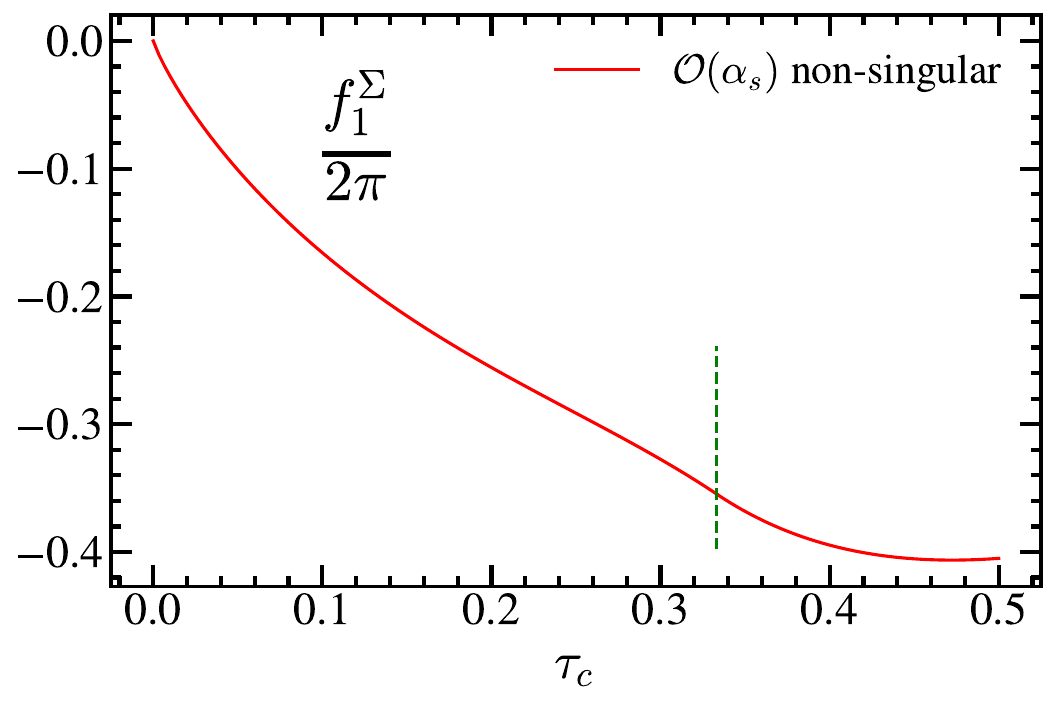}
\caption{\label{fig:fSigma1}}
\end{subfigure}
~
\begin{subfigure}[b]{0.47\textwidth}
\includegraphics[width=\textwidth]{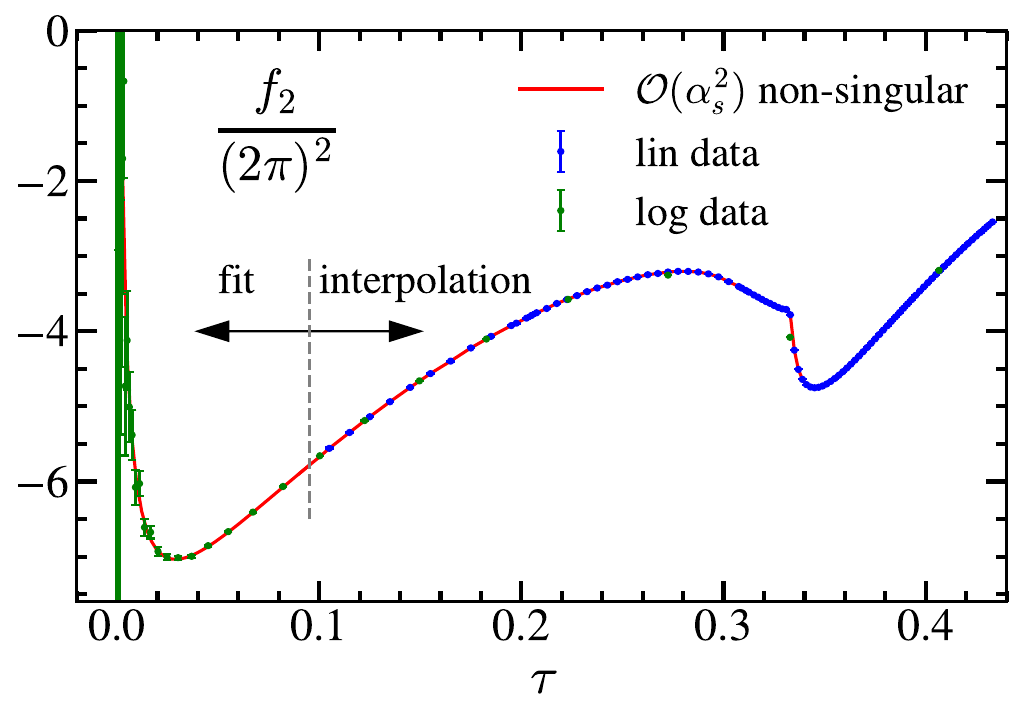}
\caption{\label{fig:NS2loop}}
\end{subfigure}
\caption{\label{fig:fSigma12}
Panel (a): $\mathcal{O}(\alpha_s)$ cumulative non-singular thrust distribution. Note that it is continuous and smooth at the $\tau=1/3$ upper limit of ${\cal O}(\alpha_s)$ fixed-order result. For $\tau>1/3$ the sum of singular and non-singular cumulative contributions add to a constant.
Panel (b): Non-singular differential distribution at $\mathcal{O}(\alpha_s^2)$. Blue and green dots with error bars show {\sc event2} data with linear and logarithmic binning, while the red curve is our ${\cal O}(\alpha_s^2)$ nonsingular result.}
\end{figure}
\begin{align}
f_1^\Sigma(\tau) \equiv\,& \int_0^\tau {\rm d}\tau' f_1(\tau')\\
=\,& \frac{2}{9} \Bigl\{9 \tau (3 \tau +4)-24\bigl[\text{Li}_2(1-\tau )+ \text{Li}_2(2 \tau )+ \,\text{Li}_2(2 \tau -1)\bigr]\nonumber\\
&-6 \log (1-2 \tau ) [6 \tau
+4 \log (2-2 \tau )-3]+36 \tau \log (\tau )+2 \pi ^2\Bigr\}\theta(1-3\tau)\nonumber\\
& + \frac{2}{9} \bigl\{6 \log (\tau ) [2 \log (\tau )+3] + 15 - 2 \pi ^2\bigr\}\theta(3\tau-1)\,,\nonumber
\end{align}
which is continuous and smooth at $\tau=1/3$.

In Refs.~\cite{Abbate:2010xh,Abbate:2012jh} the ${\cal O}(\alpha_s^2)$ non-singular function $f_2$ was determined numerically from {\sc event2}~\cite{Catani:1996jh,Catani:1996vz} histograms. Here we have improved the parametrization of $f_2$ by employing much higher statistics, using logarithmic binning for $\tau<0.1$ and linear binning (with bin size $\Delta\tau = 0.001$) for $0.1<\tau<0.414214$. We use runs with a total of $3 \times 10^{11}$ events and an infrared cutoff $y_0=10^{-8}$. In the regions of linear binning, the statistical uncertainties are quite small for $\tau\gtrsim 0.1$. So for $\tau>0.095\equiv t_{\rm cut}$ we use a numerical interpolation for $f_2(\tau)$ without uncertainties, composed of $102$ nodes which result from combining the finer binning described above. For $\tau<0.095$ we use the ansatz $f^{\rm fit}_2(\tau) = g(\tau/\tau_{\rm cut})$, with $g(\tau)=\sum_{i=0}^3 a_i \log^i \tau + a_4\,\tau \log^3\tau$ and fit the coefficients from the {\sc event2} output taking into account statistical uncertainties and including the constraint that the total fixed-order cross section reproduces the known $\ord{\alpha_s^2}$ coefficient, what permits expressing $a_0$ in terms of the rest of fit coefficients. Moreover, we fix $a_3=-8/3$ as determined by the leading logarithms at next-to-leading power in the thrust distribution~\cite{Moult:2016fqy}. For the fit we discard bins with $\tau < 6.8\times10^{-6}$ as otherwise cutoff effects become sizable, causing a rather large reduced $\chi^2$ (even though the best-fit values are barely affected). After dropping these bins we obtain $\chi^2_{\rm min}/{\rm d.o.f.}=1.22$, obtaining as a final result for the ${\cal O}(\alpha_s^2)$ non-singular function for $\tau<0.095$ the form
\begin{equation}
\label{eq:f2final}
f_2(\tau<0.095) = f^{\rm fit}_2(\tau)\,+\epsilon_2\, \delta f_2^{\rm fit}(\tau) \,.
\end{equation}
Here $f^{\rm fit}_2(\tau)$ stands for the best-fit result and $\delta f_2^{\rm fit}(\tau)$ represents the $1$-$\sigma$ error function for the fit.
The variable $\epsilon_2$ is continuously varied in the range $[-1,1]$ during our theory scans in order to account for the error in the non-singular function. The result is displayed in Fig.~\ref{fig:NS2loop}, where the error function is too small to be seen.

\begin{figure}[t!]
\centering
\begin{subfigure}[b]{0.48\textwidth}
\includegraphics[width=\textwidth]{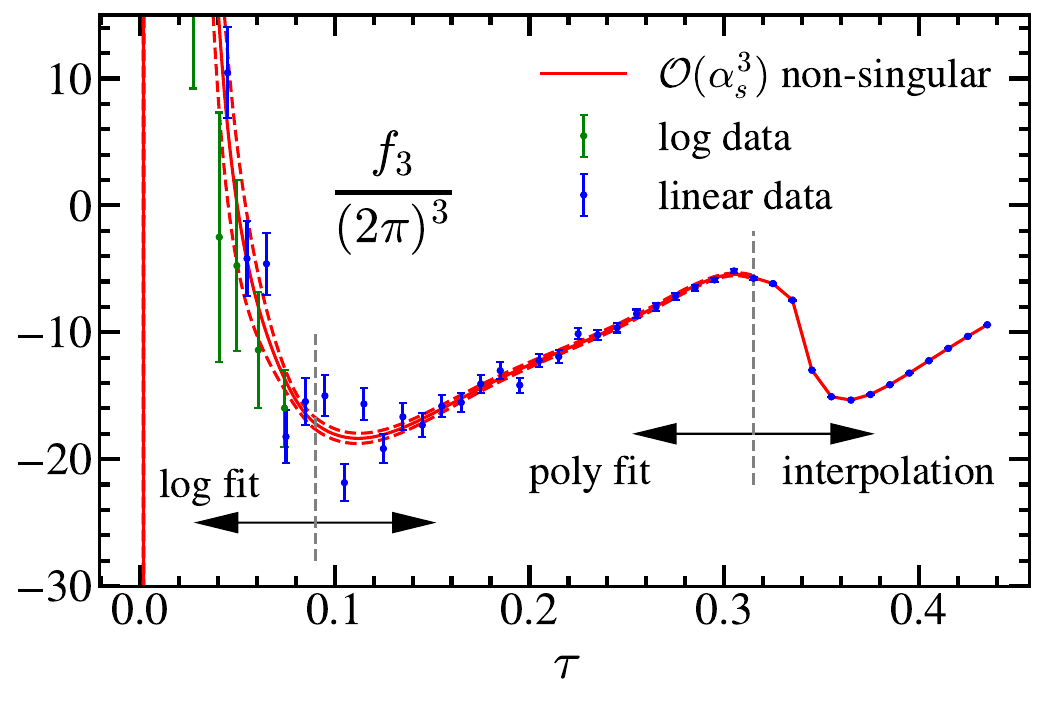}
\caption{\label{fig:fSigma2}}
\end{subfigure}
~
\begin{subfigure}[b]{0.48\textwidth}
\includegraphics[width=\textwidth]{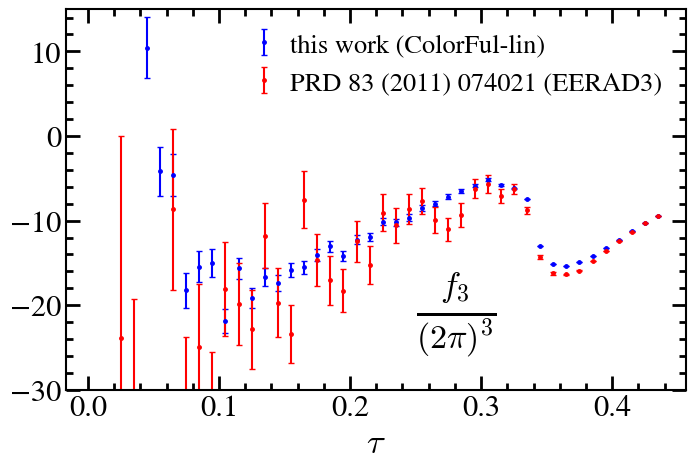}
\caption{\label{fig:NS3comp}}
\end{subfigure}
\caption{\label{fig:threeloopNS}
Non-singular thrust distribution at $\mathcal{O}(\alpha_s^3)$.
a) Fixed-order data from \mbox{\sc CoLoRFulNNLO}, given in blue and green for linear and logarithmic binning, respectively, is shown with a bin size of $0.01$ for the latter. The dashed red lines correspond to the $1$-$\sigma$ error function for the fit, employed below $\tau<0.315$. b) Comparison of the fixed-order data used in this analysis (blue) and in the 2010 extraction of Ref. \cite{Abbate:2010xh} (red).}
\end{figure}

At $\mathcal{O}(\alpha_s^3)$ there are several computer codes available to determine the fixed-order distribution: {\sc EERAD3}~\cite{Gehrmann-DeRidder:2009fgd,Gehrmann-DeRidder:2014hxk}, {\sc Mercutio}~\cite{Weinzierl:2008iv,Weinzierl:2009ms} and {\sc CoLoRFulNNLO}~\cite{DelDuca:2016ily}. In contrast to Ref.~\cite{Abbate:2010xh} were {\sc EERAD3} was employed, in the current analysis we use {\sc CoLoRFulNNLO} since it has significantly smaller statistical uncertainties than the other two programs.%
\footnote{In Ref.~\cite{Bell:2023dqs} a high-statistics sample of ${\cal O}(\alpha_s^3)$ data was created using the public {\sc EERAD3} code, and it was pointed out that it does not reproduce the $\alpha_s^3/\tau$ term in the limit $\tau\to 0$, leading to a $f_3(\tau)$ with a mismatched singular term. Interestingly, the results from this high-statistic run are incompatible with {\sc EERAD3} results in 2008 obtained form the authors of Ref.~\cite{Gehrmann-DeRidder:2007nzq}, even at $\tau=0.3$ where both statistical uncertainties are small. The {\sc CoLoRFulNNLO} results in this region are in better agreement with the {\sc EERAD3} results from 2008. The reason for this is unknown to us. We thank C.~Lee for providing us with the numerical results of their high-statistic {\sc EERAD3} run.} To obtain $f_3(\tau)$ we use the output of {\sc CoLoRFulNNLO} for the thrust distribution as provided by the authors of Ref.~\cite{DelDuca:2016ily}, which we display in logarithmic and linear bins in Fig.~\ref{fig:fSigma2}.\footnote{The CoLoRFulNNLO histograms (either linearly or logarithmically binned) cluster the thrust differential distribution times the thrust value event by event, therefore each bin $\tau_i<\tau<\tau_j$ corresponds to
\begin{equation}\label{eq:ColorFit}
{\rm bin}(\tau_i,\tau_j) = \frac{1}{\tau_j-\tau_i}\int_{\tau_i}^{\tau_j} {\rm d}\tau\,\tau \frac{{\rm d}\sigma}{{\rm d}\tau}(\tau)\approx \frac{\tau_i+\tau_j}{2}\frac{{\rm d}\sigma}{{\rm d}\tau}\biggl(\frac{\tau_i+\tau_j}{2}\biggr)\,.
\end{equation}
To obtain the best estimate for the differential thrust distribution we divide the value of the bin by the average of the bin boundaries.}
Whereas the size of the {\sc CoLoRFulNNLO} linear bins is $\Delta\tau=0.01$, the logarithmic binning used originally in Ref.~\cite{DelDuca:2016ily} had a bin size $\Delta[\log(\tau)] = 0.1$. Since for small $\tau$ statistical errors are huge, we have reclustered
neighboring bins, doubling the logarithmic bin-size. The {\sc CoLoRFulNNLO} numerical results at ${\cal O}(\alpha_s^3)$ yield larger statistical uncertainties than the {\sc event2} results at ${\cal O}(\alpha_s^2)$. We therefore employ for $\tau<t_1=0.315$ two different fit functions: a polynomial in powers of $\log(\tau/t_0)$ (to which we add the highest logarithm at next-to-next-to-leading power) for $\tau < t_0$, and a polynomial in powers of $\tau-t_0$ for $t_0<\tau < t_1$, with $t_0=0.09$, whereas for $\tau>t_1$ we use an interpolation without uncertainties.
The two fit functions smoothly join at $\tau = t_0$, adopting the following parametrization:
\begin{equation}
\label{eq:bincolorfulnnlo}
f_3^{\rm fit}(\tau) = \theta(t_0 - \tau)\Biggl[\sum_{i=0}^5 \ell_i \log^i\biggl(\frac{\tau}{t_0}\biggr)+\ell_6\, \tau\log^5\biggl(\frac{\tau}{t_0}\biggr)\Biggr]\!+
\theta(\tau - t_0) \sum_{i=0}^n p_i (\tau- t_0)^{i}\,,
\end{equation}
with $p_0=\ell_0$ and $p_1=t_0 \ell_1$ to guarantee the function and its first derivative being continuous at $\tau = t_0$. Furthermore, we fix $\ell_5 = -592/81$, which can be determined from the known expression for the leading logarithms at next-to-leading power in the thrust distribution at this order~\cite{ Moult:2018jjd,Moult:2019uhz} (see also~\cite{Beneke:2022obx}). We also express the value of $\ell_2$ in terms of the rest of fit parameters such that the known result for the fixed-order total cross section is reproduced upon integrating the singular and non-singular terms. The other coefficients, namely $\ell_3$, $\ell_4$, $\ell_6$ and $p_i$ with $0\leq i \leq n$ are determined minimizing a $\chi^2$ function that compares the fit function in Eq.~\eqref{eq:ColorFit} to logarithmic
and linear
{\sc CoLoRFulNNLO} output as shown in Fig.~\ref{fig:NS3comp}.
In both datasets, we do not consider values of $\tau<0.0074$ since these might be severely affected by cutoff artifacts. Furthermore, for the logarithmically clustered results we exclude from the fit data with $\tau > t_0$ since the resulting bin size is too large to be well approximated by the corresponding differential distribution, see Eq.~(\ref{eq:bincolorfulnnlo}).
We find that choosing
$n=6$ for the second sum in Eq.~(\ref{eq:bincolorfulnnlo}) yields a reduced $\chi^2$ of the order of $\hat \chi_{\rm min}^2 = 2.2$, producing a smooth-looking result when transitioning from the fit function to the interpolation. Varying the value of $t_0$ by an amount $\mathcal{O}(0.01)$ or the parameter $n$ by one unit does not change the result in a significant way.
All in all, the final result for the three-loop non-singular cross section function can once again be written in the form
\begin{equation}
f_3(\tau<0.315) = f_3^{\rm fit}(\tau)+\epsilon_3\, \delta f_3^{\rm fit}(\tau) \,,
\end{equation}
where $f_3^{\rm fit}(\tau)$ is the best-fit function and $\delta f_3^{\rm fit}(\tau)$ gives the $1$-$\sigma$ error function for the fit, rescaled such that the $\chi^2_{\rm min}/{\rm d.o.f.}$ is exactly one.\footnote{To be consistent, we apply this same rescaling at ${\cal O}(\alpha_s^2)$, which has very little impact on the uncertainties.}
This is a conservative approach and we find it justified because the {\sc CoLoRFulNNLO} bins for small $\tau$ might suffer from cutoff effects. The error function is also piece-wise, and if the uncertainties in the various coefficients are carefully propagated, taking into account the full covariance matrix, it turns out to be smooth at $\tau = t_0$. This is a sanity check on our determination of $\delta f_3^{\rm fit}(\tau)$. Exactly as we did for the $\ord{\alpha_s^2}$ distribution, $\epsilon_3$ is
continuously varied to generate theoretical uncertainty bands. Our final result for the $\mathcal{O}(\alpha_s^3)$ non-singular thrust distribution is also depicted in Fig.~\ref{fig:fSigma2}.

While the update on the $\mathcal{O}(\alpha_s^2)$ non-singular parametrization compared to that used in Ref.~\cite{Abbate:2010xh} (which also employed quite high statistics in the corresponding {\sc event2} runs) has very little effect on the differential cross section, the new parametrization of the $\mathcal{O}(\alpha_s^3)$ non-singular contribution based on {\sc CoLoRFulNNLO} analysis causes a more significant change to the 3-loop nonsingular result, and the statistical uncertainty on the fit-function is significantly reduced.
Furthermore, the new central curve has a more natural behavior as compared to the fit function in Ref.~\cite{Abbate:2010xh} (see their Fig.~5c), which exhibited a very pronounced dip into negative values for $\tau<0.1$. The new central curve follows the more precise {\sc CoLoRFulNNLO} binned results which show a less oscillatory behavior, see Fig.~\ref{fig:NS3comp}. The impact of the updated 3-loop non-singular parametrization on the extraction of the strong coupling is
$\Delta\alpha_s(m_Z) = +0.0001$ and $+0.0006$, where the former is for a fit in the range $\tau \in [(6\,{\rm GeV})/Q,0.33]$ and the latter for a fit in the range $\tau\in [(6\,{\rm GeV})/Q,0.15]$. These changes are compatible with the discussed changes to the ${\cal O}(\alpha_s^3)$ nonsingular distribution, which are predominantly in the small $\tau$ region.

\subsection{Non-Perturbative Corrections}
\label{subsec:nonpertCorr}
In this section, we briefly review the most important conceptual aspects on the treatment of non-perturbative effects for the thrust distribution in the dijet regime.

It can be proven that in the dijet regime the leading non-perturbative effects arise from the infrared dynamics of large-angle soft radiation which can be factorized from the hard and the collinear dynamics~\cite{Korchemsky:1999kt,Hoang:2007vb,Abbate:2010xh}. The effects of large-angle soft radiation are encoded in the thrust soft function which is defined in terms of a vacuum matrix element momentum distribution function of soft Wilson lines of the form
\begin{equation}
\label{eq:thrustSoftFunc}
S_{\tau}(k,\mu) = \frac{1}{N_c}\langle 0 | \text{tr} ~ \overline{Y}_{\bar{n}}^T Y_n
\delta (k- i \hat \partial) Y_n^{\dagger}\overline{Y}_{\bar{n}}^{*} | 0 \rangle_\mu \,,
\end{equation}
which contains perturbative as well as non-perturbative dynamics.
The subscript $\mu$ on the matrix element, indicates that renormalization is carried out in the $\overline{\rm MS}$ scheme.
In particular, the scale-dependence of the thrust soft function is described by perturbation theory for sufficiently large renormalization scales $\mu$. Here
\begin{equation}
\label{eq:ideltaHat}
i\hat{\partial} \equiv \theta (i \bar{n} \cdot \partial - i n \cdot \partial)in \cdot \partial + \theta (i n \cdot \partial - i \bar{n} \cdot \partial)i \bar{n} \cdot \partial\,.
\end{equation}

To provide a convenient practical implementation of the perturbative and nonperturbative aspects of the total soft function, one can consider the factorization ansatz~\cite{Hoang:2007vb,Korchemsky:2000kp,Ligeti:2008ac}
\begin{equation}
\label{eq:factSoftFunc}
S_{\tau}(k,\mu) = \int_0^k \text{d}k^{\prime} \, \hat S_{\tau}(k-k^{\prime}, \mu) F_\tau(k^{\prime})\,,
\end{equation}
which separates the partonic soft function $\hat S_{\tau}$, computable in perturbation theory, and the non-perturbative shape function $F_\tau$. This factorization can be derived
from Eq.~\eqref{eq:thrustSoftFunc} assuming a scale hierarchy between non-perturbative and perturbative soft radiation. The model-character of the convolution in Eq.~(\ref{eq:factSoftFunc}) lies in the functional form of the shape function, which requires some parametrization. In the peak region of the thrust distribution, where we have $k\sim Q\tau\sim {\cal O}(\Lambda_{\rm QCD})$, information on the entire form of the shape function is needed. The canonical approach to define this factorization is to strictly expand in the hierarchies and to use dimensional regularization
and minimal subtraction to handle UV divergences, thus defining coefficients and perturbative and non-perturbative matrix elements in the $\overline{\rm MS}$ scheme.
Compatibility with the OPE implies the normalization
\begin{equation}
\label{eq:zeroMoment}
\int_0^\infty \text{d}k\: F_\tau(k) = 1\,.
\end{equation}

The form of Eq.~(\ref{eq:factSoftFunc}) is particularly useful to understand how the OPE arises when moving into the dijet tail region characterized by the hierarchy \mbox{$Q\sqrt{\tau}\gg k \sim Q\tau \gg {\cal O}(\Lambda_{\rm QCD})$}. Since the interval of the convolution exceeds the size of the non-perturbative momenta one can multipole expand the shape function within the integral in the form
\begin{equation}
\label{eq:opeshape}
F_\tau(k) = \delta(k) + \sum_{i=1}^\infty 2^i \bar\Omega_i\, \delta^{(i)}(k)\,,
\end{equation}
where $\bar\Omega_i$ are non-perturbative vacuum matrix elements, defined in the $\overline{\rm MS}$ scheme, that scale as ${\cal O}(\Lambda_{\rm QCD}^i)$, and $\delta^{(i)}$ is the $i$-th derivative of the Dirac $\delta$-function. The factors of $2^i$ are conventional and motivated to quantify the non-perturbative effects arising from one of the two hemispheres that are defined from the thrust variable.

The dominant non-perturbative correction is given by the matrix element $\bar\Omega_1$, defined as
\begin{equation}
\label{eq:MSbarOmega1}
\bar{\Omega}_1 = \frac{1}{N_c} \langle 0 | \text{tr} ~ \overline{Y}_{\bar{n}}^T(0) Y_n (0) \hat {\cal E}_T(0) Y_n^{\dagger}(0)\overline{Y}_{\bar{n}}^{*}(0) | 0 \rangle\,.
\end{equation}
It is related to the first moment of the shape function, while the $\bar\Omega_i$ are related to higher moments,
\begin{equation}
\label{eq:Om1Moment}
\bar{\Omega}_i = \int_0^\infty \text{d}k \, \biggl(\frac{k}{2}\biggr)^{\!i} F_\tau(k)\,.
\end{equation}
The transverse energy flow operator $\hat {\cal E}_T$ appearing in Eq.~(\ref{eq:MSbarOmega1}), is defined by its action on states~\cite{Lee:2006nr}
\begin{equation}
\label{eq:TEnergyOperator}
\hat {\cal E}_T (\eta)| X \rangle = \sum_{i \in X} p^{\perp}_i \delta(\eta- \eta_i) | X \rangle \,.
\end{equation}
where $p_i^\perp$ and $\eta_i$ are transverse momenta and pseudo-rapidities relative to the jet axis.

Upon insertion into Eq.~\eqref{eq:factSoftFunc} this provides the OPE of the total soft function
in the dijet tail region. Up to subleading terms suppressed by ${\cal O}(\Lambda_{\rm QCD}^3/k^3)$
this yields the approximate relation
\begin{equation}
\label{eq:softFuncOPEafterMatching}
S_{\tau}(k,\mu) = \hat S_{\tau}(k,\mu)- \frac{\text{d} \hat S_{\tau}(k,\mu)}{\text{d} k} 2 \bar{\Omega}_1 + \mathcal{O}\biggl(\frac{\Lambda^2_{\rm QCD}}{k^3}\biggr)
= \hat S_{\tau}(k - 2\bar\Omega_1,\mu)+ \mathcal{O}\biggl(\frac{\Lambda^2_{\rm QCD}}{k^3}\biggr)\,.
\end{equation}
This relation is also the basis of other approaches to implement non-perturbative corrections to the thrust distribution, which are unrelated to the more rigorous shape function and the factorization formula of Eq.~(\ref{eq:factSoftFunc}).

A widely used model of this kind is the so-called dispersive or effective coupling model~\cite{Dokshitzer:1995zt}, which is roughly related to constant effective QCD coupling effects for scales below a cutoff adopted to be $2$\,GeV. The effective coupling value integrated up to the cutoff then adopts the role of the first linear power correction. Thus this dispersive coupling approach is effectively equivalent to the implementation of an IR cutoff in the perturbative QCD corrections (which also removes the running coupling effects, that are also responsible for IR renormalons~\cite{Davison:2009wzs}). In fact, dijet QCD factorization properties of the thrust distribution ensure that the leading linear sensitivity related to any modification of the perturbation thrust distribution on IR momenta can be parametrized in the form of Eq.~(\ref{eq:softFuncOPEafterMatching}), therefore also the dispersive coupling model can serve as a viable parametrization of non-perturbative corrections in the dijet thrust region~\cite{Manohar:1994kq,Webber:1994cp,Dokshitzer:1995zt,Akhoury:1995sp,Korchemsky:1994is,Nason:1995np,MovillaFernandez:2001ed,Korchemsky:1999kt,Gardi:2001ny,Hoang:2007vb,Ligeti:2008ac,Abbate:2010xh}.
It should be noticed, however, that the dispersive coupling approach or other implementations for linear power models, can be unreliable for thrust values closer to the peak position since there the OPE terms beyond ${\cal O}(\Lambda_{\rm QCD})$ become important.
Thus using the shape function convolution in Eq.~(\ref{eq:factorizationFormula}) to implement non-perturbative corrections in principle represents the more dependable approach for high precision $\alpha_s$ determinations, as it does not rely on specifying $\tau$ dijet interval, where Eq.~(\ref{eq:softFuncOPEafterMatching}) provides a reliable approximation.
We refer to the end of Sec.~\ref{subsubsec:reviewGap}, where we further elaborate on this aspect in the context of the concrete analytic ansatz we use for the shape function in our analysis.

We note that sometimes in the literature (see e.g.\ Ref.~\cite{ParticleDataGroup:2022pth}) the shape function formalism as well as the less rigorous dispersive linear power model have been collectively called `{\it analytic models for non-perturbative corrections}'. We stress that the shape function formalism is, however, more rigorous, and that its model character primarily refers to the form of the shape function and the ansatz for its parametrization. This is in analogy to the parton distribution functions. In contrast, for the dispersive coupling model the approach itself has model character since it is conceptually inequivalent to the form of Eq.~(\ref{eq:factSoftFunc}). Furthermore, the shape function formalism only provides a description of non-perturbative corrections for the observables where it can be derived from first principles, in contrast to the dispersive model, which connects hadronization corrections among completely unrelated observables.

Let us now address the particular global form of the shape function implementation shown in Eq.~(\ref{eq:factorizationFormula}). From a rigorous field theoretic perspective, the convolution of the perturbative contributions and the shape function can only be proven for the singular perturbative thrust distribution $\text{d} \hat{\sigma}_{\rm s}/\text{d}\tau$ given in detail in Eq.~(\ref{eq:singFactorizationFormula}). The global convolution, treating the non-singular distribution $\text{d} \hat{\sigma}_{\rm ns}/\text{d}\tau$ on the same footing and yielding a `hadron level' non-singular contribution of the form
\begin{equation}
\label{eq:Om1nonSing}
\frac{\text{d} \sigma_{\rm ns}}{\text{d}\tau} = \int \text{d}k \, \frac{\text{d} \hat{\sigma}_{\rm ns}}{\text{d}\tau} \biggl( \tau - \frac{k}{Q}, \frac{\mu_{\rm ns}}{Q} \biggr) F_\tau[k- 2\bar\Delta(R,\mu_s)]\,,
\end{equation}
is motivated by the practical requirement that in the far-tail region (outside the dijet-dominated regime and where the resummation in the singular distribution is smoothly switched off in our approach) the perturbative terms in the sum $\text{d} \hat{\sigma}_{\rm s} / \text{d}\tau + \text{d} \hat{\sigma}_{\rm ns} / \text{d}\tau$ must properly combine to exactly reproduce the fixed-order result. In the far-tail region, and in particular close to the large $\tau$ endpoint, the cancellation between singular and non-singular contributions is dramatic since the purely logarithmic terms contained in the former do not contain any information on that endpoint region. This approach then also implies the analogous global implementation of the ${\cal O}(\Lambda_{\rm QCD})$ renormalon gap subtraction already mentioned in Sec.~\ref{subsec:non-singular}. Interestingly, this global approach to implement the shape function (and the gap subtraction) also has the merit that the shape function does not yield any non-perturbative corrections to the total cross section obtained on integrating Eq.~(\ref{eq:factorizationFormula}) over all thrust values, up to higher order corrections. We will come back to this aspect in Sec.~\ref{sec:normchoice}.

\subsection{Renormalon Subtraction
\label{subsec:gapscheme}}

The perturbative soft function in the $\overline{\rm MS}$ scheme has infrared (IR) renormalons, which imply an equal-sign factorially divergent growing behavior of the partonic soft function at large orders of perturbation theory. Each of the IR renormalons is associated to one of the non-perturbative $\overline{\rm MS}$ matrix elements $\bar\Omega_i$ shown in Eq.~(\ref{eq:opeshape}) which compensates the partonic soft function renormalon behavior through their order-dependent values. The most important is the $\mathcal{O}(\Lambda_{\rm QCD})$ renormalon, which is linearly sensitive to non-perturbative momenta and associated to $\bar\Omega_1$. To eliminate this IR renormalon, which is known to already affect the stability of perturbation theory at lower orders, we switch to a renormalon-free scheme for $\Omega_1$. This is achieved by redefining the $\overline{\rm MS}$-scheme $\bar\Omega_1$ in terms of a renormalon-free $\Omega_1$ plus a perturbative series (called subtraction series) that exactly encodes the renormalon. Upon shifting that series into the partonic soft function the latter (and the entire partonic thrust distribution) is rendered $\mathcal{O}(\Lambda_{\rm QCD})$ renormalon-free. The first consistent implementation of such renormalon-free schemes for $\Omega_1$ was proposed in Ref.~\cite{Hoang:2007vb} and called gap formalism.

In Sec.~\ref{subsubsec:reviewGap} we review the gap formalism and revisit the relations presented in the previous section, but considering a thrust soft function free from the leading renormalon. In Sec.~\ref{subsubsec:gapSchemes} we elaborate on technical details of the implementation of a particular choice for the gap subtraction scheme and discuss different scheme choices.

\subsubsection{Review of Gap Formalism \label{subsubsec:reviewGap}}
We start by considering a particular class of non-perturbative shape functions only having support for $k \ge \Delta$, which is called the gap parameter. We can thus write a shape function of this kind in the form $F_\tau(\Delta,k)\equiv F_\tau(k - 2\Delta)$, where $F_\tau(k')$ has support for all positive values $k'>0$. We note that $F_\tau(\Delta,k)$ should not be literally considered as a shifted version of $F_\tau(k)$, but rather as a shape function form where the non-perturbative gap is made explicit by the parameter $\Delta$. Equality~(\ref{eq:Om1Moment}) for the first moment subsequently takes the particular form
\begin{equation}
\label{eq:identOm1MomentGap}
2 \bar{\Omega}_1 = 2 \Delta +\! \int_0^\infty \text{d}k \, k \, F_\tau(k)\,,
\end{equation}
where the $\mathcal{O}(\Lambda_{\rm QCD})$ renormalon can be considered to be fully contained in the gap parameter and $F_\tau(k)$ is $\mathcal{O}(\Lambda_{\rm QCD})$ renormalon-free. This can be made explicit by rewriting $\Delta$ in terms of a non-perturbative contribution, $\bar{\Delta}(R,\mu_s)$, that is free of the $\mathcal{O}(\Lambda_{\rm QCD})$ renormalon and a perturbative (subtraction) series, $\delta(R,\mu_s)$, which exactly contains the $\mathcal{O}(\Lambda_{\rm QCD})$ renormalon of the partonic soft function $\hat S_{\tau}$. Here $\mu_s$ is the renormalization scale, which is the same as the one used for the partonic soft function series, and $R$ is the subtraction scale of the series which has dimension of energy.
In this context, the scale parameter $R$ can be interpreted as an IR cutoff scale for the perturbative calculation.
Hence, we write
\begin{equation}
\label{eq:deltaGap}
\Delta = \bar{\Delta}(R,\mu_s) + \delta(R,\mu_s)\,,
\end{equation}
so that we can define a renormalon-free matrix element
\begin{equation}
\label{eq:O1renormalonfree}
\Omega_1(R,\mu_s) = \bar{\Omega}_1 - \delta(R,\mu_s)\,.
\end{equation}
Since the choice for an appropriated subtraction series is not unique, adopting a particular form of $\delta(R,\mu_s)$ implies a particular scheme for $\Omega_1(R,\mu_s)$. A discussion on different choices of gap subtraction schemes is presented in Sec.~\ref{subsubsec:gapSchemes}. While the $\overline{\rm MS}$ matrix elements $\bar\Omega_i$ and the gap parameter $\Delta$
are formally renormalization group invariant, the gap subtraction series $\delta(R,\mu_s)$ as well as the non-perturbative quantities $\Omega_1(R,\mu_s)$ and $\bar{\Delta}(R,\mu_s)$ do not have this property, and their renormalization group equations (in $R$ and potentially $\mu_s$) are related through the previous two equalities.
Hence, the $\mathcal{O}(\Lambda_{\rm QCD})$ renormalon-free first moment of $F_\tau(k)$, which is $R$- and $\mu_s$-independent, and the value of the gap parameter
\begin{equation}
\Delta_0\equiv\bar \Delta(R_0,\mu_0)\,,
\end{equation}
at some reference scales $R_0$ and $\mu_0$ also
determines the reference power correction matrix element $\Omega_1(R_0,\mu_0)$.

Considering Eq.~(\ref{eq:deltaGap}), the factorized soft function of Eq.~(\ref{eq:factSoftFunc}) can now be rewritten in the form
\begin{align}\label{eq:factSoftFuncGap}
S_{\tau}(k,\mu_s) &= \int \text{d}k^{\prime}\, \hat S_{\tau}(k-k^{\prime}-2\delta, \mu_s) F_\tau(k^{\prime}-2\bar{\Delta})
\\
&= \int \text{d}k^{\prime} \biggl[ e^{-2\delta \frac{\partial}{\partial k}}\hat S_{\tau}(k-k^{\prime}, \mu_s) \biggr] F_\tau(k^{\prime}-2\bar{\Delta})\,,\nonumber
\end{align}
where, for simplicity, the scale-dependence of $\delta$ and $\bar\Delta$ is suppressed. The exponential operator introduced in the second line of Eq.~(\ref{eq:factSoftFuncGap}) induces the perturbative subtractions in the partonic soft function that cancels the corresponding $\mathcal{O}(\Lambda_{\rm QCD})$ renormalon order-by-order upon re-expansion of all terms in the brackets. As we have already pointed out in the discussion of Eq.~(\ref{eq:Om1nonSing}), to ensure the cancellation of the singular and non-singular terms taking place in the far tail, also the non-singular partonic distribution is convolved with the shape function. This then entails the analogous conversion for the non-singular distribution leading to the expression in Eq.~(\ref{eq:nonSing}) appearing in the main factorization formula~(\ref{eq:factorizationFormula}).

For the description of the thrust distribution all renormalization scales in the factorization formula and thus also the renormalon subtraction scale $R$ become $\tau$-dependent functions, which makes $\bar{\Delta}(R,\mu_s)$ and therefore also the shape function dependent on the thrust value. To provide an unambiguous meaning to the shape function's parameters it is therefore mandatory to define the reference scales $R_0$ and $\mu_0$ for which the concrete functional parametrization of $F_\tau$, which we refer to as $F_\tau(R_0,\mu_0,k)$, is defined. To this end we write
\begin{align}
\label{eq:shapefctDeltadiff}
F_\tau[k-2\bar{\Delta}(R,\mu_s)]=
F_\tau[R_0,\mu_0;k-2\bar{\Delta}(R,R_0,\mu_s,\mu_0)]
\,,
\end{align}
where
\begin{equation}
\bar\Delta(R,R_0,\mu_s,\mu_0) \equiv \bar\Delta(R,\mu_s) - \Delta_0 \,,
\end{equation}
can be determined perturbatively from the $R$ and $\mu_s$ evolution of the subtraction series $\delta(R,\mu_s)$. Thus we also have $\int{\rm d}k\,k\,F_\tau[R_0,\mu_0,k-2\bar{\Delta}(R,R_0,\mu_s,\mu_0)] = 2\Omega_1(R,\mu_s)$ as well as
\begin{equation}
\Omega_1(R,\mu_s) - \Omega_1(R_0,\mu_0) = \bar\Delta(R,R_0,\mu_s,\mu_0)\,.
\end{equation}
Note that, since $\bar\Delta(R,R_0,\mu,\mu_0)$ is perturbative, we can in principle also consider it as part of the partonic distribution by rewriting the factorization formula~(\ref{eq:factorizationFormula}) in the form
\begin{equation}
\label{eq:factorizationFormulav2}
\frac{\text{d}\sigma}{\text{d}\tau} = \int \text{d}k \,\, \frac{\text{d} \hat{\sigma}}{\text{d}\tau} \! \biggl[\tau - \frac{k}{Q} - \frac{2\bar\Delta(R,R_0,\mu_s,\mu_0)}{Q}\biggr] F_\tau(R_0,\mu_0;k )\,,
\end{equation}
where $\text{d}\hat{\sigma}/\text{d}\tau = \text{d}\hat{\sigma}_{\rm s}/\text{d}\tau+\text{d}\hat{\sigma}_{\rm ns}/\text{d}\tau$ stands for the entire partonic differential thrust distribution.

\begin{figure}[t!]
\centering
\includegraphics[width= 0.5\textwidth]{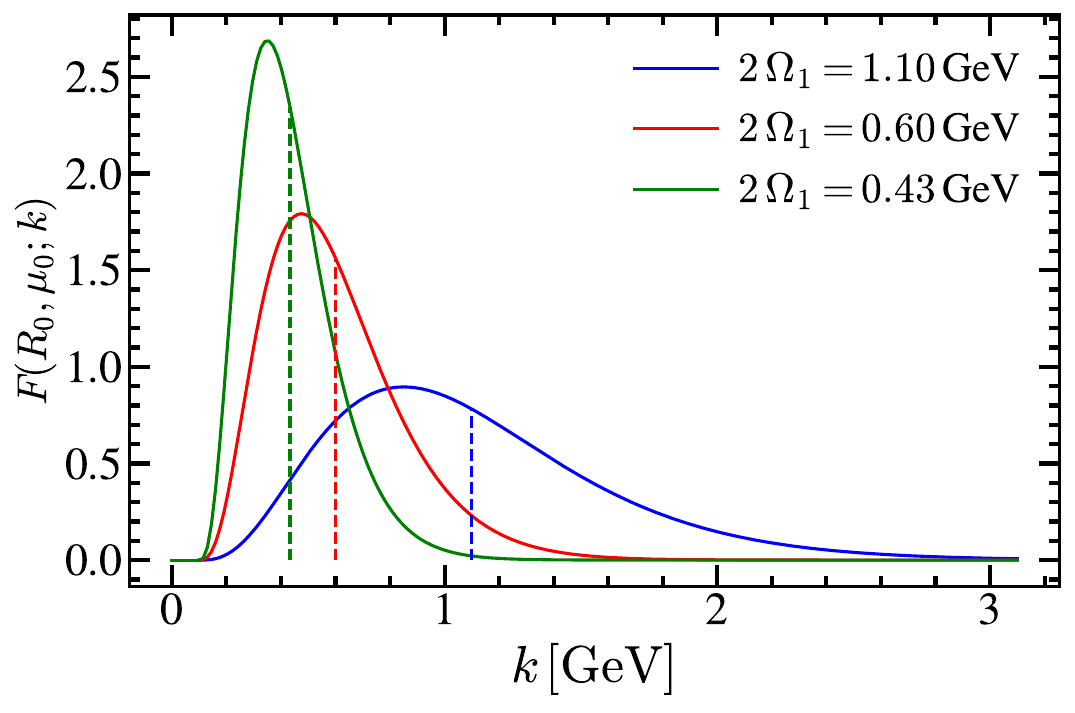}
\caption{\label{fig:Shape} Shape function used in our studies as given in Eq.~\eqref{eq:shapeFun}, for three values of $\Omega_1$: $0.55$\,GeV, $0.3\,$GeV and $0.215\,$GeV in blue, red, and green, respectively. The dashed vertical lines show the position of $2\Omega_1$, which always lies to the right of the maximum.}
\end{figure}
The concrete functional ansatz for the shape function we use in our analysis is the same as employed in the previous analyses of Refs.~\cite{Abbate:2010xh,Abbate:2012jh,Hoang:2014wka, Hoang:2015hka} and reads
\begin{equation}
\label{eq:shapeFun}
F_\tau(R_0,\mu_0;k)=\frac{128 (k-2\Delta_0)^3\, e^{-\frac{4 (k-2\Delta_0)}{\lambda}}}{3 \lambda^4}\,\theta(k-2\Delta_0)\,,
\end{equation}
where $\Delta_0=0.05$\,GeV.
This yields
\begin{align}
\label{eq:2Omega1shapefct}
2\Omega_1(R_0,\mu_0)=\lambda+2\Delta_0 \,,
\end{align}
as the expression for the first moment. Here $\lambda$ is the floating parameter in our fits, and directly yields a result for the desired fit parameter $\Omega_1(R_0,\mu_0)$. As the reference scales we take
\begin{align} \label{eq:R0mu0}
R_0=\mu_0=2\,{\rm GeV} \,.
\end{align}
The shape function ansatz, that can be visualized graphically in Fig.~\ref{fig:Shape} for three values of $\Omega_1$, corresponds to the leading term of a general expansion for an arbitrary shape function first introduced in Ref.~\cite{Ligeti:2008ac}. Since our analysis is focused on the tail region, where the contribution of the leading OPE term $\Omega_1$ dominates, see Sec.~\ref{subsec:nonpertCorr}, this truncated form is sufficient. More terms in this expansion would, however, be mandatory to carry out a proper theory description in the peak region.

Note that the partonic cross sections shown in Eqs.~(\ref{eq:singFactorizationFormula}) and (\ref{eq:nonSing}) were given within a gap subtraction scheme, indicated by the exponential derivative involving the subtraction series $\delta(R,\mu_s)$. The corresponding expression in the \MSbar scheme for the shape function's first moment are obtained by simply setting $\delta=0$ as well as $\bar \Delta(R,R_0,\mu_s,\mu_0) = 0$ in the expressions shown before. The ansatz function for the \MSbar shape function is still given in Eq.~(\ref{eq:shapeFun}), dropping the scales $R_0$ and $\mu_0$ and setting $\Delta_0=0$.

As already mentioned in Sec.~\ref{subsec:nonpertCorr} using the shape function convolution in phenomenological analyses provides in principle a more rigorous description of hadronization effects in the dijet regime in comparison to the shape-function OPE series truncated after the leading $\Omega_1$ correction. However, using the concrete functional one-parameter ansatz for the shape function in Eq.~(\ref{eq:shapeFun}) implies that all higher shape-function moments $\Omega_{n\ge 2}$ are related to $\Omega_1$ in a definite way. Here, we have
$\Omega_2=1.25\,\Omega_1^2 -0.5 \Delta_0\Omega_1+0.25 \Delta_0^2$
for the second shape-function moment, and we do not freely vary the $\Omega_{n\ge 2}$ in our fits. However, their effect is also power-suppressed in the dijet regime and can be accounted for as an uncertainty (since fitting also for $\Omega_{n\ge 2}$ in thrust tail fits is not feasible in practice anyway). The effect of varying $\Omega_2$ independently was studied in Ref.~\cite{Abbate:2010xh}, see their Secs.~IV and VII, where the range $\Omega_1^2\leq\Omega_2\leq 1.5\Omega_1^2$ was obtained from general properties of the shape function having a generic shape as shown in Fig.~\ref{fig:Shape}.
The uncertainty on a strong coupling tail fit was found to be
$[\Delta\alpha_s(m_Z)]_{\Omega_2}=0.0002$, which is much smaller than the perturbative uncertainty.

\subsubsection{Gap schemes
\label{subsubsec:gapSchemes}}

Our default gap scheme, which will be referred to simply as the {\it R-gap scheme}, was introduced in Ref.~\cite{Hoang:2008fs} and employed in the analysis of Ref.~\citep{Abbate:2010xh}. In this scheme, the subtraction series is defined from
\begin{align}
\label{eq:subSeriesHKdeltaHK}
\delta(R,\mu_s) =\,& \frac{R}{2} e^{\gamma_E} \frac{\rm d}{\text{d} \log (ix)} \Big[ \log \tilde S_{\tau}(x,\mu_s) \Big]\Big|_{x=(iRe^{\gamma_E})^{-1}}
\\
=\,& \frac{R}{2} e^{\gamma_E}
\sum_{i=1}\biggl[\frac{\alpha_s(\mu_s)}{4\pi}\biggr]^i\sum_{j=0}^{i}(\,j+1)s_{i,j+1} \log^j\biggl(\frac{\mu_s}{R}\biggr)\,,\nonumber
\end{align}
where $\tilde S_{\tau}(x,\mu_s)$ is the Fourier transform of the $\overline{\rm MS}$ partonic soft function. It can be written as
\begin{align}
\log \tilde S_{\tau}(x,\mu_s) = \sum_{i=1}\biggl[\frac{\alpha_s(\mu_s)}{4\pi}\biggr]^i\sum_{j=0}^{i+1}s_{ij}
\log^j \bigl(ie^{\gamma_E} x \mu_s \bigr)\,,
\end{align}
with $s_i=s_{i0}/2$, and satisfies the following linear RG equation:
\begin{align}\label{eq:RGsoft}
\mu_s\frac{\rm d}{\text{d}\mu_s} \Big[ \log \tilde S_{\tau}(x,\mu_s) \Big] =\,& -\!4\,\Gamma^{\rm cusp}[\alpha_s(\mu_s)] \log\bigl(ie^{\gamma_E} x \mu_s \bigr) + 2\gamma_S[\alpha_s(\mu_s)]\,,\\
\nonumber \Gamma^{\rm cusp}[\alpha_s]=\,&\sum_{n=0}^{\infty}\Gamma^{\rm cusp}_{\!n}\Bigl(\frac{\alpha_s}{4\pi}\Bigr)^{\! n+1}\,,\qquad
\gamma_S[\alpha_s]=\!\sum_{n=0}^{\infty}\gamma^S_n\Bigl(\frac{\alpha_s}{4\pi}\Bigr)^{\! n+1}\,.
\end{align}
Here $\Gamma^{\rm cusp}$ and $\gamma_S$ are the universal cusp and non-cusp soft anomalous dimensions, respectively.
It is convenient to define
\begin{equation}
d^{R}_{ij}=\frac{e^{\gamma_E}}{2}(j+1)s_{i,j+1}\,,
\end{equation}
and worth noting that, since $\gamma^S_0$ vanishes, we have $d^{R}_{10}=0$ such that $\delta(R,R)=\mathcal{O}(\alpha_s^2)$ in the R-gap scheme. All pieces needed to determine the gap subtraction up to 3 loops are analytically known.
In the R-gap scheme, $\delta(R,\mu_s)$ has a linear anomalous dimension in $R$,
\begin{align}\label{eq:RevolGap}
\gamma_R[\alpha_s(R)] =\,& R\frac{\rm d}{\text{d} R} \delta(R,R)
= \sum_{n=0}\biggl[\frac{\alpha_s(R)}{4\pi}\biggr]^{n+1}\gamma^R_n\,,\\
\gamma^R_n =\,& d^{R}_{n+1,0} -
2 \sum_{j = 0}^{n-1} (n-j)\beta_j d^{R}_{n-j,0} \,,\nonumber
\end{align}
and also has a non-zero logarithmic anomalous dimension in~$\mu$
\begin{align}\label{eq:gammaMu}
\gamma_\Delta^\mu[\alpha_s(\mu_s)] =\,&\frac{\mu_s}{R} \frac{\rm d}{\text{d} \mu_s} \delta(R,\mu_s) = -2 e^{\gamma_E}\Gamma^{\rm cusp}[\alpha_s(\mu_s)]\,.
\end{align}
This linear RG evolution in $R$ is common to any gap scheme.
We refer the reader to Sec.~II~F of Ref.~\cite{Abbate:2010xh} and to Ref.~\cite{Hoang:2008fs} for further details.
In order to avoid having large logarithms in the gap subtraction, which is
essential for a proper gap subtraction implementation, one must choose $\mu_s\approx R$. Since the soft renormalization scale $\mu_s$ depends dynamically on $\tau$ and monotonically increases until it becomes equal to the hard scale, $R$ has to become equally large and is therefore also $\tau$-dependent. If these two scales are varied, the numerical values of the gap parameter $\bar\Delta(R,\mu_s)$ and the matrix element $\Omega_1(R,\mu_s)$ vary according to their respective anomalous dimensions with respect to the reference gap parameter $\Delta_0=\bar\Delta(R_0,\mu_0)$.
We reiterate that in order for the renormalon to cancel properly between the subtraction series and the perturbative partonic soft function, one has to consistently expand the partonic cross section including the subtraction series order-by-order in $\alpha_s(\mu_s)$.

As already pointed out in Sec.~\ref{subsubsec:reviewGap}, the choice of the gap subtraction scheme is not unique. In Ref.~\cite{Bachu:2020nqn}, a general classification was provided for possible gap subtraction schemes based on the partonic position-space soft function $\tilde S_{\tau}(x,\mu_s)$. The impact of adopting different gap subtraction schemes within this classification was studied recently in Ref.~\cite{Dehnadi:2023msm} for inclusive event-shape distributions involving boosted top quark production in $e^+e^-$ annihilation and the results provided valuable information concerning the perturbative stability of the gap subtraction formalism. Their study followed the same guidelines for the gap subtraction schemes we apply in this article and yielded excellent consistency among the different renormalon-free descriptions within the remaining perturbative uncertainties. In our analysis we also study the impact of using gap subtraction schemes that differ from our default R-gap scheme defined above, see Sec.~\ref{sec:normchoice}. Following the classification of Ref.~\cite{Bachu:2020nqn}, we also consider the following class of non-derivative $\xi$-dependent gap schemes:
\begin{align}\label{eq:xiGap}
\check{\delta}(R,\xi) = \frac{R}{2\xi}
\log\biggl[ \tilde S_{\tau}\biggl(\frac{\xi}{iR}, R\biggr)
\biggr] = \frac{R}{2\xi}
\sum_{i=1}\biggl[\frac{\alpha_s(R)}{4\pi}\biggr]^i\sum_{j=0}^{i+1}s_{ij}
\log^j \bigl(e^{\gamma_E} \xi \bigr)\,.
\end{align}
As for the R-gap scheme, we find it convenient to define
\begin{align}
\check{d}_{i0}(\xi)=\frac{1}{2\xi}\sum_{j=0}^{i+1}s_{ij} \log^j \bigl(e^{\gamma_E} \xi \bigr)
\,.
\end{align}
In contrast to the R-gap scheme, the one-loop term of these non-derivative schemes is not zero. Furthermore, at $\mathcal{O}(\alpha_s^3)$ it also depends on the recently computed non-logarithmic term of the soft function $s_3$~\cite{Baranowski:2024vxg}. Since $\check\delta(R,\xi\to\infty)=0$, in this limit effectively one falls back to a setup without renormalon subtractions, that is, the $\overline{\rm MS}$ scheme. In contrast to the R-gap scheme, these non-derivative $\xi$-dependent schemes are $\mu_s$-invariant. To properly cancel the renormalon of the partonic soft function, one has to express the perturbative series of Eq.~\eqref{eq:xiGap} in powers of $\alpha_s(\mu_s)$. This is achieved by expanding $\alpha_s(R)$ in terms of $\alpha_s(\mu_s)$ and $\log(\mu_s/R)$, yielding
\begin{align}
\check{\delta}(R, \xi) =\,& R \sum_{i=1}\biggl[\frac{\alpha_s(\mu_s)}{4\pi}\biggr]^i\sum_{j=0}^{i-1} \check d_{ij}(\xi) \log^j\Bigl(\frac{\mu_s}{R}\Bigr)\,,\\
\check d_{ij}(\xi) =\, &
\frac{2}{j}\sum_{k = j}^{i-1}
k\, \check d_{k,j-1}(\xi)\,\beta_{i-k-1}\,.\nonumber
\end{align}
The last recursive relation generates all $d_{i,j>0}$ starting from the known $d_{i0}$. Relating the renormalon-free leading-power matrix element $\Omega_1$ for different scheme choices is straightforward, and the relations read
\begin{align}
\label{eq:Omega1relations}
\Omega_1^\xi(R) - \Omega_1^{\rm R}(R,R) =\,& \frac{R}{2}
\sum_{i=1}\biggl[\frac{\alpha_s(R)}{4\pi}\biggr]^i
\Biggl[e^{\gamma_E}s_{i1}-\frac{1}{\xi}\sum_{j=0}^{i+1}s_{ij}
\log^j \bigl(e^{\gamma_E} \xi \bigr)\Biggr]\,,\\
\Omega_1^{\xi_2}(R) - \Omega_1^{\xi_1}(R) =\,& \frac{R}{2}
\sum_{i=1}\biggl[\frac{\alpha_s(R)}{4\pi}\biggr]^i
\sum_{j=0}^{i+1}s_{ij}\Biggl[
\frac{\log^j \bigl(e^{\gamma_E} \xi_1 \bigr)}{\xi_1} -
\frac{\log^j \bigl(e^{\gamma_E} \xi_2 \bigr)}{\xi_2}\Biggr]\,.
\nonumber
\end{align}
Here we have expressed both series in powers of the strong coupling at the same scale so that they are renormalon-free.
The R-anomalous dimension for the non-derivative gap scheme can be obtained from Eq.~\eqref{eq:RevolGap} replacing $d_{i0}^{R}\to \check d_{i0}(\xi)$.
For the non-derivative gap schemes the $\mu$-anomalous dimension vanishes.

Note that for the analyses carried out below, when comparing different gap subtraction schemes along with the $\overline{\rm MS}$ scheme, we always convert to the reference value given in the R-gap scheme,
\begin{align}
\Omega_1^R\equiv \Omega_1(R_0,\mu_0) \,,
\end{align}
with the scales given in \eq{R0mu0}. In this regard, we emphasize the fact that when converting $\Omega_1$ from the $\overline{\rm MS}$ scheme to the R-gap scheme the series has a renormalon. This implies that the value for $\Omega_1$ depends more strongly on the scale at which the fixed-order conversion is carried out and the order we are working at, in contrast with the conversion between two different renormalon-free schemes.

It is useful to consider different values of $\xi$ as independent gap schemes, and we shall explore a few values of $\xi$ to test the dependence of our results on the scheme used to define $\Omega_1$. The phenomenological analysis of Ref.~\cite{Bachu:2020nqn}, which was based on the peak of the distribution for boosted tops, employed $\xi=1$. Ref.~\cite{Dehnadi:2023msm} studied the R-gap scheme and the non-derivative schemes for $\xi=1$ as well as $\xi=e^{5\gamma_E}$ for calibrating the top quark mass parameters for different Monte Carlo event generators.
For the massless quark tail analysis carried out in this article, it turns out those values of $\xi$ are too small, implying artificially enhanced 1-loop subtraction and R-evolution.\footnote{This is because in the application to boosted top quark production in the resonance region the interplay between large-angle soft and the ultra-collinear radiation differs from thrust in the dijet tail region for massless quark production.} Here we explore the three (larger) values: $\xi_1 = e^{10\gamma_E}$, $\xi_2 = e^{10\gamma_E}/2$ and $\xi_3 = e^{10\gamma_E}/3$. When comparing results between different gap schemes below, we will use the convention: $\xi_1$-gap, $\xi_2$-gap, $\xi_3$-gap denoting these three values. The numerical values for the coefficients $\check d_{i0}(\xi)$ for $i=1,2,3$ in these three gap schemes, together with the values for $d_{i0}^{R}$, are listed in Table~\ref{tab:di0}.

\begin{table}[t!]
\centering
\begin{tabular}{c | c | c | c | c}
& R-gap & $\xi_1$-gap & $\xi_2$-gap & $\xi_3$-gap \\ [0.5ex]
\hline
$d_{10}$ & $0$ & $-0.69$ & $-2.21$ & $-1.43$ \\
$d_{20}$ & $-43.954$ & $-26.96$ & $-78.35$ & $-47.93$ \\
$d_{30}$ & $-1954.45$ & $-1569.44$ & $-2401.11$ & $-4782.98$ \\
\end{tabular}
\caption{Values of $d_{i0}$ for $i=1,2,3$ in the different gap schemes we consider.
All four schemes have coefficients whose growth is consistent with a ${\cal O}(\Lambda_{\rm QCD})$ renormalon.}
\label{tab:di0}
\end{table}

In Ref.~\cite{Bell:2023dqs} the impact of adopting different gap subtraction schemes was first discussed in the context of $\alpha_s$ fits. They employed the R-gap scheme of Eq.~(\ref{eq:subSeriesHKdeltaHK}) in two different ways. One, called \mbox{$R$-scheme} in Ref.~\cite{Bell:2023dqs}, follows our standard of systematically summing all logarithms. The other is based on the $\mu_s$-independent subtraction series
\begin{equation} \label{eq:deltaRR}
\delta(R,R) = \frac{R}{2} e^{\gamma_E}
\sum_{i=2}\biggl[\frac{\alpha_s(R)}{4\pi}\biggr]^is_{i1} \,,
\end{equation}
which is the R-gap subtraction series for $\mu_s=R$. To cancel the renormalon one needs to re-expand $\alpha_s(R)$ in terms of $\alpha_s(\mu_s)$ and $\log(\mu_s/R)$, which generates a single power of the logarithm at $\mathcal{O}(\alpha_s^3)$. For this subtraction, the authors of Ref.~\cite{Bell:2023dqs} use a constant $R$ scale in the tail region, which leads to an increasing scale separation between $R$ and $\mu_s$ for large~$\tau$. This use of \Eq{deltaRR} together with the choice of the $R(\tau)$ profile was referred to as the R$^*$-scheme in Ref.~\cite{Bell:2023dqs}. They considered this implementation of the R-gap scheme as a viable alternative to avoid an apparent increase of the non-perturbative corrections in the R-gap scheme with $\tau$ due to the (predominantly linear) $R$-dependence of $\bar\Delta(R,R_0,\mu_s,\mu_0)$. However, as also illustrated in the form of the factorization theorem displayed in Eq.~(\ref{eq:factorizationFormulav2}), this $R$-dependence is a completely perturbative feature emerging from the removal of the soft-function renormalon through a scheme-dependent subtraction, and should not be considered as an $R$-dependence of the genuine non-perturbative QCD corrections. Formally, the $R$-dependence cancels from the prediction up to higher order terms dropped by the perturbative truncation. Within this R$^*$-scheme Ref.~\cite{Bell:2023dqs} found a quite sizable variation of the theoretical predictions with respect to the $R$-scheme when using the 2010 profiles introduced in Ref.~\cite{Abbate:2010xh}, which, for reasons explained in the next section, we do not employ in the analysis of this article. Therefore, for our analysis of scheme dependence we will stick to comparisons of the R-gap scheme with the three \mbox{$\xi_i$-gap} schemes.
Since the summation of large logarithmic terms is essential for a proper convergence of the theory description, the R$^*$-scheme is in general disfavored as compared to our scheme choices.

\subsection{Profile Functions
\label{subsec:profiles}}

The factorization formula for the singular partonic thrust distribution is governed by three renormalization scales: the hard $\mu_H$, jet $\mu_J$, and soft $\mu_s$ scales. In order to avoid having large logarithms in the peak and tail regions, while simultaneously maintaining the cancellation of singular and non-singular terms in the far-tail region, the renormalization scales must satisfy the following canonical scaling constraints
\begin{align}
\text{peak:}& ~~~ \mu_H \sim Q, ~ \mu_J \sim \sqrt{\Lambda_{\rm QCD}Q}, ~ \mu_s \gtrsim \Lambda_{\rm QCD}\,,\\
\text{tail:}& ~~~ \mu_H \sim Q, ~ \mu_J \sim Q \sqrt{\tau}, \qquad~ \mu_s \sim Q\tau\,,\nonumber\\
\text{far-tail:}& ~~~ \mu_H = \mu_J = \mu_s \sim Q\,.\nonumber
\end{align}
For a given region, the phrase canonical scales refers to replacing the above $\sim$'s and $\gtrsim$ by $=$, and $\Lambda_{\rm QCD}$ by $1.1\,{\rm GeV}$.
Note that there is very little freedom in these constraints; they are not model dependent and any valid resummation approach must fulfill them with \mbox{$\tau$-dependent} renormalization scales $\mu_i(\tau)$.
However, to satisfy these requirements across the entire thrust distribution, the three constraints must be smoothly combined, and there is some freedom in how and where this joining is done, which introduces parameters and leads to the concept of {\it profile functions} $\mu_i(\tau)$. It is important to note that the resummed cross section, even if formally renormalization-scale independent, exhibits residual dependence on the details associated to the scale choices order-by-order in resummed perturbation theory.
To assess the perturbative uncertainty in the various regions, we therefore vary both the absolute scales in canonical regions and the way in which we join canonical regions.
The history of profile functions in the literature includes Ref.~\cite{Ligeti:2008ac} where they were used to handle the peak-tail transition in $B\to X_s\gamma$, and the thrust analysis in
Ref.~\cite{Abbate:2010xh} where the joining for all three regions was carried out.
The original profiles of Ref.~\cite{Abbate:2010xh} suffered from a deficiency, in that the canonical scaling in the tail region was only approximately satisfied, at the benefit of having less joining parameters. This issue was rectified for the thrust and C-parameter event shapes in
Ref.~\cite{Hoang:2014wka}, with the introduction of profile functions that satisfy the canonical scaling constraints in all regions, and whose form we also use here.
In the following, we briefly review the functional form of the hard, jet and soft scales encoded in Eq.~(\ref{eq:singFactorizationFormula}).

The hard scale is defined as
\begin{equation}
\label{eq: hardScale}
\mu_H = e_H Q\,,
\end{equation}
where $e_H$ is one of the parameters we vary when determining the perturbative uncertainty in our analysis. The range of variation for $e_H$, as well as the range of variation for all other parameters encoded in the definition of our profile functions are summarized in Table~\ref{tab:profParams}.

The soft scale obeys the functional form
\begin{eqnarray}
\label{eq:softScale}
\mu_s(\tau) = \left\{\begin{array}{lrcl}
\mu_s(0)& 0 &\le& \tau < t_0
\\
\zeta(\mu_s(0),0,0,r_s \mu_H,t_0,t_1,\tau) \qquad &t_0 &\le& \tau < t_1
\\
r_s \mu_H \tau & t_1 &\le& \tau < t_2
\\
\zeta(0,r_s \mu_H,\mu_H,0,t_2,t_s,\tau)\quad &t_2 &\le& \tau < t_s
\\
\mu_H & t_s &\le& \tau < 0.5
\end{array}
\right.
,
\end{eqnarray}
where the first, third, and fifth are canonical scaling regions, and the second and fourth involving the function $\zeta$ are where the transitions occur.
For our analysis we fix $\mu_s(0) = 1.1$\,GeV.
Here $t_0$ defines the boundary between the non-perturbative region (where all of the $\Omega_i$ are important) and the beginning of the transition to the tail resummation region (where an OPE for the $\Omega_i$ corrections applies), and $t_1$ denotes the end of this transition.
In the tail region $t_1 < \tau < t_2$, the parameter $r_s$ corresponds to the linear slope with which the soft scale rises.
Here, $t_2$ controls the start of the transition between the tail resummation and fixed-order regions, and $t_s$ defines the value of $\tau$ where this transition ends and all renormalization scales become equal to a common scale.
The function $\zeta(a_1,b_1,a_2,b_2,t_1,t_2,t)$, whose definition can be found in Ref.~\cite{Hoang:2014wka}, connects the canonical patches of the spectrum in a smooth manner.
This function smoothly joins two linear curves $\ell_1$ and $\ell_2$ defined for $t\leq t_1$ and $t\geq t_2$ with slopes $b_1$ and $b_2$, and intersects $a_1$ and $a_2$. The connection is achieved by means of two second-order polynomials with support on $t_1\leq t \leq (t_1+t_2)/2$ and $(t_1+t_2)/2 \leq t \leq t_2$ which are smoothly connected to each other at $(t_1+t_2)/2$ and to the two linear curves at $t_1$ and $t_2$.

The functional form we use for the jet scale is
\begin{eqnarray}
\label{eq:jetScale}
\mu_J(\tau) = \left\{\begin{array}{ll}
\left[ 1+e_J (\tau-t_s)^2 \right] \sqrt{\mu_H \mu_s(\tau)} &\quad \tau \le t_s
\\
\mu_H & \quad \tau > t_s
\end{array}
\right.
,
\end{eqnarray}
where $e_J$ is one of the parameters we vary when determining the theoretical uncertainty in our fits. Due to its dependence on $\mu_s(\tau)$ it also inherits the proper behavior for the different regions shown in \Eq{softScale}.

In addition to the three scales defined above, there are two additional scales that require a description. The first is $R(\tau)$, which denotes the gap subtraction scale. It agrees with the soft renormalization scale $\mu_s(\tau)$ in the resummation region to avoid large logs in the subtraction series given by the second line of Eq.~\eqref{eq:subSeriesHKdeltaHK}, but it can differ from the soft scale in the non-perturbative region.
The corresponding profile function reads
\begin{eqnarray}
\label{eq:subtractionScale}
R(\tau) = \left\{\begin{array}{lrcl}
R(0) & 0 &\le& \tau < t_0
\\
\zeta(R(0),0,0,r_s \mu_H,t_0,t_1,\tau) \quad &t_0 &\le& \tau < t_1
\\
\mu_s(\tau) & t_1 &\le& \tau \le 0.5
\end{array}
\right.
,
\end{eqnarray}
with $R(0)=0.7\,$GeV.
The values of $\mu_s(0)$ and $R(0)$ are kept fixed in our fit analysis since they only affect the peak region, which is outside our fit window.

Finally, there is also the non-singular renormalization scale $\mu_{\rm ns}$,
whose dependence in the cross section independently decreases order-by-order. In the far-tail region it must be equal to the other scales to ensure the proper (large) cancellation between singular and non-singular terms, which are thus treated as a common entity rather than distinct contributions. In the peak and dijet tail regions the non-singular scale can differ from the other scales, and its variation provides an estimate for missing higher-order perturbative non-singular terms.
We use
\begin{equation}\label{eq:NSScale}
\mu_{\rm ns}(\tau) = \mu_H - \frac{n_s}{2} \big[ \mu_H - \mu_J(\tau) \big] \,.
\end{equation}

\begin{figure}[t!]
\centering
\begin{subfigure}[b]{0.475\textwidth}
\includegraphics[width=\textwidth]{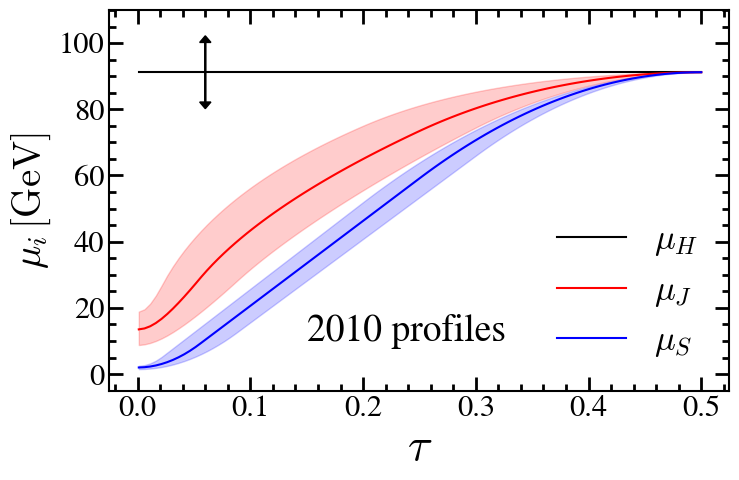}
\vspace*{-0.75cm}
\caption{\label{fig:oldProfs}}
\end{subfigure}
~
\begin{subfigure}[b]{0.475\textwidth}
\includegraphics[width=\textwidth]{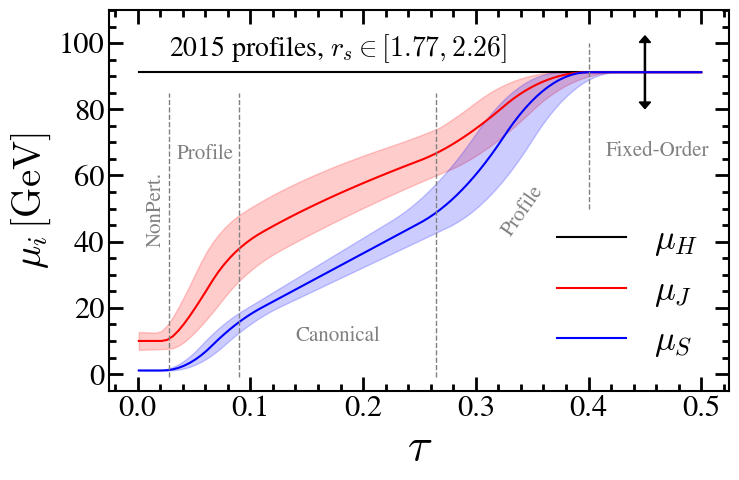}
\vspace*{-0.75cm}
\caption{\label{fig:newProfs}}
\end{subfigure}

\begin{subfigure}[b]{0.47\textwidth}
\includegraphics[width=\textwidth]{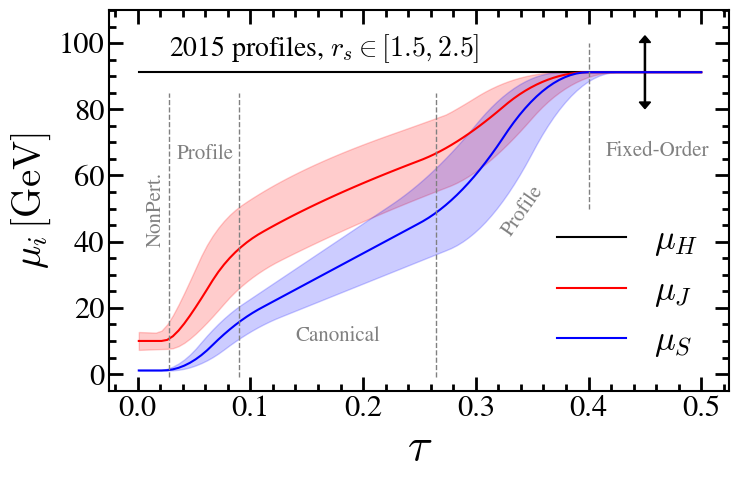}
\vspace*{-0.75cm}
\caption{\label{fig:profs2024}}
\end{subfigure}
~
\begin{subfigure}[b]{0.5\textwidth}
\includegraphics[width=\textwidth]{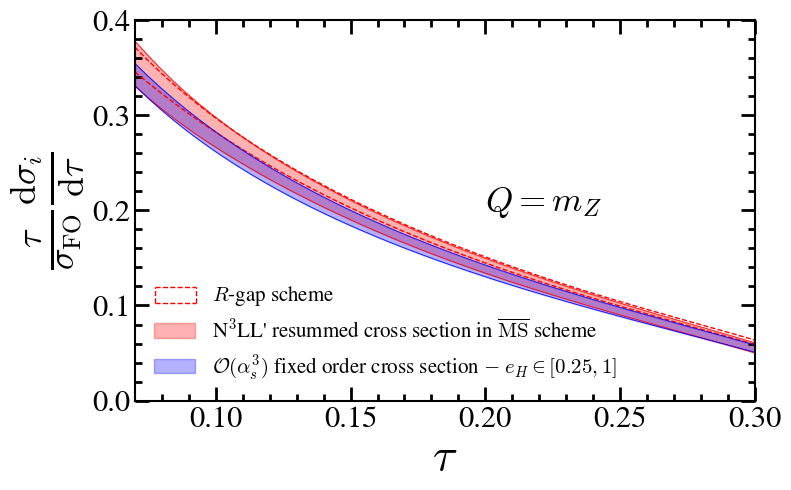}
\vspace*{-0.75cm}
\caption{\label{fig:xsCompDefault}}
\end{subfigure}
\caption{\label{fig:profs}
Comparison of the profile functions used in Ref.~\cite{Abbate:2010xh}, panel (a), in Ref.~\cite{Hoang:2015hka}, panel (b), and in this article, panel (c), for the center-of-mass energy $Q=m_Z$. To avoid cluttering the figure, the variation of the scales induced by changing the hard scale parameter $e_H$ are only indicated by the vertical $\updownarrow$ arrows.
The range of variation for the various profile parameters used in this article are given in Table~\ref{tab:profParams}. Panel~(d) shows a comparison between the N$^3$LL$^{\prime}$ resummed (solid red) and fixed-order (solid blue) cross sections in the $\overline{\rm MS}$ scheme including perturbative uncertainties. Also shown is the N$^3$LL$^{\prime}$ resummed cross section in the R-gap scheme (dashed red curves).}
\end{figure}

Since it is relevant to our analysis, we elaborate further on the differences between the profiles used in the 2010~\cite{Abbate:2010xh} and 2015~\cite{Hoang:2014wka} analyses, displaying a comparison between the two in Figs.~\ref{fig:profs} for $Q=m_Z$. The most relevant difference is that the 2015 profiles implement canonical scale setting in the tail region of the distribution, with a slope fixed by the $r_s$ parameter, which for this $Q$ value occurs for $\tau$ between about $0.1$ and $0.25$, see the second panel.
This exact linear growth is absent in the 2010 profile functions shown in the first panel, where the slope is instead determined by the parameters which specified how the scales merge into the peak and far-tail canonical regions.
The impact this has on the stability of the predictions and the fits is discussed in Sec.~\ref{subsec:stability}.
In addition, the 2015 profiles are in general more flexible compared to their predecessors, the parameters that constitute the profile functions control different features of the profile functions in a more independent way, and the scales merge already by the smaller value $\tau=0.4$ where a fixed-order prediction is applicable rather than considering dijets.
Note that even though the variation bands overlap at certain values of $\tau$, for a given set of profile parameters the required scale hierarchy associated to the three renormalization scales is, by construction, always respected. Also note that the bands shown in Figs.~\ref{fig:profs} do not include the hard scale $e_H$ variation to avoid clutter. A variation of $e_H$ between $0.5$ and $2$, results in significant vertical movement of the black $\mu_H$ line indicated by the up-down arrow, and also drags with it the jet and soft scales. Since we consider simultaneous variations in our uncertainty analysis this leads to a significant broadening of the variation bands.

\begin{table}[t!]
\centering
\begin{tabular}{c c c}
parameter & default value & range of values \\ [0.5ex]
\hline
$\mu_s(0)$ & 1.1\,GeV & - \\
$R(0)$ & 0.7\,GeV & - \\
$n_0$ & 2 & [1.5, 2.5] \\
$n_1$ & 10 & [8.5, 11.5] \\
$t_2$ & 0.25 & [0.225, 0.275] \\
$t_s$ & 0.4 & [0.375, 0.425] \\
$r_s$ & 2 & [1.5, 2.5] \\
$e_J$ & 0 & [$-1.5$, 1.5] \\
$e_H$ & 1 & [0.5, 2.0] \\
$n_s$ & 0 & [$-1$, 1] \\
\hline
$\eps_2$ & 0 & [$-1$, 1] \\
$\eps_3$ & 0 & [$-1$, 1] \\
\end{tabular}
\caption{Profile function parameters and corresponding ranges in which they are varied during the random scan.
Here the parameters $t_0=(1\,{\rm GeV}) n_0/Q$ and $t_1=(1\,{\rm GeV})n_1/Q$.}
\label{tab:profParams}
\end{table}

At this point, a discussion on the variation intervals of the profile function parameters we use in this article, and which are shown in Table~\ref{tab:profParams}, is in order. The only difference with respect to the 2015 analysis in Ref.~\cite{Hoang:2014wka} is that we increase the interval for the canonical slope parameter $r_s$ from $[1.77, 2.26]$ to $[1.5, 2.5]$. We will refer to these as 2024 profiles. This leads to increased variation bands visible in the third panel of Fig.~\ref{fig:profs} compared to the second panel of the same figure. The increase is motivated by two reasons. Firstly, as can be seen in Fig.~\ref{fig:profs}, the 2015 profiles have smaller variation than the 2010 profiles, whereas the increase obtained by the 2024 profiles represents a more conservative uncertainty estimate. Secondly, an updated study of the perturbative convergence of the thrust distribution normalized to the fixed-order total cross section shows that a wider variation of $r_s$ is preferable (in comparison to using the integral over the thrust distribution for normalizing, which is our default and shows excellent convergence already with the smaller $r_s$ variation). This is elaborated on below.

In Fig.~\ref{fig:xsCompDefault} we display a comparison between the N$^3$LL$^{\prime}$ resummed cross section, based on the 2024 profiles, and the $\mathcal{O}(\alpha_s^3)$ fixed-order cross section, both in the $\overline{\rm MS}$ scheme. It becomes apparent that already at $\tau$$\sim 0.25$ both descriptions largely coincide, demonstrating the smooth transition from the resummed prediction to the purely fixed-order based cross section. In addition to the $\overline{\rm MS}$ results, we also show the N$^3$LL$^{\prime}$+${\cal O}(\alpha_s^3)$ resummed cross section in the R-gap scheme, which
yields a more precise cross section that sits in the upper part of the uncertainty band of the $\overline{\rm MS}$ result.
This feature will become important when analyzing the fit results in Sec.~\ref{subsec:gapFit}, as a larger cross section in the tail region yields lower values for the strong coupling.

In Ref.~\cite{Bell:2023dqs} it was argued that an even wider scale variation may be mandatory to achieve reliable perturbative uncertainty. They used a larger variation for the ranges of profile parameters, by wishing to ensure that the same profiles could be employed to achieve overlapping uncertainty bands for a wide range of angularity observables, which depend on the continuous $a$ and for which a thrust-like observable emerges for $a=0$.
Our view is that it is appropriate to analyze the uncertainties for each value of $a$ separately as angularities for different $a$ are observables on their own, in the same way that it is appropriate to motivate the use of somewhat different profile functions and parameter variations for thrust, \mbox{C-parameter}, and heavy jet mass. Note that the increased range that we motivate and use for $r_s$ does go in the direction of increasing the perturbative uncertainties. Since we advocate analyzing the convergence on an observable-by-observable basis, and the bands for thrust overlap well and yield uncertainties that are appropriate to the very high order N$^3$LL$^\prime$+${\cal O}(\alpha_s^3)$ analysis being performed, we choose not to increase the range of variations further to avoid artificially inflating the perturbative uncertainties without any justification particular to thrust.

\subsection{Total Rate OPE Compatible Implementation of Profiles}
\label{subsec:opeprofiles}

As discussed in the previous subsection, in order to sum up large logs, the renormalization scales $\mu_H,\mu_J,\mu_s$ and $R$ become $\tau$-dependent profile functions. Since this yields an additional source of \mbox{$\tau$-dependence} in the convolution with the non-perturbative shape function $F_\tau$, see Eq.~(\ref{eq:factorizationFormulav2}), one has to make a decision on how to treat the profile function's $\tau$ dependence in this convolution. To explain this issue, let us write the partonic cross section of Eq.~(\ref{eq:factorizationFormulav2}), which contains all the renormalization and gap subtraction scale dependence of the factorization theorem, as
\begin{equation}
\label{eq:partonicrewritten}
\frac{\text{d} \hat{\sigma}}{\text{d}\tau} \!\bigl(\tau;\{\mu_i,R\}\bigr) \,\equiv \,\frac{\text{d} \hat{\sigma}}{\text{d}\tau} \! \biggl[\tau - \frac{2\bar \Delta(R,R_0,\mu_s,\mu_0)}{Q}\biggr]\,,
\end{equation}
making all renormalization scales $\{\mu_i,R\}=\{\mu_H,\mu_J,\mu_s,R\}$ it depends on explicit. The analogue definition for the $\overline{\rm MS}$ scheme reads
\begin{equation}
\label{eq:partonicrewrittenMSbar}
\frac{\text{d} \hat{\sigma}}{\text{d}\tau}\bigl(\tau;\{\mu_i\}\bigr) \, \equiv \, \frac{\text{d}\hat{\sigma}}{\text{d}\tau}(\tau)\,.
\end{equation}
The first option is to treat the implicit $\tau$-dependence from the profile functions $\mu_i(\tau)$ independent of the explicit $\tau$-dependence, i.e.\ the profile functions are employed after the convolution, already at the hadron level. This choice was used in Refs.~\cite{Abbate:2010xh,Abbate:2012jh,Hoang:2014wka,Hoang:2015hka}
and various other analyses.
It yields
\begin{equation}
\label{eq:factorizationFormulav3}
\frac{\text{d}\sigma}{\text{d}\tau} =\! \int \text{d}k \,\, \frac{\text{d} \hat{\sigma}}{\text{d}\tau} \! \biggl[\tau - \frac{k}{Q};\{\mu_i,R\}(\tau)\biggr] F_\tau(R_0,\mu_0;k )\,,
\end{equation}
within a gap scheme, and
\begin{equation}
\frac{\text{d}\sigma}{\text{d}\tau} = \!\int\! \text{d}k \,\frac{\text{d} \hat{\sigma}}{\text{d}\tau} \! \biggl[\tau - \frac{k}{Q};\mu_i\bigl(\tau \bigr)\biggr] F_\tau(k) \,,
\end{equation}
in $\overline{\rm MS}$. However,
this choice violates the property that all $\tau$ dependence of the partonic cross section is convolved with the shape function.
In effect, one should recall that the $\tau$-dependent scales are together responsible for yielding the correct $\tau$ dependence in the {\it partonic} cross section, in particular its large logarithms of $\tau$, and this places the implicit and explicit $\tau$ dependence at the same level. When they are treated separately for the inclusion of hadronization effects, as in \Eq{factorizationFormulav3}, this
means that the normalization condition in Eq.~\eqref{eq:zeroMoment} is no longer sufficient to yield exact independence of the integrated total cross section on the shape function form. Predominantly this implies that there will be a residual dependence on the value of the first moment $\Omega_1$, that signals a residual linear infrared sensitivity.

\begin{figure}[t!]
\centering
\begin{subfigure}[b]{0.48\textwidth}
\includegraphics[width=\textwidth]{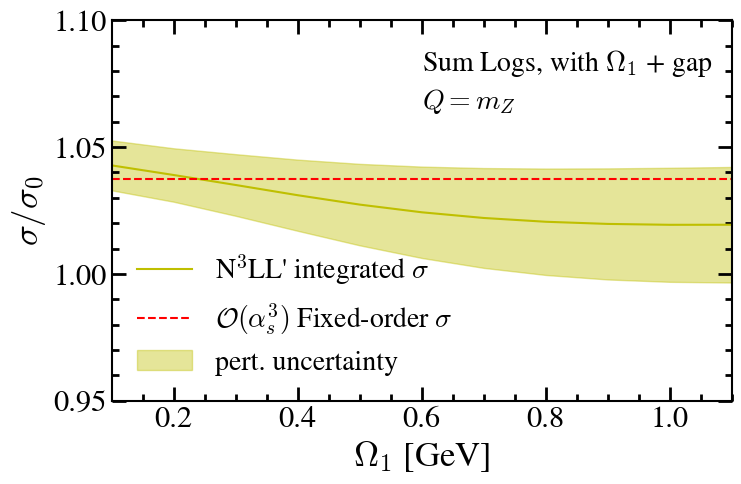}
\caption{\label{fig:intNormOld}}
\end{subfigure}
~
\begin{subfigure}[b]{0.48\textwidth}
\includegraphics[width=\textwidth]{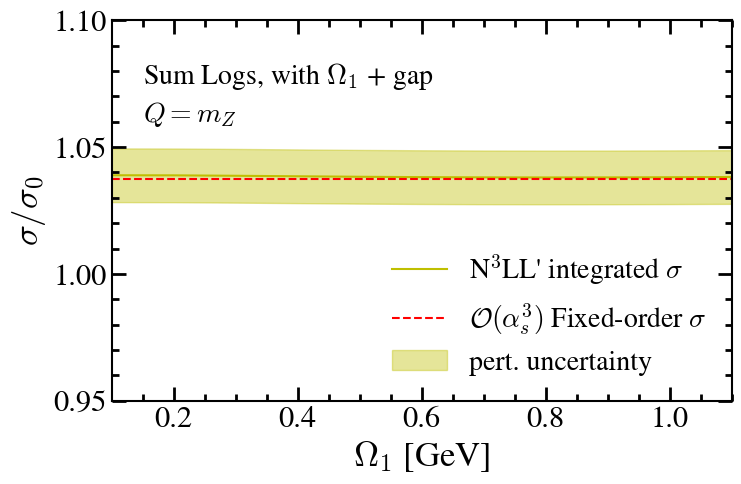}
\caption{\label{fig:intNormNew}}
\end{subfigure}

\caption{\label{fig:intNormOmega1}
Comparison of the total integrated cross section at N$^3$LL$^\prime$+${\cal O}(\alpha_s^3)$ order in the R-gap scheme (solid yellow) to the fixed order cross section at ${\cal O}(\alpha_s^3)$ (red dashed) as a function of the first moment $\Omega_1$. These are normalized by the Born cross section $\sigma_0$. The yellow solid line uses the default values for the profile function while the yellow band displays the perturbative uncertainty associated to varying profile function parameters. In the left panel we use the older method of \Eq{factorizationFormulav3}, while the right panel uses the improved approach from Eq.~\eqref{eq:hadronleveldistrv1}.}
\end{figure}

In Fig.~\ref{fig:intNormOld} we show the total integrated cross section obtained using \Eq{factorizationFormulav3} for our N$^3$LL$^{\prime}+{\cal O}(\alpha_s^3)$ thrust distribution in the R-gap scheme for $\alpha_s(m_Z)=0.1142$ and $Q=m_Z$ as a function of the shape function's first moment $\Omega_1$. (We obtain completely analogous results for the other gap schemes.)
The yellow solid line corresponds to the result for the default profile functions and the dashed green lines indicate the perturbative uncertainty estimate from the profile function variations. The dashed red line is the result for the fixed-order total cross section at ${\cal O}(\alpha_s^3)$. The integrated total cross section is compatible with the more precise fixed-order cross section within uncertainties at N$^3$LL$^{\prime}+{\cal O}(\alpha_s^3)$ order. This was also the case in Refs.~\cite{Abbate:2010xh,Abbate:2012jh}, which had similar uncertainties for the integrated cross section using the 2010 profiles, but where only the fit value of $\Omega_1(2\, {\rm GeV})=0.323\,$GeV was examined. In Fig.~\ref{fig:intNormOld} we observe a clear dependence on $\Omega_1$. This residual shape function dependence is consistent within the perturbative uncertainty, but it is still undesirable since the total hadronic $e^+e^-$ cross section is known to have a more strongly suppressed quartic infrared sensitivity.

The alternative entails that the implicit $\tau$-dependence be treated just
like the explicit one, and yields a strictly shape-function independent integrated total cross section. The implementation of such a choice is, however, not quite practical when scanning over a large number of profile function sets. There is, however, an approximate implementation of this alternative by replacing the convolution over the implicit $\tau$-dependence by a shift of the shape function's first moment. For the $\overline{\rm MS}$ scheme this leads to
\begin{equation}
\frac{\text{d}\sigma}{\text{d}\tau} = \!\int\! \text{d}k \,\frac{\text{d} \hat{\sigma}}{\text{d}\tau} \! \biggl[\tau - \frac{k}{Q},\mu_i\biggl(\tau - \frac{k}{Q}\biggr)\biggr] F_\tau(k)
\simeq \!\int\! \text{d}k \,\frac{\text{d} \hat{\sigma}}{\text{d}\tau} \! \biggl[\tau - \frac{k}{Q},\mu_i\biggl(\tau - \frac{2\bar\Omega_1}{Q}\biggr)\biggr] F_\tau(k)\,,
\end{equation}
which eliminates at least all linear sensitivity of the total integrated cross section on the shape function. The generalization of this approximation when using a gap scheme reads
\begin{align}
\label{eq:hadronleveldistrv1}
\frac{\text{d}\sigma}{\text{d}\tau} =&
\!\int\! \text{d}k
\frac{\text{d} \hat{\sigma}}{\text{d}\tau} \! \biggl[\tau - \frac{k}{Q};\{\mu_i,R\}\!\biggl(\!\tau - \frac{2[\Omega_1(R_0,\mu_0)+\delta(R_0,\mu_0)]}{Q}\biggr)\!\biggr]\!F_\tau(R_0,\mu_0,k)
\,.
\end{align}
This is the choice to treat the $\tau$-dependence of the profile function for the convolution with the shape function we adopt in the analyses we carry out in this article.

In Fig.~\ref{fig:intNormNew} we show the total integrated cross section obtained in our modified OPE-compatible approach in \Eq{hadronleveldistrv1}, using otherwise the same setup as in the left panel. The result shows a substantially improved shape function independence which is now negligible from the perspective of the remaining perturbative uncertainties. For all practical purposes the shape function dependence of the integrated total cross section can thus be ignored for all aspects of our analyses. The fixed-order ${\cal O}(\alpha_s^3)$ cross section $\sigma/\sigma_0=1.038$ also agrees very well with our central prediction for the integrated N$^3$LL$^\prime+{\cal O}(\alpha_s^3)$ cross section and is well within the uncertainty band.
The recent calculation for $s_3$ significantly contributes to this improved agreement, giving the displayed $\sigma/\sigma_0=1.04\pm 0.01$.
Note that if we had instead used $s_3 =0\pm 500$ as in Ref.~\cite{Abbate:2010xh} then the agreement from the integrated cross section would be less accurate, $\sigma/\sigma_0 = 1.06 \pm 0.02$.
In what follows we self-normalize our differential cross section, which
ensures that the normalized cross section integrates to unity. (We remark that using the fixed-order cross section for normalization, rather than the integrated cross section, would lead to a value of $\alpha_s$ that is larger by only $\Delta \alpha_s(m_Z)=0.0002$,
well within our other uncertainties.)

\subsection{Impact of Resummation on the Cross Section
\label{subsubsec:ResummationXSlevel}}

The systematic resummation of large logarithms with matching based on the factorization theorem in Eq.~(\ref{eq:factorizationFormula}) is essential in the dijet region, where $\tau$ is small, as has already been pointed out by detailed analyses in Refs.~\cite{Abbate:2012jh,Hoang:2014wka}. However, recently in Ref.~\cite{Nason:2023asn} the need and even the reliability of summing up higher-order logarithmic corrections in the tail region for $\tau\gtrsim 0.1$ was questioned, and it was argued that fixed-order results should be employed in that region. Within our factorization approach the transition of the dijet to the three-jet region occur in the interval $t_2<\tau<t_s$ (with default values $t_2=0.25$ and $t_s=0.4$ for the two transition points) where the profile functions approach the hard scale, see Fig.~\ref{fig:profs}. In the analysis of Ref.~\cite{Nason:2023asn}, it is assumed that the three-jet region reaches much lower $\tau$ values and that the resummation of logarithms is potentially unreliable already for $\tau>0.1$, advocating for the use of fixed-order.
We come back to analyze the arguments given in Ref.~\cite{Nason:2023asn} in
more detail in Sec.~\ref{sec:argumentationDiTrijet}. In view of recent results provided in Refs.~\cite{Luisoni:2020efy,Caola:2021kzt,Caola:2022vea} we also believe
a more detailed estimate of the impact of the three-jet configurations in hadronization is warranted. However, we also stress that currently there is no coherent physical understanding of the non-perturbative corrections in the transition between the dijet- and three- or multijet-dominated thrust regions. A discussion along these lines was also not provided in Ref.~\cite{Nason:2023asn}. Such an understanding is, however, mandatory, both for a restricted dijet or a dedicated three-jet analysis, since it is practically unavoidable that thrust data used for these analyses covers this transition region either on the upper or on the lower side of the respective fit ranges. Nevertheless, we believe that the level of perturbative convergence of either using our resummed factorization approach or a purely fixed-order treatment should be able to tell us about the reliability, as least from the perspective of the perturbative contributions.

\begin{figure}[t!]
\centering
\begin{subfigure}[b]{0.48\textwidth}
\includegraphics[width=\textwidth]{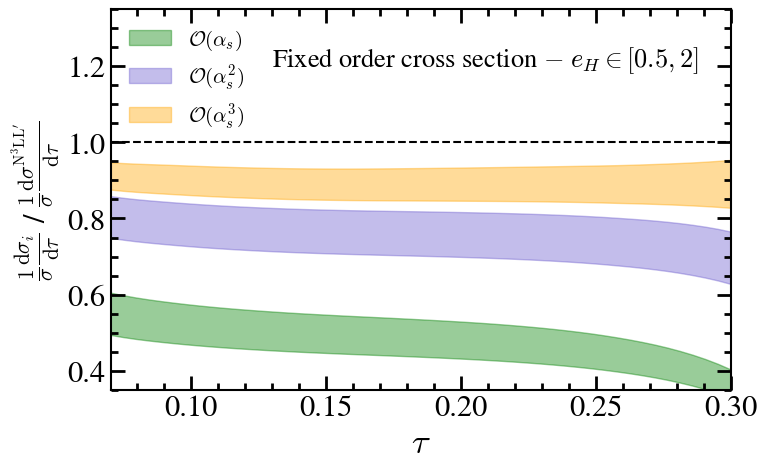}
\caption{\label{fig:ratioFO}}
\end{subfigure}
~
\begin{subfigure}[b]{0.48\textwidth}
\includegraphics[width=\textwidth]{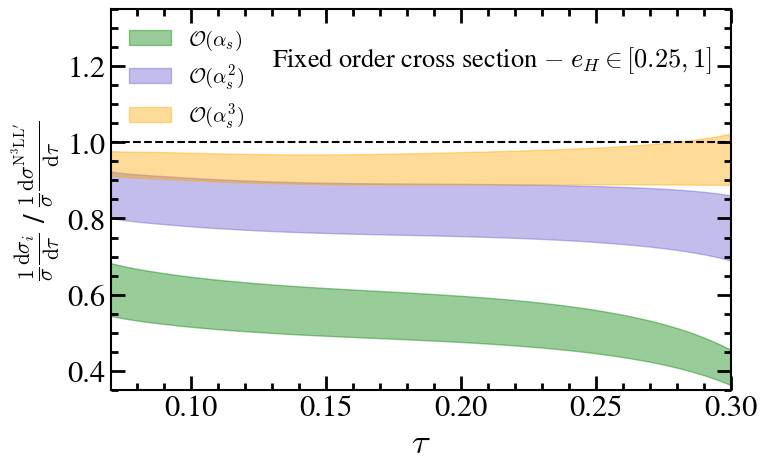}
\caption{\label{fig:ratioFONZ}}
\end{subfigure}

\begin{subfigure}[b]{0.48\textwidth}
\includegraphics[width=\textwidth]{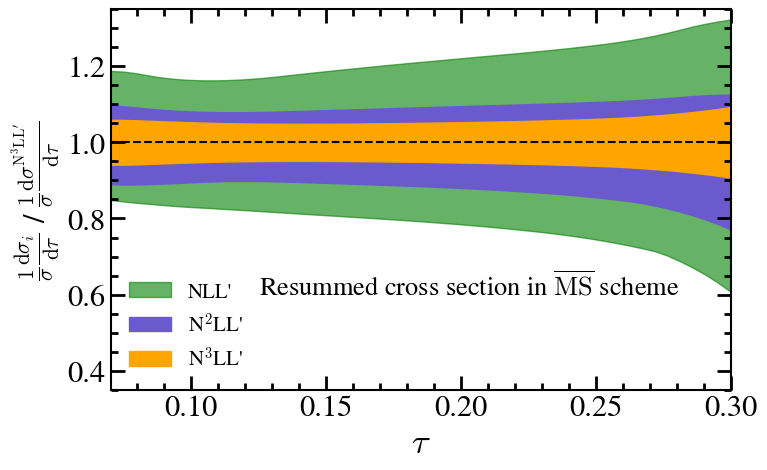}
\caption{\label{fig:ratioResummed}}
\end{subfigure}
~
\begin{subfigure}[b]{0.48\textwidth}
\includegraphics[width=\textwidth]{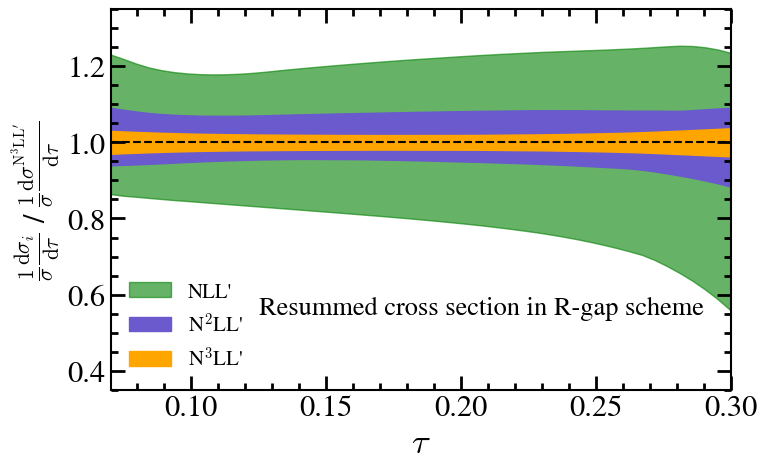}
\caption{\label{fig:ratioResummedgap}}
\end{subfigure}
\caption{\label{fig:XSRatio}
Ratios of the fixed-order (top panels) and resummed (bottom panels) differential distributions at various orders for $Q=m_Z$. In the top panels fixed-order scale uncertainties are obtained by varying $e_H=\mu/Q$ in two different ranges, while in the bottom panels the resummed scale uncertainties arise from varying the profile function parameters. In panels (a), (b) and (c) we use $\bar \Omega_1=0.3\,{\rm GeV}$ and normalize with the N$^3$LL$^\prime$+${\cal O}(\alpha_s^3)$ resummed distribution in $\overline{\rm MS}$, while panel (d) removes the leading renormalon ambiguity with R-gap subtractions, uses $\Omega_1^R=0.3\,{\rm GeV}$, and is normalized with the N$^3$LL$^\prime$+${\cal O}(\alpha_s^3)$ R-gap resummed distribution.}
\end{figure}

In this section we therefore provide a renewed analysis of the convergence properties of our resummed approach based on Eq.~(\ref{eq:factorizationFormula}) using our updated setup, compared to pure fixed-order. In Fig.~\ref{fig:XSRatio} the thrust distribution in different approaches (and at different orders) is displayed normalized by the
N$^3$LL$^\prime$+${\cal O}(\alpha_s^3)$ resummed distribution. Panels (a) and (b) show fixed-order results with two different ranges for the scale variation $\mu=e_H Q$,
while panel (c) shows resummed results. Panels (a), (b) and (c) use $\bar\Omega_1$ in the $\overline{\rm MS}$ scheme and are normalized to the default N$^3$LL$^\prime$+${\cal O}(\alpha_s^3)$ result with $\bar\Omega_1$ (corresponding to the dashed black horizontal line). Panel (d) shows resummed results with $\Omega_1^R$ in the R-gap scheme, normalized to the default N$^3$LL$^\prime$+${\cal O}(\alpha_s^3)$ result in the R-gap scheme. For the fixed-order distributions, the uncertainty bands arise from varying the single renormalization scale involved in the range $\mu \in [Q/2,2Q]$ in panel (a), and $\mu \in [Q/4,Q]$ in panel (b), respectively. For the cross sections including resummation, the profile function parameters are varied according to the ranges displayed in Table~\ref{tab:profParams}. The results in (c) and (d) exhibit a very stable perturbative behavior with systematically order-by-order overlapping and
decreasing uncertainty bands. As anticipated, the renormalon-free R-gap scheme yields smaller uncertainty bands. For the fixed-order cross sections we observe that all the uncertainty bands are quite similar regardless of the order and that all higher-order corrections constitute positive contributions. Furthermore, for large parts of the region displayed in Fig.~\ref{fig:XSRatio}, no overlap of the uncertainty bands is visible for both $\mu$-variation intervals. Interestingly, we see that the ${\cal O}(\alpha_s^3)$ fixed-order distribution is closer to the N$^3$LL$^\prime$+${\cal O}(\alpha_s^3)$ resummed result for the lower \mbox{$\mu$-variation} interval $e_H\in [0.25,1]$, indicating that this choice reduces the effects of the higher-order logarithms involving the separated soft, jet and hard scales. The behavior of the uncertainty bands in comparison to the default N$^3$LL$^\prime$+${\cal O}(\alpha_s^3)$ resummed prediction (corresponding to the dashed black horizontal line) implies that the yet unknown full ${\cal O}(\alpha_s^4)$ fixed-order corrections may very likely again provide another positive shift that provides an even better agreement with the N$^3$LL$^\prime$+${\cal O}(\alpha_s^3)$ resummed result. It is also likely that this happens for $\tau$ values up to $0.3$.\footnote{We note that the discrepancy between the ${\cal O}(\alpha_s^3)$ fixed-order results and the N$^3$LL$^\prime$+${\cal O}(\alpha_s^3)$ resummed distribution in the R-gap scheme shown in panels (a) and (b) of Fig.~\ref{fig:XSRatio} varies with $Q$ and is relevant for $\alpha_s$ extractions from thrust data for different c.m.\ energies. In addition, we emphasize that the conclusions one can draw from both Figs.~\ref{fig:XSRatio} and \ref{fig:ratioNorms},
are independent of the choice of $\alpha_s(m_Z)$ and $\bar \Omega_1$.}

In any case, based on Fig.~\ref{fig:XSRatio}, a strong-coupling determination based on fixed-order perturbation theory alone yields large order dependence already based on the known corrections up to ${\cal O}(\alpha_s^3)$. In contrast, the resummed predictions exhibit an ideal order-by-order convergent behavior, irrespective of considering the cross section in the $\overline{\rm MS}$ or R-gap schemes. We observe order-by-order decreasing uncertainty (estimate) bands and very nice stability of the central prediction. Following the same pattern, this can be expected to yield a significantly more stable order-by-order behavior for strong coupling determinations. This will be reconfirmed
explicitly by carrying out fits presented in Sec.~\ref{subsec:stability}. Furthermore, the ratio of the ${\cal O}(\alpha_s^3)$ fixed-order predictions over the N$^3$LL$^\prime$+${\cal O}(\alpha_s^3)$ resummed result is flat across the entire $\tau$ range up to $0.3$ without any indication of an anomalous behavior of the resummation effects for increasing $\tau$ values. Overall, no patterns of instabilities are visible for the entire relevant range of $\tau$ values, reinforcing the conclusion that resummed predictions are reliable and favored over fixed-order results.

\subsection{Normalization Choice and Gap Scheme Dependence}
\label{sec:normchoice}

\begin{figure}[t!]
\centering
\begin{subfigure}[b]{0.48\textwidth}
\includegraphics[width=\textwidth]{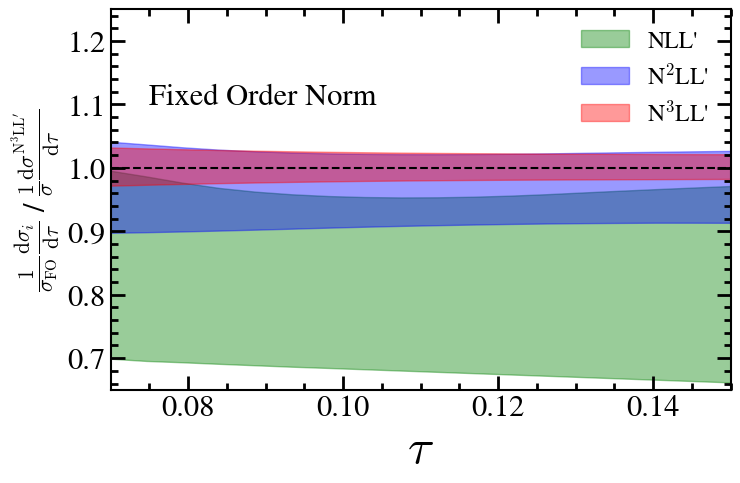}
\caption{\label{fig:ratioFOconvergence}}
\end{subfigure}
~
\begin{subfigure}[b]{0.48\textwidth}
\includegraphics[width=\textwidth]{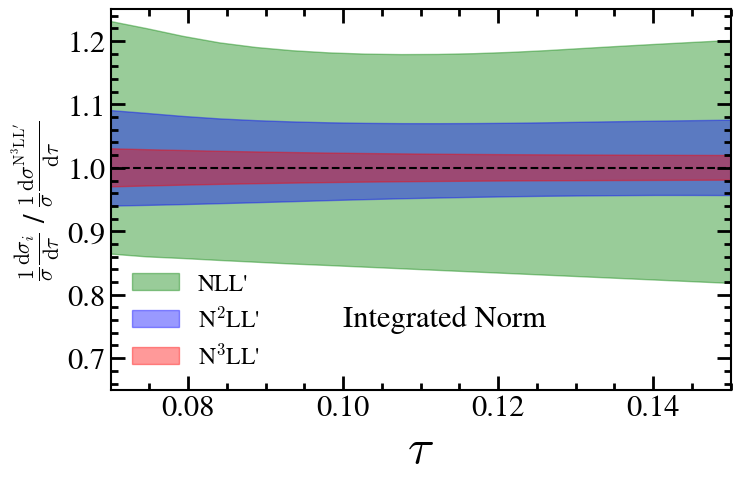}
\caption{\label{fig:ratioResummedConvergence}}
\end{subfigure}

\caption{\label{fig:ratioNorms}
Ratios of the resummed cross section in the R-gap scheme using both fixed-order normalization $\sigma_{\rm FO}$ (left panel), and self-normalization $\sigma$ (right panel), at various orders, over the default self-normalized N$^3$LL$^\prime$+${\cal O}(\alpha_s^3)$ resummed distribution. In order to obtain these results we use $\alpha_s(m_Z)=0.114$ and $\Omega_1^R=0.3\,{\rm GeV}$.}
\end{figure}

As already stated at the end of Sec.~\ref{subsec:opeprofiles},
in the following sections we self-normalize our factorization based prediction for the differential distribution,
which ensures that at each order it integrates to unity, which is also true for the experimental data to which we will eventually compare.
In this section we investigate further aspects of our normalization choice focusing, once more on the perturbative convergence, and on the gap subtraction scheme dependence, and explaining our motivation for the increased $r_s$ variation range mentioned at the end of Sec.~\ref{subsec:profiles}.

In Fig.~\ref{fig:ratioNorms} we display our resummed distribution in the R-gap scheme at various orders normalized by the corresponding fixed-order total cross section (left panel) and by the integrated differential cross section (right panel). Each of the distributions is divided again by our default self-normalized N$^3$LL$^\prime$+${\cal O}(\alpha_s^3)$ resummed distribution in the R-gap scheme (corresponding to the dashed-black horizontal line at unity). The colored bands are the perturbative uncertainties estimated from our profile function variation. We observe order-by-order convergent behavior for both normalizations, but a significantly better convergence pattern can be attributed to the self-normalized distributions. This feature also arises when using the 2015 profiles with the smaller $r_s$ variation range, but the overlap of the uncertainty bands for the fixed-order normalization is larger with the variation of $r_s$ used in the 2024 profiles.
This same conclusion favoring self normalization was also drawn in the 2010 analysis~\cite{Abbate:2010xh}.

\begin{figure}[t!]
\centering
\includegraphics[width= 0.8\textwidth]{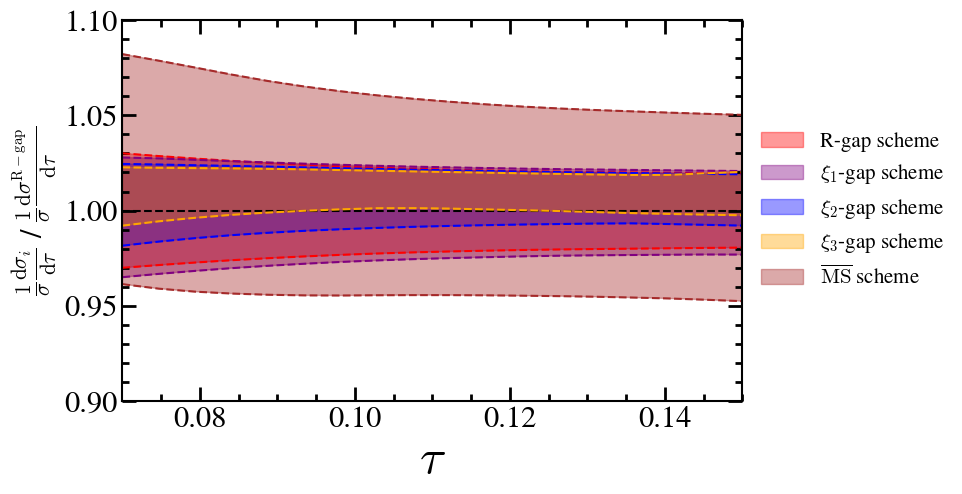}
\caption{\label{fig:gapSchemesXS}
Ratios of self-normalized N$^3$LL$^\prime$+${\cal O}(\alpha_s^3)$ resummed distribution in four different gap schemes and in $\overline{\rm MS}$, over self-normalized N$^3$LL$^\prime$+${\cal O}(\alpha_s^3)$ resummed distribution in the \mbox{R-gap} scheme. We use $\alpha_s(m_Z)=0.114$ and \mbox{$\Omega_1^R = 0.3\,$GeV}. The corresponding values for $\Omega_1$ in the other three other schemes read: \mbox{$\Omega_1^{\xi_1}(2\,\mbox{GeV})=0.31\,$GeV}, \mbox{$\Omega_1^{\xi_2}(2\,\mbox{GeV})=0.34\,$GeV}, \mbox{$\Omega_1^{\xi_3}(2\,\mbox{GeV})=0.37\,$GeV}. At ${\cal O}(\alpha_s^3)$ the moment in the $\overline{\rm MS}$ scheme reads \mbox{$\bar\Omega_1=\Omega_1^R+0-0.04-0.04=0.22\,$GeV}, where we have shown the subsequent terms in the conversion series in $\alpha_s$ which contains an ${\cal O}(\Lambda_{\rm QCD})$ renormalon. }
\end{figure}

At this point it is prudent to also examine the gap-subtraction scheme dependence of our resummed distribution. In Ref.~\cite{Bell:2023dqs}, a sizable variation of the resummed predictions was found for the 2010 profile functions (which we do not use in our current analysis) between the R-gap and R$^*$ schemes. As we already explained at the end of Sec.~\ref{subsubsec:gapSchemes}, we do not consider the R$^*$ scheme as a viable implementation for subtracting the ${\cal O}(\Lambda_{\rm QCD})$ renormalon arising from the perturbative large-angle soft radiation as it leaves large logarithms associated to the renormalon subtraction unresummed. However, the study of the gap subtraction scheme dependence
advocated by Ref.~\cite{Bell:2023dqs}
is certainly warranted to scrutinize the stability of the results, so we analyze this scheme dependence in detail here.

In Sec.~\ref{subsubsec:gapSchemes} we discussed four different gap subtractions, the R-gap scheme, which was already employed in Ref.~\cite{Abbate:2010xh} and is our default, and the three non-derivative $\xi_{1,2,3}$-gap schemes. Stability with respect to the choice of gap subtractions should yield predictions that are consistent within the respective uncertainties (for values of $\Omega_1^{\xi_i}$ and $\bar\Omega_1$ which have been obtained by scheme conversion starting from a common $\Omega_1^{\rm R}$). In Fig.~\ref{fig:gapSchemesXS} the self-normalized N$^3$LL$^\prime$+${\cal O}(\alpha_s^3)$ resummed distributions with perturbative uncertainty bands in the four gap schemes are shown, again divided by the default N$^3$LL$^\prime$+${\cal O}(\alpha_s^3)$ distribution in the R-gap scheme. We also display the result in \MSbar where the renormalon is not subtracted.
We first observe the significant reduction of the perturbative uncertainty bands for all the renormalon subtracted predictions (by a factor of two or more) compared to the \MSbar scheme. This improvement, which was already discussed at length in Ref.~\cite{Abbate:2010xh} (and even earlier in Ref.~\cite{Davison:2009wzs}), represents one of the main reasons for the high precision of our cross section and highlights the advantage of employing the renormalon subtraction over the unsubtracted \MSbar scheme.
Furthermore, the results in the four gap schemes show very good mutual consistency. Interestingly, the $\xi_2$- and $\xi_3$-gap schemes exhibit somewhat smaller uncertainty bands compared to the $R$- and $\xi_1$-gap schemes, but this should be considered accidental. The larger uncertainty bands obtained in the $\xi_1$- and R-gap schemes are essentially equivalent, and reconfirm that adopting the latter as our default is adequate.

\subsection{Hadron Mass Effects}

For partonic final states it is justified to use massless final state particles since the gluon is massless, and light quarks have a mass much smaller than $\Lambda_{\rm QCD}$. (For a discussion of heavy quark mass effects we refer the reader to the next section.) In contrast, for the experimentally accessible hadronic cross sections, it is never justified to neglect the mass of the hadrons, since this is entangled with the $\Lambda_{\rm QCD}$ hadronization effects that one is aiming to capture.
For example, one can compare $\tau$ in Eq.~\eqref{eq:tdef} with the following observables
\begin{align}\label{eq:EPschemes}
\tau_2 = \min_{\vec{n}} \frac{\sum_j \bigl(E_j - | \vec{p}_j \cdot \vec{n} |\bigr)}{Q}\,,
\qquad
\tau_E= \min_{\vec{n}} \frac{\sum_j E_j\bigl(1 - | \hat{p}_j \cdot \vec{n} |\bigr)}{Q}\,,
\end{align}
where $E_j$ is the energy of the final-state particle $j$ and $\hat{p}_j=\vec{p}_j/|\vec{p}_j|$ a unit 3-vector pointing in the direction of $\vec{p}_j$. For massless partons all three observables are identical, whereas all three differ for massive hadrons.

The effect of hadron masses on event-shape distribution was first pointed out in Ref.~\cite{Salam:2001bd}, and later studied using a quantum-field formalism in Ref.~\cite{Mateu:2012nk}. For very energetic hadrons the difference between $E_j$ or $|\vec{p}_j|$ is small. The hadronization effects for collinear particles correspond to a subleading power correction of ${\cal O}(\Lambda_{\rm QCD}^2/(Q^2\tau))$, which we neglect. On the contrary,
the hadron mass effects for soft particles with momenta $\sim \Lambda_{\rm QCD}$ are the same order as the leading hadronization corrections at ${\cal O}(\Lambda_{\rm QCD}/(Q\tau))$, causing the dominant difference between measuring $\tau$, $\tau_2$, or $\tau_E$.

Thus hadron mass effects in the dijet region are captured by nonperturbative contributions to the soft function, that is the shape function in \eq{factSoftFunc}.
One of the advantages of having an operator definition for $\Omega_1$
is that relations for different observables $\tau_i$ in the dijet region can be derived from first principles in QCD.
In Ref.~\cite{Lee:2006nr} it was shown that the boost and rotational symmetry of the matrix element defining $\Omega_1$ enable the universality of this parameter among different event shapes to be established. These symmetries play a crucial role when accounting for hadron masses and enable one to match observables $\tau_i$ to the appropriate universality class definition of $\Omega_1$~\cite{Mateu:2012nk}.\footnote{In our formalism the hadronization effects incorporated in moments $\Omega_n$ are formulated on the asymptotic celestial sphere at $\infty$, which implies that we assume that all particles that are unstable under strong interactions have decayed. This properly incorporates the mass gap $\Lambda_{\rm QCD}$ that is present in real QCD. In this context, the decay scheme that replaces massive QCD hadrons like the $p$ and $\pi$ by massless final state particles is not related to any theory of nature, and should not be considered for estimating uncertainties in $\alpha_s$ fits.}
Note that this theoretical treatment of hadron mass effects has not yet been properly incorporated in dispersive model based approaches.
Since we consider a single event shape with a unique universality class in our study, the differences between hadron mass effects in different universality classes is not relevant to us here.\footnote{Ref.~\cite{Mateu:2012nk} also showed that the leading power correction has a non-trivial, hadron-mass-dependent anomalous dimension. The running of $\Omega_1$ was studied phenomenologically in Ref.~\cite{Hoang:2015hka} and it was concluded that for thrust (and C-parameter) the effect is very small. Therefore, we do not implement it in this article.}

In the tail region in which the OPE is applicable the treatment of hadron masses in measurements will fix the appropriate definition of $\Omega_1$ and hence, provided all experiments use the same treatment, the dominant hadron mass effects are absorbed when fitting for this parameter. The fact that such hadron mass scheme dependence can be absorbed into the definition of $\Omega_1$ was confirmed for the first moment of event shapes generated
in both the \Pythia and \Herwig
Monte Carlos in Ref.~\cite{Mateu:2012nk}.
We will assume that the hadron mass dependence for $\tau$ as defined in Eq.~(\ref{eq:tdef}) are correctly accounted for across the experimental measurements used in our analysis, and therefore we do not assign any additional uncertainty for these effects.

\subsection{QED, Electroweak, and Bottom Mass Corrections
\label{subsec:additionalUncertainties}}
In this section we briefly review the known theoretical effects that will be neglected in our analysis, and comment on the impact they would have in the determination of the strong coupling. All of these have been studied in Ref.~\cite{Abbate:2010xh}, but in subsequent analyses have mostly been ignored.

We start with perturbative contributions. Like the vast majority of phenomenological analyses carried out for event shapes, our theoretical expression accounts only for QCD effects and assumes massless primary and secondary quarks. While electroweak corrections are quite small~\cite{Denner:2010ia}, QED effects are more important.
In Ref.~\cite{Abbate:2010xh} QED effects were analyzed, properly accounting for the distinction between virtual effects, initial state radiation, and final state radiation in accordance with the experimental treatment.
It was found that these QED effects lead to a decrease of $\alpha_s(m_Z)$ by $\Delta\alpha_s(m_Z)_{\rm QED} = - 0.0005$, see their Fig.~14.
This can be understood intuitively since the inclusion of soft and collinear final state photons adds a bit of strength to the QCD radiation, and thus lowers the value of $\alpha_s(m_Z)$ in the fit.
We will take the attitude that this is a known effect that can included if higher precision is desired. However, to make our results easier to compare with others in the literature that perform fits purely based on QCD corrections, we will leave them out in our default analysis. We also remark that the size of this correction is less than half of our total uncertainty in $\alpha_s(m_Z)$.
If this shift was taken as an additional uncertainty and added in quadrature, our total uncertainty would only increase by $0.0001$.

The effect of primary massive bottom quarks was also studied in Ref.~\cite{Abbate:2010xh}, accounting for such effects at N$^2$LL+$\mathcal{O}(\alpha_s)$.\footnote{The analysis of Ref.~\cite{Abbate:2010xh} made the assumption that the momenta of heavy hadrons is not reconstructed, that is, assuming the 2-jettiness measurement can be used on heavy-quark momenta. This corresponds to making measurements on heavy hadron decay products.
If the standard thrust measurement is used on heavy-quark momenta, the effect on $b$-quark masses on the cross section is even smaller, see Refs.~\cite{Lepenik:2019jjk,Bris:2020uyb}.}
This effect is also rather small since electroweak factors dampen the production of bottom quarks, rendering its influence on $\alpha_s(m_Z)$ at the per-mil level. As shown in Table 5 of Ref.~\cite{Abbate:2010xh} this effect slightly increases $\alpha_s(m_Z)$ by $\Delta\alpha_s(m_Z)_{m_b}\approx 0.0001$.
Inclusion of the bottom mass has a larger effect on the value of $\Omega_1$, causing a shift of $\Delta \Omega_1 = -0.022\,{\rm GeV}$. However, when including both bottom masses and QED radiation, the combined shift is only
$\Delta \Omega_1 = -0.009\,{\rm GeV}$, which is also negligible when compared to the total uncertainty on extractions of $\Omega_1$.

Even though top quarks cannot be produced at LEP energies, they enter through virtual diagrams. These appear for the first time at $\mathcal{O}(\alpha_s^2)$ in the form of the axial-anomaly due to the large bottom-top mass splitting and modify both the hard function and the non-singular distribution. In Ref.~\cite{Abbate:2010xh}, these corrections have also been studied and were found to be smaller than the $b$-mass effects, see their Sec.~II.C.

All in all, we can conclude that given the accuracy of both experimental data and our theoretical description, it is safe to neglect the majority of the corrections discussed in this section,
and to include the $\Delta\alpha_s(m_Z)_{\rm QED}\simeq - 0.0005$ shift from including QED corrections only when one wants to add the dominant electroweak correction beyond a pure QCD analysis.

\section{Data Selection and Fit Procedure
\label{sec:expData}}

Experimental data for the thrust distribution exist for different values of center-of-mass energies in the range \mbox{$Q\in[14,207]\,$GeV}. The datasets, corresponding to a certain value of $Q$, have been obtained from the following experimental collaborations:\footnote{This is the same dataset used in Ref.~\cite{Abbate:2010xh}.} TASSO with $Q=$ \{35, 44\}\,GeV~\cite{TASSO:1990cdg}, JADE with $Q=$ \{35, 44\}\,GeV~\cite{MovillaFernandez:1997fr}, AMY with $Q=$ 55.2\,GeV~\cite{AMY:1989feg}, SLC with $Q=$ 91.2\,GeV \cite{SLD:1994idb}, L3 with $Q=$ \{41.4, 55.3, 65.4, 75.7, 82.3, 85.1, 91.2, 130.1, 136.1, 161.3, 172.3, 182.8, 188.6, 194.4, 200, 206.2\}\,GeV \cite{L3:1992nwf,L3:2004cdh}, DELPHI with $Q=$ \{45, 66, 76, 89.5, 91.2, 93, 133, 161, 172, 183, 189, 192, 196, 200, 202, 205, 207\}\,GeV \cite{DELPHI:2003yqh,DELPHI:2000uri,DELPHI:1999vbd}, OPAL with $Q=$ \{91, 133, 161, 172, 177, 183, 189, 197\}\,GeV \cite{OPAL:2004wof,OPAL:1997asf,OPAL:1999ldr} and ALEPH with $Q$~=~\{91.2, 133, 161, 172, 183, 189, 200, 206\}\,GeV~\cite{ALEPH:2003obs}.

For our analysis we will restrict the choice of datasets to the range \mbox{$Q\in[35,207]\,$GeV}. Lower energies are excluded, as we defer from including bottom mass corrections (along with QED final-state radiation) in our theory description, and lower energies would require a more refined treatment of these effects (see Ref.~\cite{Bris:2020uyb} for a detailed discussion). As will be argued in Sec.~\ref{subsec:stability}, our default dataset is $(6\,{\rm GeV})/Q\le\tau\le0.15$, where we exclude bins that extend more than 50$\%$ outside the fit range on either side. We proceed in the exact same manner for every fit range chosen throughout our analysis. Carrying out a global fit to the available data for all $Q\geq 35\,$GeV is important given the strong degeneracy between $\alpha_s$ and $\Omega_1$ in the tail region of the thrust distribution, which is lifted when including multiple values of $Q$ at the same time into the fit.

To perform our fits, for a given set of profile parameters we construct a $\chi^2$ function which includes both the statistical and systematic experimental errors. As there is no available information on the correlation for the systematic experimental uncertainties we rely on a model. As in Ref.~\cite{Abbate:2010xh}, we implement the minimal overlap model \cite{OPAL:2004wof,ALEPH:2003obs}, which yields a positive correlation of these systematic uncertainties:
\begin{equation}
\delta \sigma^{\rm sys}_{ij} = {\rm min}[\sigma^{\rm sys}_{i},\sigma^{\rm sys}_{j}]^2\delta_{Q_i,Q_j}\delta_{{\rm exp}_i, {\rm exp}_j}\,,
\end{equation}
that is, the correlation is zero among different experiments and also within the same experiment for different values of $Q$. Adding up both statistical (uncorrelated) and systematic (correlated) experimental errors yields the experimental covariance matrix for the fit parameters for a given dataset. From this matrix one can infer the experimental uncertainties and the correlation between $\alpha_s$ and $\Omega_1$.

To obtain the perturbative theoretical error for a given dataset we proceed as follows. We select $500$ random sets of profile function parameters, which will be referred to as carrying out a random scan for the remainder of this article, and perform one fit per random point. The ranges for the individual profile function parameters are shown in Table~\ref{tab:profParams} and the random points are generated with a flat distribution. The theoretical (or perturbative) error is obtained by taking half of the difference between the maximum and minimum best-fit value of the $500$ random points. The theoretical correlation between $\alpha_s$ and $\Omega_1^R$ is computed as that of the $500$ pairs of best-fit values. From this correlation coefficient and the theoretical errors, the theoretical covariance matrix is built up. Each of the $500$ fits yields a different (experimental) $2\times2$ covariance matrix. From the average of each of the entries of these matrices we obtain the overall experimental covariance matrix, out of which the experimental uncertainties of the fit parameters, along with their experimental correlation, are obtained. The total covariance matrix is obtaining adding up the theoretical and experimental ones. Using this matrix, the total error ellipse can be drawn using the best-fit values for $\alpha_s$ and $\Omega_1$ as its center, which are obtained as the average of the maximum and minimum best-fit values obtained in the random scan for each parameter.

All of the analyses carried out are based on a new \texttt{C++}~\cite{gcccompileprocess} code, which we cross-checked against the former \texttt{Fortran 77}~\cite{gfortran} program used in Refs.~\citep{Abbate:2010xh,Hoang:2015hka} and the updated \texttt{Fortran 2008} version employed in Ref.~\cite{Dehnadi:2023msm}. We have checked that the new and old codes agree on the differential and cumulative thrust distributions up to seven decimal places. The new \texttt{C++} code uses the \texttt{gsl} library~\cite{galassi2018scientific} for numerical integration (based on \texttt{QUADPACK}~\cite{quadpackpp}, same set of routines already used in the previous codes), interpolation and special functions, and has several advantages. Since we are able to access the code using a \texttt{Python}~\cite{Rossum:1995:PRM:869369} wrapper (generated using \texttt{SWIG}~\cite{swig}), it is easy to use and suitable for parallelized runs on computer clusters. Due to the efficient implementation in a fast computer language, we have been able to restrict the performance capacities needed for our analysis to the Cluster of the Particle Physics Group at the University of Vienna.

\section{Stability of the Predictions in Fits to Data \label{sec:results}}

In Secs.~\ref{subsubsec:ResummationXSlevel} and \ref{sec:normchoice} we have discussed the impact of resummation and gap-scheme choice on the theoretical prediction for the thrust distribution for particular values of the strong coupling and $\Omega_1^R$. However, in fits to experimental data the values of $\alpha_s$ and $\Omega_1$ themselves are order-dependent and may depend on the way in which fits are carried out. We therefore examine in this section whether the observations we could make in these previous sections are also imprinted on the fit results. In particular we investigate the impact of resummation on the fit outcome focusing on variations of the dataset (i.e.\ the thrust intervals used in the fits) and on the choice of the gap subtraction scheme. Furthermore, we discuss the improvements in stability concerning variations of thrust ranges used for the fits employing the improved 2015 profile functions~\cite{Hoang:2014wka,Hoang:2015hka} compared to the ones used in the original 2010 thrust analysis of Ref.~\cite{Abbate:2010xh}.
We emphasize that the discussions in this section do not yet constitute our final strong coupling determination, which is presented in Sec.~\ref{sec:inputUpdate}. They are still conceptual, focusing primarily on the overall stability and reliability of our resummed approach with gap subtractions and the 2015 profiles (with increased $r_s$ variation) and without yet specifying the uncertainties from profile function (or renormalization scale) variations.

\subsection{Stability of Results and Impact of Resummation
\label{subsec:stability}}

\begin{figure}[t!]
\centering
\begin{subfigure}[b]{0.47\textwidth}
\includegraphics[width=\textwidth]{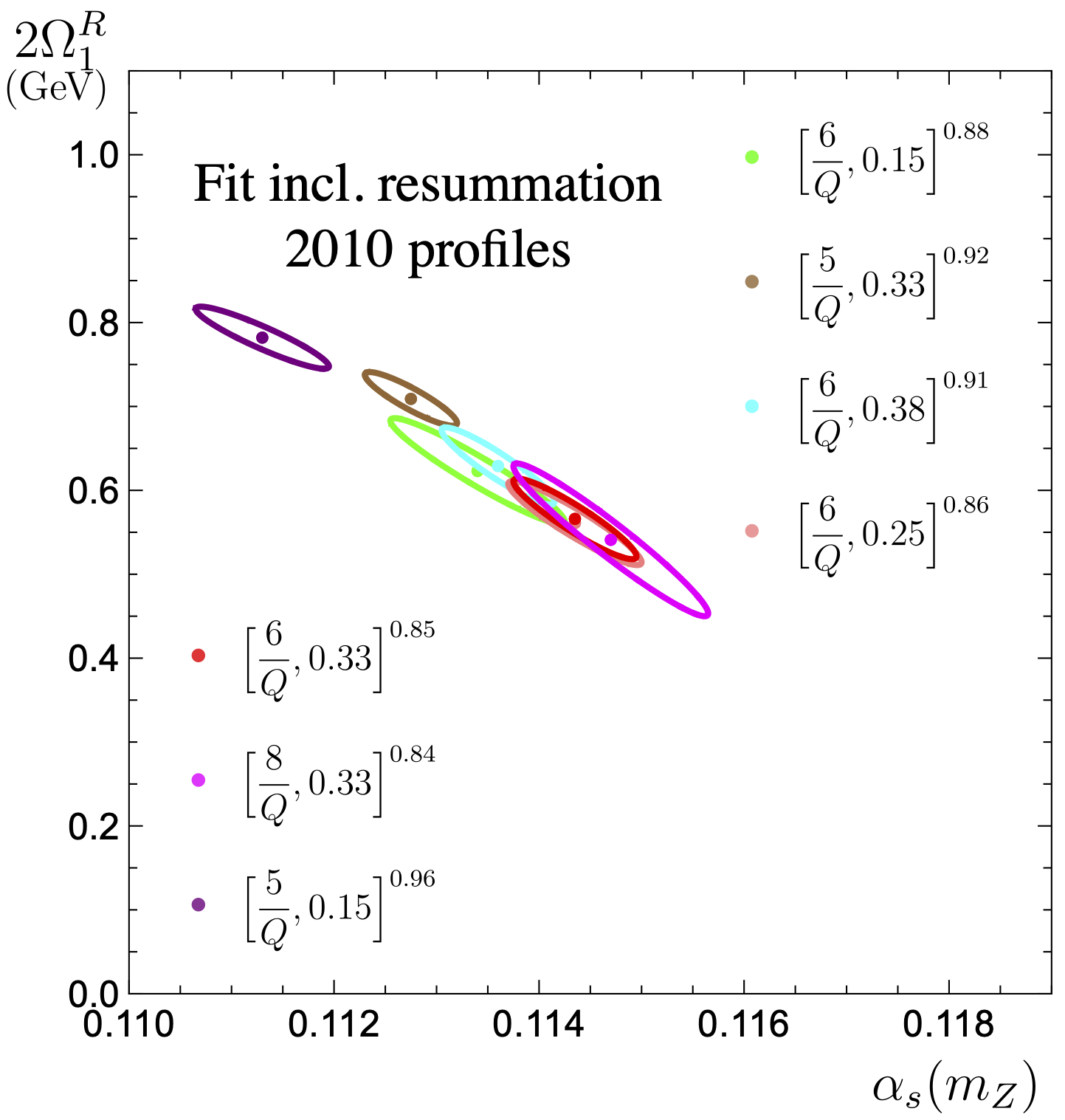}
\vspace*{-1.0cm}
\caption{\label{fig:thrust2010}}
\end{subfigure}
~
\begin{subfigure}[b]{0.47\textwidth}
\includegraphics[width=\textwidth]{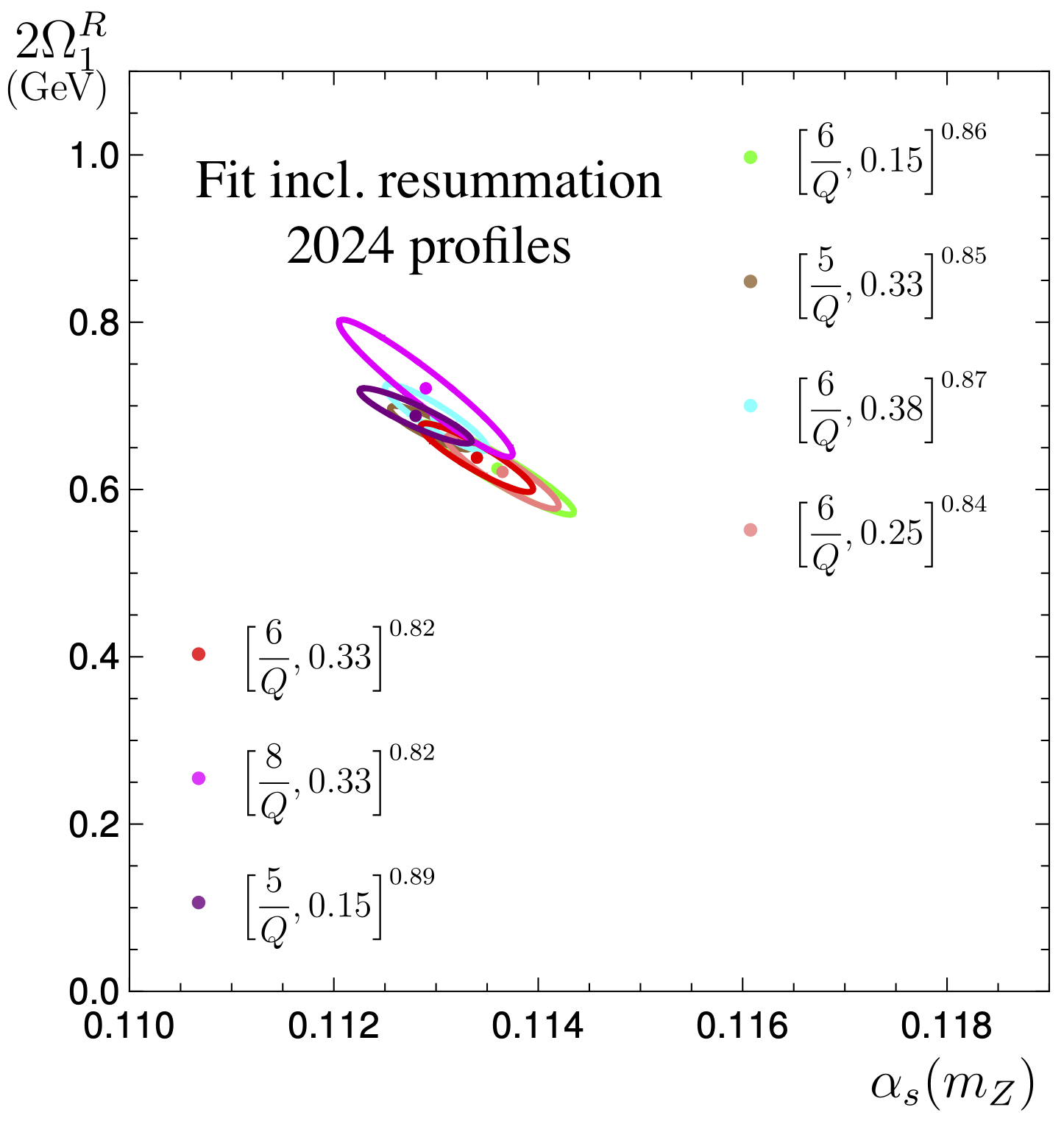}
\vspace*{-1.0cm}
\caption{\label{fig:thrust2024}}
\end{subfigure}
\caption{\label{fig:thrustStability}
Comparison of the stability of the fit results with respect to variations of the dataset. Only experimental uncertainty ellipses are shown, which correspond to $39\%$ confidence level in two parameters, and 68\% confidence level when projected onto one parameter. Panel~(a)
depicts the results using the 2010 profile functions of Ref.~\cite{Abbate:2010xh}, whereas panel~(b)
utilizes the more canonical set of profile functions based on Ref.~\cite{Hoang:2014wka}. The intervals represent the $\tau$ range of a given dataset, being the reduced $\hat \chi^2=\chi^2/{\rm dof}$ shown as a superscript: $[\tau_{\min},\tau_{\max}]^{\hat \chi^2}$. }
\end{figure}

In the analysis of Ref.~\cite{Abbate:2010xh}, using the old 2010 profile functions, a sizable dependence for the $\alpha_s$ and $\Omega_1^R$ fit results on the $\tau$ intervals employed for the fits was observed. In this analysis different fit-ranges $[\tau_{\min},\tau_{\rm max}]$ with $\tau_{\rm min}=(5,6,8){\rm GeV}/Q$ and $\tau_{\rm max}=0.25-0.38$ were considered, and a marginal consistency was found if only the experimental uncertainties were accounted for; see Fig.~17 in Ref.~\cite{Abbate:2010xh}, where the small ellipses represent only the 68\% CL (1-$\sigma$) experimental uncertainties for $\alpha_s$. (This 68\% CL for one dimension corresponds to 39\% CL for two dimensions, in the $\alpha_s$-$\Omega_1^R$ plane.)\footnote{The purpose of not showing the theoretical and in particular the renormalization scale uncertainties was to better expose the stability aspects of the fit results.}
The origin of this variation was considered of theoretical nature, since a perfect theory prediction valid for the considered $\tau$ values would yield equivalent results. Thus, the fit range variation can be seen as a representative of the theoretical uncertainty. Indeed, including theoretical uncertainties, the different results for $\alpha_s$ were statistically consistent at the 68\% CL.

We reanalyze in this subsection the $\tau$ interval dependence of the fit results using the 2015 profile functions (with the increased $r_s$ variation range). These new profiles are an improvements over the 2010 ones, as the jet and soft renormalization scales exhibit a more consistent canonical scale behavior for the dijet tail region above the peak. In Fig.~\ref{fig:thrustStability} we show the fit outcome for $\alpha_s(m_Z)$ and $\Omega_1^R$ for different intervals in analogy to Fig.~17 of Ref.~\cite{Abbate:2010xh}.
All results (and in particular the center of the ellipses) are obtained with full profile function (or renormalization scale) variations as outlined in Sec.~\ref{sec:expData}, but following the discussion of Fig.~17 in Ref.~\cite{Abbate:2010xh}
only the experimental uncertainties representing 69\% CL (1-$\sigma$) for $\alpha_s$ are displayed.

In Fig.~\ref{fig:thrust2010} results for different $\tau$ intervals, with $\tau_{\rm min}=(5,6){\rm GeV}/Q$ and $\tau_{\rm max}=0.15-0.38$, are shown for the 2010 profile functions based on our N$^3$LL$^\prime$+${\cal O}(\alpha_s^3)$ resummed distribution in the R-gap scheme. We reproduce the sizable dataset dependence, with the ellipses covering $\alpha_s(m_Z)$ values between $0.111$ and $0.114$ located along the degeneracy line in the $\alpha_s$-$\Omega_1^R$ plane already observed in Ref.~\cite{Abbate:2010xh}.
In Fig.~\ref{fig:thrust2024} the corresponding results are shown with our updated theory expressions, with the difference being entirely caused by the use of the 2015 profile functions (other changes, such as the new $\alpha_s^3$ non-singular, use of the constants $s_3$, $j_3$, $\Gamma_4$, and OPE compatible implementation of the profiles, play no role in this discussion of fit-range stability).
We observe a significant improvement in stability and fit quality. The latter can be inferred from the general reduction of the minimal reduced $\chi^2$ shown as the superscript of the fit-range information in the panel's legends. With the 2015 profile functions all error ellipses are perfectly compatible and centered at $\alpha_s(m_Z)$ around $0.1133$, ranging from $0.1128$ to $0.1137$. This improvement demonstrates the importance of resummation with the proper choice of the profile functions.
Taking the fit-range dependence as indicative of the expected theoretical uncertainty,
we obtain $\Delta\alpha_s(m_Z)=\pm 0.0005$, which is compatible with the theoretical uncertainty assessed by scale variation, which gives $\Delta\alpha_s(m_Z)=\pm 0.0008$.
In Sec.~\ref{sec:inputUpdate} we will discuss this fit range analysis further in the context of our final $\alpha_s$ extraction.

It should be pointed out that in the results displayed in Figs.~\ref{fig:thrust2010} and \ref{fig:thrust2024}, we also show two fit-ranges which were not considered in Ref.~\cite{Abbate:2010xh}, $[5\,{\rm GeV}/Q,0.15]$ and $[6\,{\rm GeV}/Q,0.15]$, which are clearly dominated by the dijet resummation region. With the 2010 profiles, the $[5\,{\rm GeV}/Q,0.15]$ fit-range gives a particularly small $\alpha_s(m_Z)$.
With the 2015 profiles this fit-range yields a result which is perfectly consistent with all other fit ranges, including those having larger $\tau_{\rm max}$.

Overall, with the 2015 profile functions we see no sign of instability concerning the dataset choice and also obtain consistent results if we restrict the data to be more dijet dominated. The fact that this consistency is observed even without considering any of the theoretical uncertainties, which of course are included in the final analysis of Sec.~\ref{sec:inputUpdate}, underlines the improvement achieved using the 2015 profiles rather than the 2010 profiles.

\begin{figure}[t!]
\centering
\begin{subfigure}[b]{0.47\textwidth}
\includegraphics[width=\textwidth]{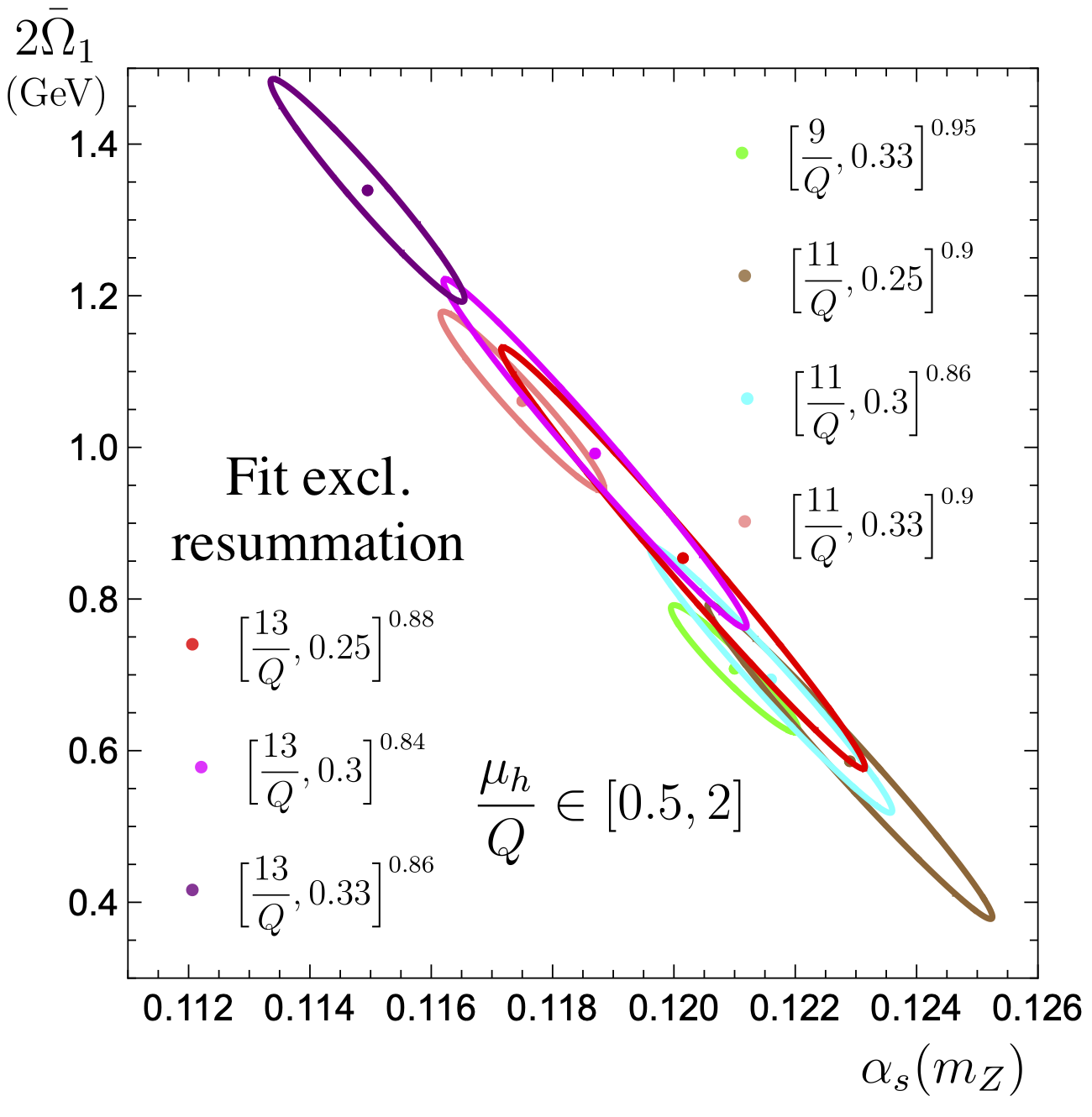}
\vspace*{-1.0cm}
\caption{\label{fig:thrustF0}}
\end{subfigure}
~
\begin{subfigure}[b]{0.475\textwidth}
\includegraphics[width=\textwidth]{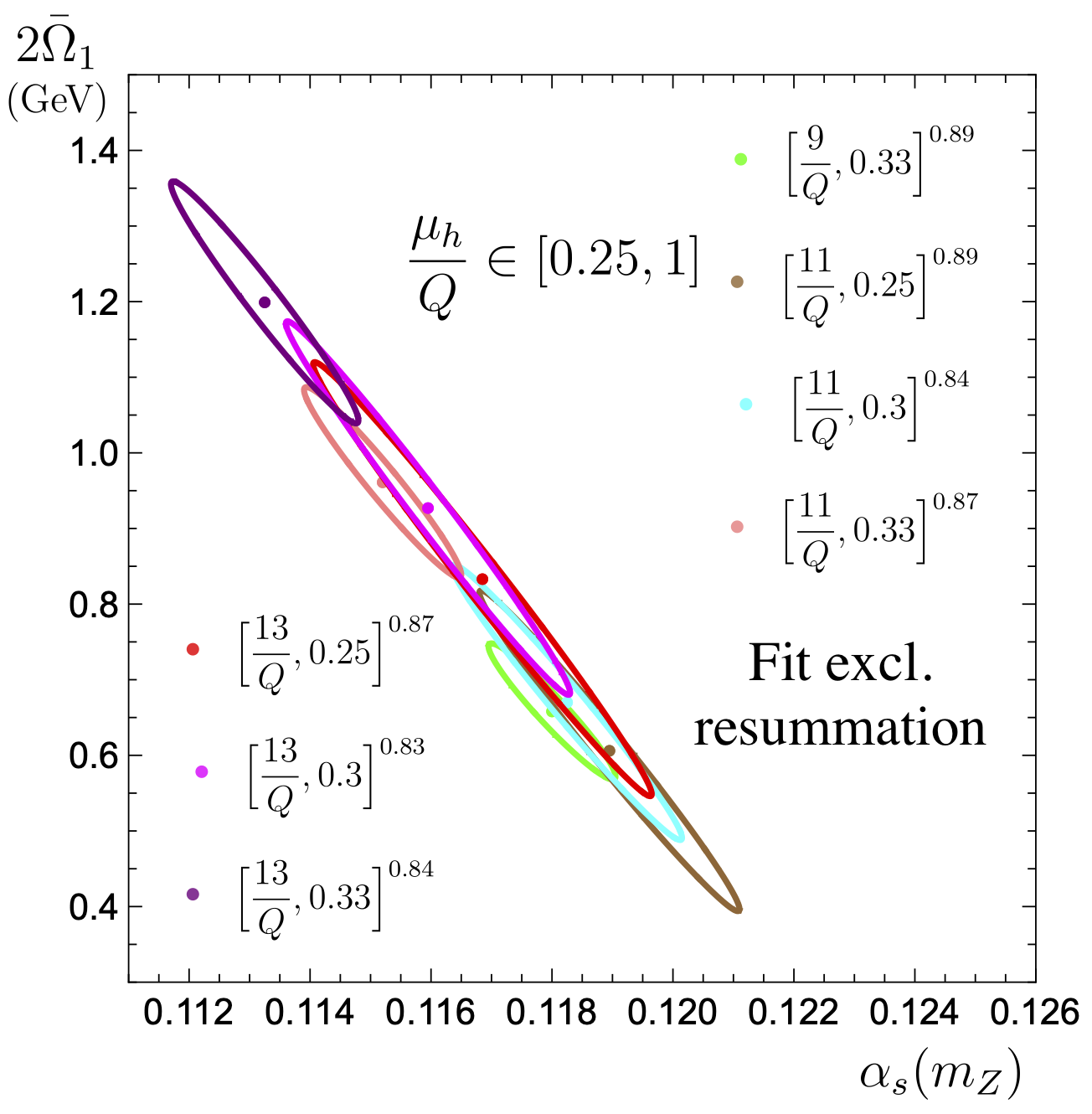}
\vspace*{-1.0cm}
\caption{\label{fig:thrustF0NZ}}
\end{subfigure}
\caption{\label{fig:thrustStabilityFOnoGap}
Comparison of the stability of the fit results with respect to variations of the dataset. Only experimental ellipses are shown, which correspond to $39\%$ confidence level in two parameters. Here we show the outcome of the fits when using a pure fixed-order theoretical description with shape function in the $\overline{\rm MS}$ scheme with different ranges for the renormalization scale variation $\mu_h$ in panels (a) and (b). The intervals represent the $\tau$ range of a given dataset, being the reduced $\hat \chi^2=\chi^2/{\rm dof}$ shown as a superscript: $[\tau_{\min},\tau_{\max}]^{\hat \chi^2}$.}
\end{figure}

As already mentioned, in Ref.~\cite{Nason:2023asn} the need and the reliability of summing up higher order logarithmic corrections in the tail region for $\tau\gtrsim 0.1$ was questioned, and it was argued that fixed-order results should be employed in that region. However, from the analysis shown in in Fig.~\ref{fig:XSRatio} and conducted in Sec.~\ref{subsubsec:ResummationXSlevel} we already found that up to ${\cal O}(\alpha_s^3)$ the fixed-order results at consecutive orders consistently approach the highest order N$^3$LL$^\prime$+${\cal O}(\alpha_s^3)$ resummed results [which are fully matched to the ${\cal O}(\alpha_s^3)$ fixed-order distribution] for $\tau<0.3$. This indicates that the yet unknown ${\cal O}(\alpha_s^4)$ fixed-order corrections yield even better agreement with the resummed predictions in this region. In any case, there is no evidence that the resummed distributions may yield inconsistent predictions for this range of $\tau$. As an additional response to the claim of Ref.~\cite{Nason:2023asn} we consider it warranted to explore the stability of fits that are based on the pure fixed-order predictions with respect to the data range. In Fig.~\ref{fig:thrustStabilityFOnoGap} we show the results obtained using the ${\cal O}(\alpha_s^3)$ fixed-order theoretical prediction in the $\overline{\rm MS}$ scheme.\footnote{If we use renormalon subtractions in the fixed-order cross section, i.e.\ $\Omega_1^R$ in the R-gap scheme, we find that ellipses are centered at smaller values, $\alpha_s(m_Z)\simeq 0.114$, but with a similar spread in results, showing sensitivity to the fit range. In this case, the fit range yielding the rightmost ellipse stretches up to $\alpha_s(m_Z)\simeq 0.118$, while the fit range for the leftmost ellipse reaches down to $\alpha_s(m_Z)\simeq 0.110$.} The fits account for scale variation with $\mu = \xi Q$ in the range $\xi \in [0.5,2]$ in Fig.~\ref{fig:thrustF0} and $\xi \in [0.25,1]$ in Fig.~\ref{fig:thrustF0NZ}.
In contrast to our analysis based on the N$^3$LL$^\prime$+${\cal O}(\alpha_s^3)$ resummed distribution, we only employ fit-ranges with a larger lower bound, i.e.\ $\tau_{\rm min}=(9,11,13){\rm GeV}/Q$, such that $\tau>0.1$ for the $Z$-pole data following Ref.~\cite{Nason:2023asn} and their argumentation concerning the potential irrelevance of resummation in that regime. Overall, due to the reduced statistics, the experimental error ellipses are somewhat larger. The important observation is that we see a very large fit interval dependence, where the best-fit $\alpha_s(m_Z)$ values range between $0.1150$ and $0.1229$ in Fig.~\ref{fig:thrustF0}. Furthermore, there is a tendency for lower $\alpha_s(m_Z)$ when $\tau_{\rm min}$ is increased, but the best-fit value of $\alpha_s(m_Z)$ can also depend significantly on the $\tau_{\rm max}$ value. For the lower renormalization scales used in Fig.~\ref{fig:thrustF0NZ}, where the fixed-order description is closer to the resummed cross section, with best-fit $\alpha_s(m_Z)$ values between $0.1133$ and $0.1190$, indicating somewhat improved stability. This again goes in the direction of supporting our conclusion of having improved stability with the inclusion of resummation.

Since the fit-range dependence can be taken as indicative of the expected theoretical uncertainty, this leads to an estimate of $\Delta\alpha_s(m_Z)\simeq \pm 0.0028$ for the ${\cal O}(\alpha_s^3)$ fixed-order results using $\mu=Q/2$. This can be compared with the result from scale variation $\mu\in [Q/4,Q]$, which gives a compatible estimate $\Delta\alpha_s(m_Z)\simeq \pm 0.0026$.
This factor of five increase in theoretical uncertainty compared to the resummed result should be attributed to the lack of important higher-order logarithmic contributions, even for datasets where dijet events may be less dominating.
Averaging the central results in Fig.~\ref{fig:thrustF0NZ} gives
$\alpha_s(m_Z) \big|_{{\rm FO\: at\: } {\cal O}(\alpha_s^3)} \simeq 0.1160 \pm 0.0026$, which is consistent with the world average, but also equally consistent with smaller values of $\alpha_s(m_Z)$.

\subsection{Gap Scheme Dependence of Fit Results \label{subsec:gapFit}}

\begin{figure}[t!]
\centering
\includegraphics[width= 0.5\textwidth]{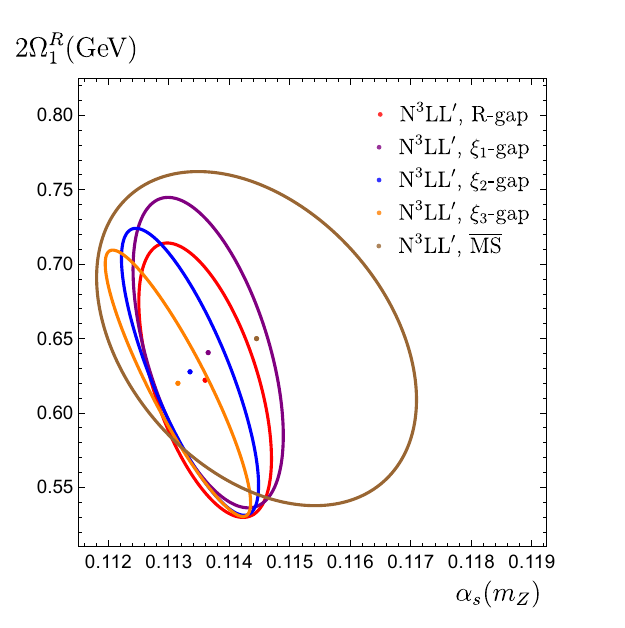}
\caption{\label{fig:gapIntNormFit}
Best-fit results in the $\alpha_s(m_Z)$-$2\Omega_1^R$ plane for different gap subtraction schemes and in $\overline{\rm MS}$. Ellipses show the combined experimental and perturbative theoretical uncertainties at the 39\%-confidence level in two dimensions. The best-fit values for $\Omega_1$ in the various schemes considered have been converted to the R-gap scheme $\Omega_1^R$.}
\end{figure}

In Fig.~\ref{fig:gapSchemesXS} we showed the excellent compatibility of our N$^3$LL$^\prime$+${\cal O}(\alpha_s^3)$ resummed thrust distribution with respect to different choices of the gap subtraction scheme. We also observed that all gap scheme predictions accumulate within the much larger $\overline{\rm MS}$ uncertainty band. We now examine if this behavior is also imprinted on the results obtained in fits. In Fig.~\ref{fig:gapIntNormFit} we show the outcome for the fit-range $\tau \in [(6\,{\rm GeV})/Q,0.15]$ based on the N$^3$LL$^\prime$+${\cal O}(\alpha_s^3)$ resummed distribution in the R-gap, $\xi_{1,2,3}$-gap and $\overline{\rm MS}$ schemes for $\Omega_1$, including experimental and theoretical uncertainties as described in Sec.~\ref{sec:expData}. In all cases, the best-fit values for $\Omega_1$ are converted to the reference R-gap $\Omega_1^R$ (with $R_0=\mu_0=2$\,GeV), as described in Sec.~\ref{subsubsec:gapSchemes}, at ${\cal O}(\alpha_s^3)$. If variations of the gap subtraction scheme lead to consistent results (with the scheme choice as a subdominant uncertainty), we should obtain consistent results for $\alpha_s(m_Z)$ and $\Omega_1^R$ from these different fits.

We see in Fig.~\ref{fig:gapIntNormFit} that this is indeed the case as very little variation is observed on the fit results in the $\alpha_s(m_Z)$-$\Omega_1^R$ plane, having ellipses centered at $(\alpha_s(m_Z),\Omega_1^R)\approx (0.1135,0.32\,{\rm GeV})$, each with uncertainties $(\delta\alpha_s(m_Z),\delta\Omega_1^R)\approx \pm (0.0013,0.10\,{\rm GeV})$.
The variation of the best fit results among the four gap subtraction schemes corresponds to an uncertainty in $\alpha_s(m_Z)$ of $0.0003$. Even fits with the N$^3$LL$^\prime$+${\cal O}(\alpha_s^3)$ resummed distribution for the $\overline{\rm MS}$ $\overline\Omega_1$ (converted perturbatively to the $\Omega_1^R$ scheme at ${\cal O}(\alpha_s^3)$), yield a result for $\Omega_1^R$ fully consistent with those obtained with gap subtractions. In $\overline{\rm MS}$ the best-fit value for $\alpha_s(m_Z)$ is located at $\alpha_s(m_Z)\approx 0.1145$ with an uncertainty of $\delta\alpha_s(m_Z)\approx 0.0025$. We see that the results follow precisely the pattern observed in the theoretical distribution shown in Fig.~\ref{fig:gapSchemesXS}.
The result in Fig.~\ref{fig:gapIntNormFit} shows in particular that we can get reliable results using solely the R-gap scheme and that uncertainties from variations in the gap scheme are already covered by the profile-function variations, which thus properly quantify the perturbative uncertainty.

\section{On the Validity of the Dijet Factorization Approach
\label{sec:argumentationDiTrijet}}

In the previous sections we have reviewed and made improvements to the N$^3$LL$^\prime$+${\cal O}(\alpha_s^3)$ factorized and resummed description of the thrust distribution. The resummation of logarithms and the treatment of the non-perturbative corrections are based on the leading power
factorization of the collinear and large-angle soft dynamical modes that emerge in the small $\tau$ dijet region. We have made a particular effort to combine all known ingredients (resummed singular jet, soft and hard matching functions, non-singular contributions, non-perturbative corrections, renormalon subtractions) such that our best prediction can be applied for the description of the entire thrust spectrum even including $\tau>0.3$, which are away from the dijet regime. In Sec.~\ref{subsec:comparison} we demonstrate that the resulting theoretical description provides an excellent description of the experimental data, not only within the thrust ranges used for the $\alpha_s$ fit analysis, but also for all thrust values and for all energies $Q$. This was the spirit of the 2010 analysis in Ref.~\cite{Abbate:2010xh} using the interval $\tau\in[(6\,{\rm GeV})/Q,0.33]$ as the default fit range, and we note that similar ranges were used in other thrust fits such as Refs.~\cite{Dissertori:2007xa,Becher:2009th,Davison:2009wzs,Gehrmann:2012sc}.
An essential aspect of our default formalism is that the effects of resummation are smoothly switched off for $\tau\gtrsim 0.25$ and that the description becomes purely fixed-order for $\tau\gtrsim 0.4$. In Secs.~\ref{subsubsec:ResummationXSlevel} and \ref{subsec:stability} we have also addressed the concern brought forward in Ref.~\cite{Nason:2023asn} that the resummation of logarithms based on the dijet factorization may be unreliable for $\tau\gtrsim 0.1$, finding no evidence for any untrustworthy behavior of the resummed plus fixed-order distribution, relative to the purely fixed-order results, for all $\tau$ values up to $0.3$, i.e.\ covering the range of thrust values typically used for $\alpha_s$ determinations. In fact, for $\tau<0.3$ the fixed-order results up to ${\cal O}(\alpha_s^3)$ clearly converge towards the N$^3$LL$^\prime$+${\cal O}(\alpha_s^3)$ resummed distribution, and it seems quite likely that this will continue to happen at even higher orders in the fixed-order expansion,
see the upper panels of Fig.~\ref{fig:XSRatio}.

Apart from their reservation concerning the reliability of the summation of logarithms, Ref.~\cite{Nason:2023asn} also expressed serious doubts concerning the validity of the parametrization of the hadronization corrections based on the dijet factorization (in terms of a shape function) for thrust values $\tau\gtrsim 0.1$. They argued that the impact of hard three-jet configurations would render the dijet treatment of non-perturbative corrections ---\,and the associated error estimates\,--- invalid. This section is dedicated to address these concerns. Their objections concerning hadronization corrections is of course tightly connected to their criticism of the summation of logarithms as both arise from the dijet factorization theorem, but we find it useful to discuss both aspects separately. First, in Sec.~\ref{ubsec:rangeDijet} we review arguments concerning the validity of the dijet factorization approach based on scaling arguments and based on the concrete structure of the fixed order series in $\alpha_s$. In Ref.~\cite{Nason:2023asn}, building on earlier work carried out in Refs.~\cite{Caola:2021kzt,Caola:2022vea}, a parametric calculation of three-jet non-perturbative power corrections was presented to support their concern. In Sec.~\ref{subsec:3jetarguments} we therefore
discuss the general structure of linear power corrections, putting in perspective the model results of Ref.~\cite{Nason:2023asn} for the thrust distribution accounting for the dijet, the three-jet and the transition regimes. This allows to see more clearly
the assumptions made in Ref.~\cite{Nason:2023asn} for these calculations. We reveal that their interpretation of the results for small $\tau$ values close to $0.1$ is tied to the implicit assumption that non-perturbative modes can resolve hard three-jet configurations
for any thrust value, including $\tau\ll 0.1$.
However, this assumption is in sharp contradiction to the existence of the dijet region and even the validity of Monte Carlo generators like Pythia, Herwig and PanScales, which are based on summing the same leading power logarithms, and we disagree strongly.

However, in view of the recent insights provided in Refs.~\cite{Luisoni:2020efy,Caola:2021kzt,Caola:2022vea}, we do acknowledge that it is possible that a strict dijet treatment of hadronization effects misses
hadronization corrections from three-jet events at larger values of $\tau$.
Such three-jet events can yield additional sources of hadronization corrections, whose uncertainty should be assessed, which we will do below.
Currently there are no concrete first-principles insights on how to treat the
{\it transition} of non-perturbative effects between the dijet and the three-jet regions,
or on the dominant non-perturbative parameters in the transition or three-jet region of thrust (effectively we lack the analog of the QCD matrix element defining $\Omega_1$ and higher-order matrix elements that are derived using field theory for the dijet region).
So far these effects have been assessed in models that only have a single hadronization parameter. The predictions from these models imply concrete relations between the hadronization corrections in the dijet, transition, and trijet regions which, however, have no clear justification from QCD.

In view of all relevant aspects in our discussion, which are summarized in Sec.~\ref{subsec:intermediate},
for our final $\alpha_s$ extraction in this article, we
develop a modified treatment of the hadronization corrections to account for additional uncertainties related to the potential deviations with respect to the strict dijet treatment in our default factorization approach. This includes uncertainties from both three-jet hadronization corrections and also subleading power dijet non-perturbative corrections (which were not estimated in Refs.~\cite{Nason:2023asn}).
We also discuss how to reduce the thrust dataset so that dijet events clearly dominate. This is the content of Sec.~\ref{subsec:3jetuncertainties}.

\subsection{Range of the Dijet Region}
\label{ubsec:rangeDijet}

We start by discussing the validity of the dijet factorization approach of Eq.~(\ref{eq:factorizationFormula})
from the parametric point of view. In the dijet region, apart from the (local) hard modes with momentum scaling\footnote{We use the usual light-cone momentum decomposition $p^\mu=(p^+,p^-,p_\perp)=p^+ \bar n^\mu/2+p^- n^\mu/2+p^{\mu}_\perp$, with $n^\mu=(1,\vec{n})$, $\bar n^\mu=(1,-\vec{n})$, where $\vec{n}$ is the thrust axis.} $p^\mu_{\rm hard}
\sim Q(1,1,1)$, we have the dynamical $n$- and $\bar n$-collinear modes with momenta $p^\mu_{{\rm c},n}\sim Q(\tau,1,\sqrt{\tau})$ and $p^\mu_{{\rm c},\bar n}\sim Q(1,\tau,\sqrt{\tau})$, respectively, and the large-angle soft perturbative modes with $p^\mu_{\rm s}=Q(\tau,\tau,\tau)$. These dijet modes are distinct and represent the relevant modes to consider for the validity of Eq.~(\ref{eq:factorizationFormula}) as long as the hierarchy
\begin{equation}
\label{eq:tauhierarchy}
\tau\ll \sqrt{\tau}\ll 1\,,
\end{equation}
holds. The leading hadronization effects when the inequality~(\ref{eq:tauhierarchy}) and thus the dijet factorization approach are valid, arise from the thrust hemisphere constraints imposed on non-perturbative large-angle soft modes with momenta $p^\mu_{s,{\rm np}} \sim \Lambda_{\rm QCD}(1,1,1)$.
These leading hadronization effects are fully encoded in the shape function $F_\tau$. For $Q\tau \sim \Lambda_{\rm QCD}$, which is associated to the peak region, the detailed form of the shape function $F_\tau$ is mandatory for the theoretical distribution, while for $Q\tau \gg \Lambda_{\rm QCD}$, i.e.\ in the dijet tail region, the OPE description mentioned in Sec.~\ref{subsec:nonpertCorr} is valid. There are also non-perturbative collinear modes with momenta $p^\mu_{{\rm np},n}\sim \Lambda_{\rm QCD}(\sqrt{\tau},1/\sqrt{\tau},1)$ and $p^\mu_{{\rm np},\bar n}\sim \Lambda_{\rm QCD}(1/\sqrt{\tau},\sqrt{\tau},1)$. However, their leading contributions are only quadratically sensitive to $\Lambda_{\rm QCD}$ and furthermore suppressed compared to the non-perturbative large-angle soft modes due to the larger virtuality $Q\sqrt{\tau}$ of the perturbative collinear modes. The effects arising from the non-perturbative collinear modes can therefore be neglected. This has been reconfirmed recently in Ref.~\cite{Gracia:2021nut} by computations of the thrust distribution in the large-$\beta_0$ approximation, and in Refs.~\cite{Hoang:2018zrp,Hoang:2024zwl} through an analytic and numerical examination of the parton shower cutoff dependence of the thrust distribution obtained from the angular-ordered parton shower of the Herwig event generator.

\begin{figure}[t!]
\centering
\includegraphics[width= 0.6\textwidth]{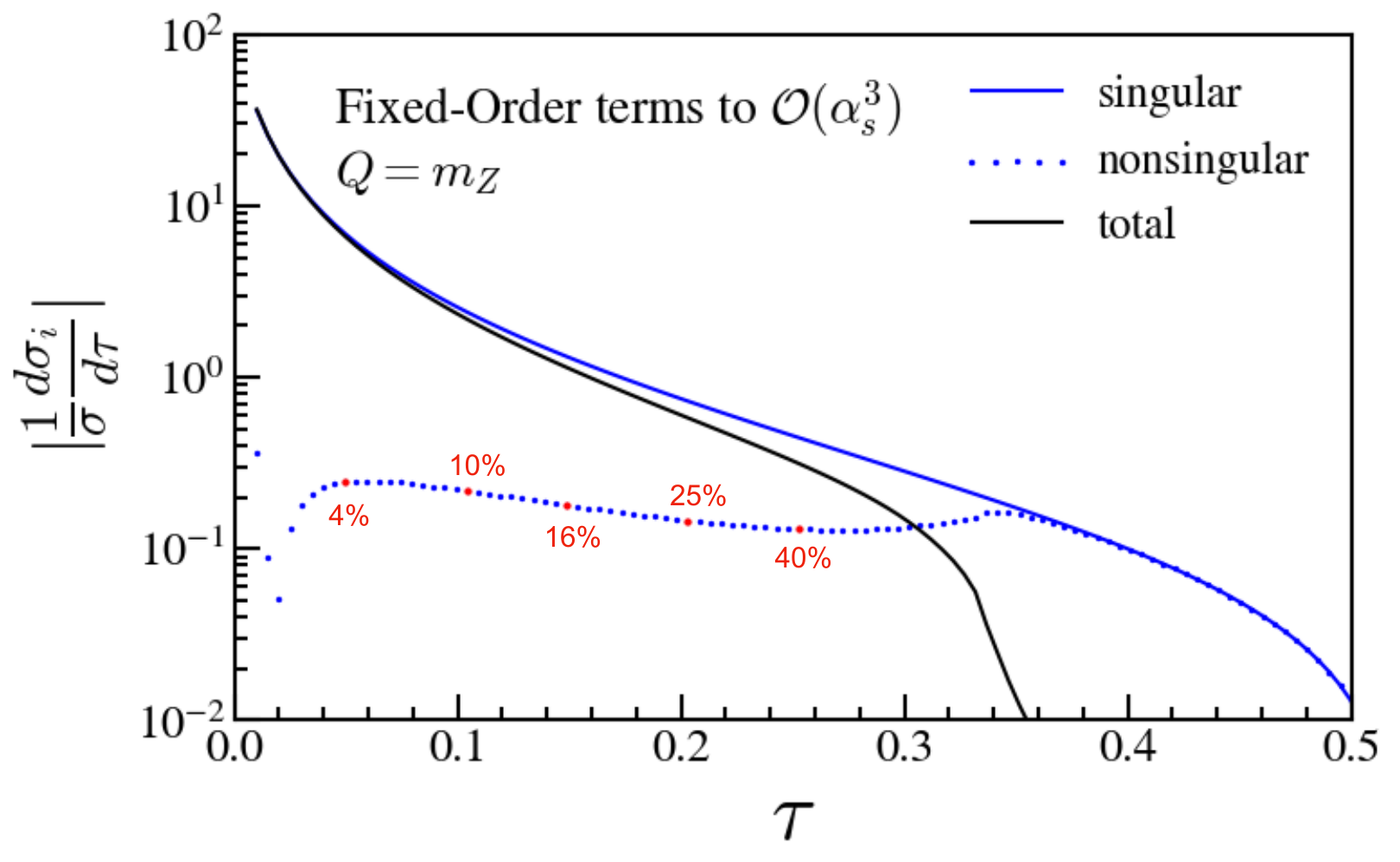}
\caption{\label{fig:XScontributions}
The size of the singular (blue solid line) and non-singular (blue dotted line) components of the ${\cal O}(\alpha_s^3)$ fixed-order cross section for $\mu=Q=m_Z$. The red numbers denote the relative size of the non-singular contributions to the total differential fixed-order cross section at $\tau=\{0.05,0.1,0.15,0.2,0.25\}$.}
\end{figure}

It is now natural to ask what in practice is the upper bound on $\tau$ so that the inequality~(\ref{eq:tauhierarchy}) is satisfied. Assuming that the inequality $a\gg b$ concretely means $a>3 b$ we obtain
\begin{equation}
\label{eq:taudijetbound}
\tau < 0.11\,,
\end{equation}
which is consistent with our examinations concerning the validity of the resummation of logarithms in Secs.~\ref{subsubsec:ResummationXSlevel} and \ref{subsec:stability}. Nevertheless, if correct, it should also be true that the singular contributions in the known fixed-order corrections of the thrust distribution, obtained from $\mathrm{d}\hat{\sigma}_{\rm s}/\mathrm{d}\tau$ in Eq.~(\ref{eq:singFactorizationFormula}) upon expansion, should largely dominate over the non-singular contributions obtained from $\mathrm{d}\hat{\sigma}_{\rm ns}/\mathrm{d}\tau$ in Eq.~(\ref{eq:nonSing}). This is analyzed in Fig.~\ref{fig:XScontributions} where the singular (blue solid line) and the (modulus of the) non-singular (blue dashed line) contributions in the fixed-order thrust distribution up to $\mathcal{O}(\alpha_s^3)$ are displayed for $\mu=Q=m_Z$. The solid black line represents the total fixed-order result.\footnote{A similar figure is shown in Fig.~7 of Ref.~\cite{Abbate:2010xh}, however there the singular N$^3$LL$^\prime$+${\cal O}(\alpha_s^3)$ resummed thrust distribution was compared to the
fixed order non-singular contribution. This contrasts to the plot shown here where both components are fixed order.} At $\tau=0.1$ the non-singular contributions only amount to 10\% of the total result, i.e.\ the singular contributions are $11$ times larger than the non-singular ones. At $\tau=0.2$ the non-singular contributions amount to 25\% of the total result, i.e.\ the singular contributions are still 5 times larger than the non-singular ones. We believe that this refutes any claim that the dijet factorization were unreliable and untrustworthy in the region around $\tau=0.1$.
This strongly favors the conclusion that the dijet treatment of non-perturbative corrections is perfectly valid and important not only for $\tau < 0.11$, but even for thrust values up to $0.2$. Thus, once again, we do not find any evidence that would support the claims made in Ref.~\cite{Nason:2023asn} regarding the dijet treatment of non-perturbative corrections (and the summation of logarithms) becoming invalid in the region around and above $\tau=0.1$.

For our analysis we consider the thrust range $\tau<0.11$ as the region where the dijet treatment of non-perturbative corrections is strictly valid beyond any reasonable doubt. For $\tau>0.11$ the dijet treatment is still valid, but modifications on the parametrization of non-perturbative effects (that increase with $\tau$) could arise due to the growing size of the non-singular contributions.

\subsection{Model for Three-Jet Power Corrections
\label{subsec:3jetarguments}}
In Ref.~\cite{Caola:2021kzt,Caola:2022vea,Nason:2023asn} a model for the dominant non-perturbative power correction in the thrust tail region was devised. It was based on a computation of the leading linear infrared sensitivity of the NLO QCD corrections to the process $e^+e^-\to q\bar q\gamma$ due to a fictitious gluon mass $\lambda$ and the additional assumption that the corresponding corrections arising from (anti)quark-gluon dipoles for $e^+e^-\to q\bar q g$, which involve the triple gluon interaction, can be obtained by a modification of the color factor for $e^+e^-\to q\bar q \gamma$.\footnote{In Ref.~\cite{Caola:2021kzt,Caola:2022vea,Nason:2023asn}, these results based on working in the linear approximation for $\lambda$, were also used to quantify the ambiguity caused by the corresponding ${\cal O}(\Lambda_{\rm QCD})$ infrared renormalon. From the perspective of quantifying the parametric dependence of non-perturbative effects, the gluon mass and the renormalon computations are equivalent.} For a given $\tau$ value of the hard $q\bar q g$ configuration the model quantifies the leading linear non-perturbative correction. Their analysis differs from the previous discussions in the literature on the leading linear non-perturbative power correction for the thrust tail region that were based on the pure dijet $q\bar q$ final state~\cite{Korchemsky:1994is,Gardi:2001ny,Lee:2006nr,Mateu:2012nk}. In this subsection we explain the connection between both approaches, an issue not discussed in Ref.~\cite{Nason:2023asn}, and we also clarify why their model cannot be used to make any statements on the validity of the dijet treatment of non-perturbative corrections.

As already mentioned before, in the dijet factorization theorem the dominant non-perturbative corrections arise from low-momentum large-angle soft radiation acting on the (unresolved) harder perturbative dijet configuration that determines the plane between the two thrust axis hemispheres.
It should be remembered that the definition of jet depends on specifying an angular jet size, the so-called jet radius,
and that the same final state particles can be resolved as a dijet event with a larger jet radius, or as a trijet event with a smaller jet radius.
These harder configurations, described by the perturbative collinear and soft radiation, mostly constitute two jets, but may sometimes also have three or even more visible jets arising from the collinear and harder soft dynamics. This does not contradict the dijet treatment of the non-perturbative effects, since the essential point is that the non-perturbative large-angle soft modes, in most cases, cannot resolve such multijet configurations if they are collimated together within one hemisphere, see e.g.~Ref.~\cite{Bauer:2011uc} for the theoretical description of collinear subjects within a larger jet and Ref.~\cite{Larkoski:2017jix} for a review of jet substructure, which provides a number of tools for such calculations.
A similar feature arises for boosted top-antitop pair production, where large-angle soft radiation in the c.m.\ frame cannot resolve the top quark decay~\cite{Fleming:2007qr,Fleming:2007xt}.

For thrust values (sufficiently) above the thrust peak location, the leading linear non-perturbative power correction in the dijet OPE is then related to a shift in the partonic thrust distribution $\mathrm{d}\hat\sigma/\mathrm{d}\tau$ proportional to the first moment of the shape function $\Omega_1$, see Eq.~(\ref{eq:Om1Moment}):
\begin{align}
\label{eq:shiftedthrust}
\frac{\mathrm{d}\sigma}{\mathrm{d}\tau}(\tau,Q,\Lambda_{\rm QCD})
=
\frac{\mathrm{d}\hat\sigma}{\mathrm{d}\tau}\biggl(\tau-\frac{2\Omega_1}{Q},Q\biggr)
+\mathcal{O}\biggl(\frac{\Lambda^2_{\rm QCD}}{Q^2 \tau^3}\biggr)\,.
\end{align}
Note that on the LHS we have made the dependence of the hadron-level distribution $\mathrm{d}\sigma/\mathrm{d}\tau$ on the c.m.\ energy and non-perturbative effects explicit by the arguments $Q$ and $\Lambda_{\rm QCD}$, respectively. As this relation applies for any gap subtraction (as well as for the $\overline{\rm MS}$ scheme $\overline{\Omega}_1$) we also suppress the scheme-dependence of $\Omega_1$.
Defining the cumulative thrust distribution by
$\Sigma(\tau) \equiv \int_0^{\tau} \mathrm{d}\tau^\prime
\frac{\mathrm{d}\sigma}{\mathrm{d}\tau}(\tau^\prime)$
this leads to the following relation between the hadronic and partonic cumulative distributions for $\tau$ values above the peak, as long as the dijet parametrization of the non-perturbative corrections is applicable, see inequalities~(\ref{eq:tauhierarchy}) and (\ref{eq:taudijetbound}):
\begin{align}
\label{eq:cumiulantdiffference}
\frac{\Sigma(\tau,Q,\Lambda_{\rm QCD})-\hat\Sigma(\tau,Q)}{\frac{\mathrm{d}\hat\sigma}{\mathrm{d}\tau}(\tau,Q)}
\stackrel{\rm dijet}{=} -\frac{2\Omega_1}{Q} +
\mathcal{O}\biggl(\frac{\Lambda^2_{\rm QCD}}{Q^2 \tau}\biggr).
\end{align}
The generalization of this formula for all $\tau$-values can now also be easily written down in the form
\begin{align}
\label{eq:cumiulantdiffferencev2}
\frac{\Sigma(\tau,Q,\Lambda_{\rm QCD})-\hat\Sigma(\tau,Q)}{\frac{\mathrm{d}\hat\sigma}{\mathrm{d}\tau}(\tau,Q)}
&= -\frac{2\Omega_1}{Q}\, h\biggl(\tau, \frac{\Lambda_{\rm QCD}}{Q} \biggr)\,,
\end{align}
where we defined the function $h(\tau,\Lambda_{\rm QCD}/Q)$. The dependence on both its arguments is in principle non-perturbative, and at this time has not been quantified from first principles away from the dijet region. Therefore, even if we had a non-perturbative calculation of $h(\tau,\Lambda_{\rm QCD}/Q)$, for which we can take $Q\to \infty$,
then the resulting function $h(\tau,0)$ is still non-perturbative. However, we know that $h(\tau,\Lambda_{\rm QCD}/Q)$ is essentially constant and very close to unity in the dijet region, i.e.\ $h(\tau,\Lambda_{\rm QCD}/Q)\approx 1$, which is physically related to the non-perturbative soft modes not being able to resolve the harder perturbative modes, as already mentioned above. Mathematically this corresponds to taking the limit of small $\tau$ in $h(\tau,\Lambda_{\rm QCD}/Q)$.

\begin{figure}[t!]
\centering
\includegraphics[width=0.5\textwidth]{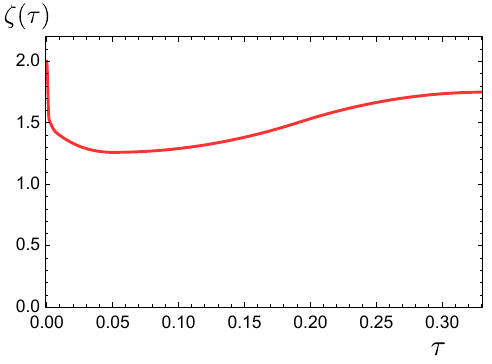}
\caption{ \label{fig:modelNZ}
Plot of the model function $\zeta(\tau)$ of Ref.~\cite{Nason:2023asn}.}
\end{figure}

Away from the dijet region, where three-jet configurations dominate and can be resolved by non-perturbative modes, we do not know anything about this function for thrust from first principles.
This can be compared to the model calculation obtained in Ref.~\cite{Nason:2023asn}, which can be written as
\begin{align}
\label{eq:cumiulantdiffferencev3}
\frac{\hat\Sigma(\tau,Q)-\Sigma(\tau,Q,\Lambda_{\rm QCD})}{\frac{\mathrm{d}\hat\sigma}{\mathrm{d}\tau}(\tau,Q)}
\stackrel{\rm three-jet}{=} -\frac{\tilde H_{\rm NP}}{Q}\, \zeta(\tau)
+ {\cal O}\biggl( \frac{\Lambda_{\rm QCD}^2}{Q^2}\biggr)
\,,
\end{align}
where the model enables an explicit computation of $\zeta(\tau)$. Ref.~\cite{Nason:2023asn} carried out calculations for thrust and a number of other event-shape and $e^+e^-$ observables. Technically, they obtained $\zeta(\tau)$ for all $\tau$ values by an expansion in the gluon mass $\lambda$, assuming it is the smallest scale regardless of the value of $\tau$, and keeping the linear $\lambda$ term. For thrust, the model states that $h(\tau,\Lambda_{\rm QCD}/Q)\approx \tilde H_{\rm NP}/(2\Omega_1)\, \zeta(\tau)$ in this three-jet regime. This provides a parametrization with a different non-perturbative parameter $\tilde H_{\rm NP}$ and thus a new non-perturbative parameter in the ratio $\tilde H_{\rm NP}/(2\Omega_1)$.
The form of the function $\zeta(\tau)$ is shown in Fig.~1 of Ref.~\cite{Nason:2023asn}, and for convenience of the reader we display this function in Fig.~\ref{fig:modelNZ}. We see that, except for extremely small thrust values, it takes values between $1.25$ (around $\tau\approx 0.05$) and $1.75$ (for $\tau\gtrsim 0.3$) and has a very smooth and overall rather flat behavior.

The conspicuous aspect of the function $\zeta(\tau)$ is that it exhibits a very sharp rise for extremely small $\tau$ values, where there are no events in the experimental data. This behavior differs sharply from the flat behavior of $h(\tau,\Lambda_{\rm QCD}/Q)$ in the dijet regime. They also found that eventually $\zeta(\tau)$ approaches the value $2$ for $\tau\ll 0.01$, which is what one expects if \Eq{cumiulantdiffferencev3} holds for all $\tau$. Analogous observations were made in Ref.~\cite{Nason:2023asn} for other event-shapes. Based on this, they concluded that the $\zeta(\tau)$ must be interpreted as a good model function for $2 h(\tau,\Lambda_{\rm QCD})$ for almost all $\tau$ values, and that the identity $\tilde H_{\rm NP}=2\Omega_1$ holds, which implies that the leading dijet and three-jet power corrections are associated to the exact same non-perturbative parameter. As a consequence of the sharp rise at very small $\tau$, they infer that the traditional dijet treatment of non-perturbative corrections based on Eq.~(\ref{eq:cumiulantdiffference}) is unreliable for the thrust values that have been commonly used for strong-coupling determinations, including the 2010 analysis in Ref.~\cite{Abbate:2010xh}.

However, we should point out that in the calculation of the function $\zeta(\tau)$ in Ref.~\cite{Nason:2023asn}, the exact range of validity in $\tau$ for the small $\lambda$ expansion has not been analyzed, so that the actual range of validity of Eq.~(\ref{eq:cumiulantdiffferencev3}) in the dijet regime is unknown. Physically, their expansion is associated to the assumption that the $q\bar q g$ final state is resolved as 3 jets by the non-perturbative modes for any $\tau$ value.%
\footnote{Technically, the validity of the expansion in the small gluon mass $\lambda$ in a fixed-order calculation for the thrust distribution for fixed $\tau$, breaks down in the dijet regime due to the appearance of large power corrections such as $[\lambda/(Q\tau)]^n\sim [\lqcd/(Q\tau)]^n$. Therefore in the dijet region the correct approach is to apply the dijet approximation first and only then expand in the small gluon mass. The same statement applies in applications of infrared renormalon calculus. To reliably study the transition region, a gluon mass calculation without any expansion would be necessary.} In addition, their power correction model based on the finite gluon mass $\lambda$, yields relations for different kinds of non-perturbative effects, which are unrelated in full QCD. This is also true for the dijet and the three-jet power corrections, where the dijet matrix element $\Omega_1$ is defined from the matrix element in Eq.~(\ref{eq:MSbarOmega1}), with Wilson lines for two light-like directions. In contrast, the non-perturbative matrix element for the dominant three-jet power correction will depend on Wilson lines for three light-like directions, see Refs.~\cite{Stewart:2014nna,Bhattacharya:2022dtm}. The non-perturbative three-jet Wilson line matrix element depends also on additional parameters, which are the angles between the direction of the lines. Only in the limit where an angle goes to zero and Wilson lines can be recombined, can one derive a connection between the three-jet non-perturbative function and the dijet non-perturbative parameter~\cite{Stewart:2014nna}. Away from this limit the angular dependence is non-perturbative. Since there is no other known relation between the two non-perturbative matrix elements, the assumption that $\tilde H_{\rm NP}=2\Omega_1$ is unreliable.

Finally, we also point out that in a recent analytic and numerical study of the coherent branching algorithm in the Herwig~7.2 event generator~\cite{Hoang:2018zrp,Hoang:2024zwl}, the cumulative difference of Eq.~\eqref{eq:cumiulantdiffferencev2} was studied for different parton-shower cutoff values. This plays a similar role as the finite gluon mass concerning the parametric sensitivity to non-perturbative effects. In that analysis, the flat behavior of $h(\tau,\Lambda_{\rm QCD}/Q)$ follows Eq.~\eqref{eq:cumiulantdiffference} up to thrust values of $0.2$, see Fig.~1 in Ref.~\cite{Hoang:2024zwl}. This is consistent with the dijet treatment being a good approximation for $\tau$ values up to $0.2$.

We therefore conclude that the model calculations presented in Ref.~\cite{Nason:2023asn} is only relevant for discussing non-perturbative corrections when three-jet configurations dominate. However, their function $\zeta(\tau)$ cannot be applied in the dijet regime. In particular, the function $\zeta(\tau)$ neither provides any information on the dijet regime itself or its $\tau$ range of validity, nor does it give insight on the important transition region between the dijet and three-jet regions. Furthermore, the knowledge of their function $\zeta(\tau)$ does not specify the size of the three-jet region where $\zeta(\tau)$ is valid in a strict sense (at least within the model).

Nevertheless, assuming ---\,hypothetically\,--- that the form of Eq.~(\ref{eq:cumiulantdiffferencev3}) is the correct leading non-perturbative power correction for all $\tau$ values above the peak, we can test what the impact is on the outcome of an $\alpha_s$ fit. The implementation of a $\tau$-dependent power correction in the form of Eq.~(\ref{eq:cumiulantdiffferencev3}) is straightforward in our factorization approach starting from the expression of the hadron-level thrust distribution in Eq.~(\ref{eq:hadronleveldistrv1}), where the parton-level distribution and the shape function $F_\tau(R_0,\mu_0,k)$ at the reference scales $R_0$ and $\mu_0$ are fully separated. To achieve this we apply the replacement $2\Omega_1(R_0,\mu_0)\to \zeta(\tau) \tilde H_{\rm NP}$. Given the analytic ansatz of the shape function in Eq.~(\ref{eq:shapeFun}), which yields the relation~(\ref{eq:2Omega1shapefct}) for the shape function's first moment, we can implement the three-jet model of Ref.~\cite{Nason:2023asn}, imposing the condition $2\tilde H_{\rm NP}= \lambda+2\Delta_0$, by applying the replacements \mbox{$\lambda\to\frac{1}{2}\lambda \zeta(\tau)$} and $\Delta_0\to\frac{1}{2}\zeta(\tau) \Delta_0 $ in the analytic expression of the shape function $F_\tau(R_0,\mu_0,k)$ of Eq.~(\ref{eq:shapeFun}). Furthermore, for setups in which the renormalon is subtracted, we also need to rescale $\delta(R,\mu_s)\to \frac{1}{2}\zeta(\tau)\delta(R,\mu_s)$ and $\bar\Delta(R,\mu_s)\to \frac{1}{2}\zeta(\tau)\bar\Delta(R,\mu_s)$.

\begin{figure}[t!]
\centering
\begin{subfigure}[b]{0.49\textwidth}
\includegraphics[width=\textwidth]{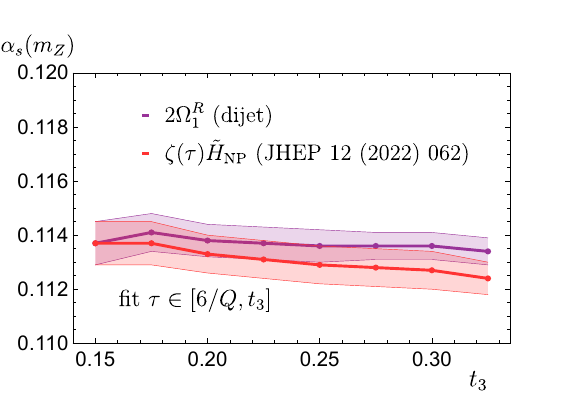}
\vspace*{-1cm}
\caption{\label{fig:alphasPC}}
\end{subfigure}
\begin{subfigure}[b]{0.49\textwidth}
\includegraphics[width=\textwidth]{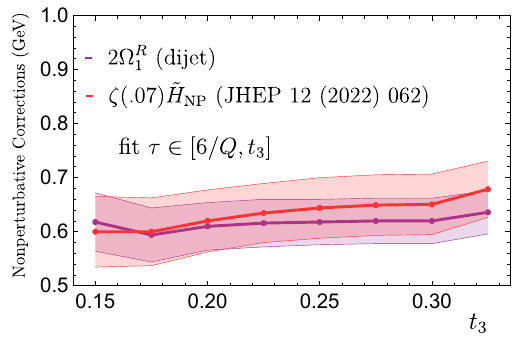}
\vspace*{-1cm}
\caption{\label{fig:pureOmega1PC}}
\end{subfigure}
\\
\caption{\label{fig:PCmodelComparison}
Comparison of results obtained using the dijet power correction prediction for the power correction $\Omega_1$ (purple curves) in the whole spectrum with those obtained when using the treatment of power correction model in Refs.~\cite{Caola:2021kzt,Caola:2022vea} (red curves).
We show results for $\alpha_s(m_Z)$ in panel (a) and those for the hadronization correction in panel (b), where we vary the upper limit of the fit range $t_3$.}
\end{figure}

We carried out the following comparative toy analysis using our (best) N$^3$LL$^\prime$+${\cal O}(\alpha_s^3)$ resummed distribution in the R-gap scheme for the default profile function (i.e.\ not accounting for any perturbative uncertainties) using data in the thrust intervals $[(6\,{\rm GeV})/Q,t_3]$ for $0.15\leq t_3\leq 0.3$. We determined $\alpha_s(m_Z)$ and $\Omega_1^R$ using our dijet treatment for non-perturbative corrections, and with a different fit also $\alpha_s(m_Z)$ and $\tilde H_{\rm NP}$ using the three-jet model of Ref.~\cite{Nason:2023asn} for the non-perturbative power correction. The outcome of both analyses for $\alpha_s(m_Z)$ as a function of $t_3$ is shown in Fig.~\ref{fig:alphasPC}, where the red dots correspond to the best fits for the three-jet model of Ref.~\cite{Nason:2023asn} and the purple dots indicate the best fit values for the strict dijet treatment. The colored bands represent the corresponding experimental uncertainties, while theoretical uncertainties are not displayed. Interestingly, we find perfectly compatible values for $\alpha_s(m_Z)$ around $0.114$ for both fits. There is a slight tendency of decreasing $\alpha_s(m_Z)$ for the three-jet model with increasing $t_3$, but the effect is not very significant. The reason for the consistency of $\alpha_s(m_Z)$ values obtained with both approaches is that the function $\zeta(\tau)$ in Fig.~\ref{fig:modelNZ} is fairly flat, with values between $1.3$ and $1.4$ for the lower part of the fit interval
and reaching up to $1.75$ when $\tau$ approaches $0.3$.
The dominant effect when using the three-jet model is associated to a rescaling
by $\zeta(0.07)/2$ for the non-perturbative correction, which has no effect
on the physical description. Thus the entire effect on the value of $\alpha_s$ is small.
This behavior can be seen in Fig.~\ref{fig:pureOmega1PC}, where we display the fit results for $2\Omega_1^R$ in the strict dijet treatment and $\zeta(0.07)\,\tilde H_{\rm NP}\approx 1.3\,\tilde H_{\rm NP}$ obtained from the fit with the three-jet model. Here the factor $\zeta(0.07)\approx 1.3$ is picked as the typical value of the function $\zeta(\tau)$ in the lower part of the fit interval, which has the highest weight in the fits. The fit results for both quantities are again mutually compatible.
For completeness, we have we also carried out the fit-range stability analysis from Sec.~\ref{subsec:stability} using the $\tau$-dependent shape function based on the function $\zeta(\tau)$ determined in Ref.~\cite{Nason:2023asn}. We find that all the observations concerning stability remain true with minimal modifications.
\begin{figure}[t!]
\centering
\includegraphics[width=0.5\textwidth]{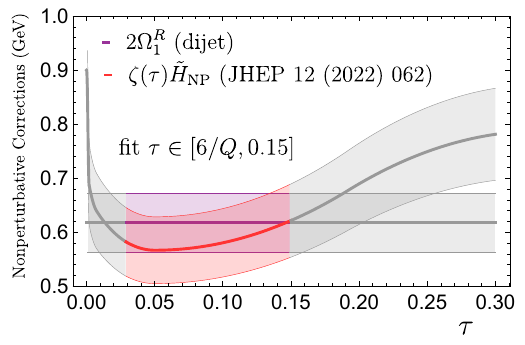}
\caption{\label{fig:zeatOmega1}
Fit results for $2\Omega_1^R$ or $\zeta(\tau)\Tilde{H}_{\rm NP}$, as a function of $\tau$ using data in the range $\tau \in [(6\,{\rm GeV})/Q,0.15]$. We see that, irrespective of the treatment of the non-perturbative corrections entering the fit, a consistent behavior of the power corrections is observed in the fit window, indicated by the colored area.}
\end{figure}
Another way of looking at these results is shown in Fig.~\ref{fig:zeatOmega1}, where we display
the outcome of $2\Omega_1^R$ and $\zeta(\tau)\,\tilde H_{\rm NP}$ for fits within the dijet-dominated fit region $[(6\,{\rm GeV})/Q,0.15]$.
We see that within the actual fit range, shaded red and purple, the results for $2\Omega_1^R$ and $\zeta(\tau)\,\tilde H_{\rm NP}$ are very similar taking values around $0.62$\,GeV.
By examining the $t_3$ dependence in Fig.~\ref{fig:PCmodelComparison}, we see that the mild 30\% rise between $\zeta(0.15)$ and $\zeta(0.3)$ does not have a strong impact on the fit results.

\subsection{Intermediate Summary and Discussion}
\label{subsec:intermediate}

Let us briefly summarize the situation before elaborating on the updated fit strategy we will apply given all the insights we gained in the previous sections:
\begin{itemize}
\item[(a)] There is no evidence that the summation of logarithms based on the dijet factorization is unreliable for thrust values up to $0.3$ (see Secs.~\ref{subsubsec:ResummationXSlevel} and \ref{subsec:stability}). The use of gap subtractions yields an improved perturbative stability as it removes the effects of the most important ${\cal O}(\Lambda_{\rm QCD})$ infrared renormalon arising from the large-angle soft modes. Furthermore, we find consistency between different gap subtraction schemes and the associated conversion of the obtained results for gap scheme dependent $\Omega_1$ values (see Secs.~\ref{sec:normchoice} and \ref{subsec:stability}). We therefore employ dijet resummation of logarithms and gap subtractions in the context of the factorization formula~(\ref{eq:factorizationFormula}) for the final fits, which was also the basis of the $\alpha_s$ determination from 2010 of Ref.~\cite{Abbate:2010xh}. The factorization formula includes smooth matching to the fixed-order regime for $\tau>0.3$ supplemented by self-normalization for the total cross section which, with the new result for $s_3$ obtained in Ref.~\cite{Baranowski:2024vxg}, is in perfect agreement with the fixed-order result. Finally, we have updated the 2- and 3-loop non-singular contributions using more precise numerical determinations.
\item[(b)] The treatment of non-perturbative corrections based on the dijet factorization is valid for $\tau<0.11$, the region where the singular perturbative contributions dominate over the power-suppressed non-singular terms by a factor of $10$ or more. However, for $\tau>0.11$ the relative size of the non-singular contributions smoothly increases (becoming as large as the singular contribution at $\tau\approx 0.35$) indicating that non-perturbative corrections related to three-jet configurations should become more relevant as we increase $\tau$. At this time we do not have any concrete and reliable information on the size and form of the three-jet-type non-perturbative power corrections. For the final strong-coupling determination it is therefore conservative to allow for a deviation of the dijet treatment of non-perturbative corrections for $\tau>0.11$ due to the potential impact of hadronization effects related to three-jet configurations.
\item[(c)] Within the dijet region, the non-perturbative corrections related to subleading-power effects in $\tau$ and associated to the non-singular cross section contributions are known to partly arise from the leading-order shape function $F_\tau$, and partly from subleading shape functions and collinear hadronization~\cite{Feige:2017zci,Moult:2018jjd,Moult:2019mog,Beneke:2022obx}.
These are not covered by our default formula in Eq.~(\ref{eq:Om1nonSing}), that was constructed so that the singular and non-singular perturbative distributions merge to the fixed-order result in the far-tail region. Such subleading power dijet non-perturbative effects have a different character than the three-jet power corrections which affect singular and non-singular as a whole, but there is currently also no concrete knowledge on their form, even though they are expected to be rather small. We carry out a quantitative analysis of these corrections in Sec.~\ref{subsec:3jetuncertainties}.
\end{itemize}

\subsection{Non-Perturbative Uncertainties from Beyond the Leading Dijet Corrections
\label{subsec:3jetuncertainties}}

To reduce the impact of the unknown three-jet non-perturbative corrections as much as possible, and in the absence of an exact knowledge on the size of the dijet to three-jet transition region (which may extend up to $\tau\sim 0.35$) it is conservative to significantly restrict the dataset used for $\alpha_s$ extractions toward small values of $\tau$, in order to have a higher fraction of events in the dijet regime, but maintaining at the same time a reasonable amount of data statistics. This should limit the impact of the three-jet non-perturbative corrections so that they can be simply modeled and then accounted for as an additional uncertainty. In a similar manner, we also analyze the uncertainty arising from modeling the deviation of the dijet non-singular power correction from our default treatment in Eq.~(\ref{eq:Om1nonSing}).

\begin{figure}[t!]
\centering
\begin{subfigure}[b]{0.44\textwidth}
\includegraphics[width=\textwidth]{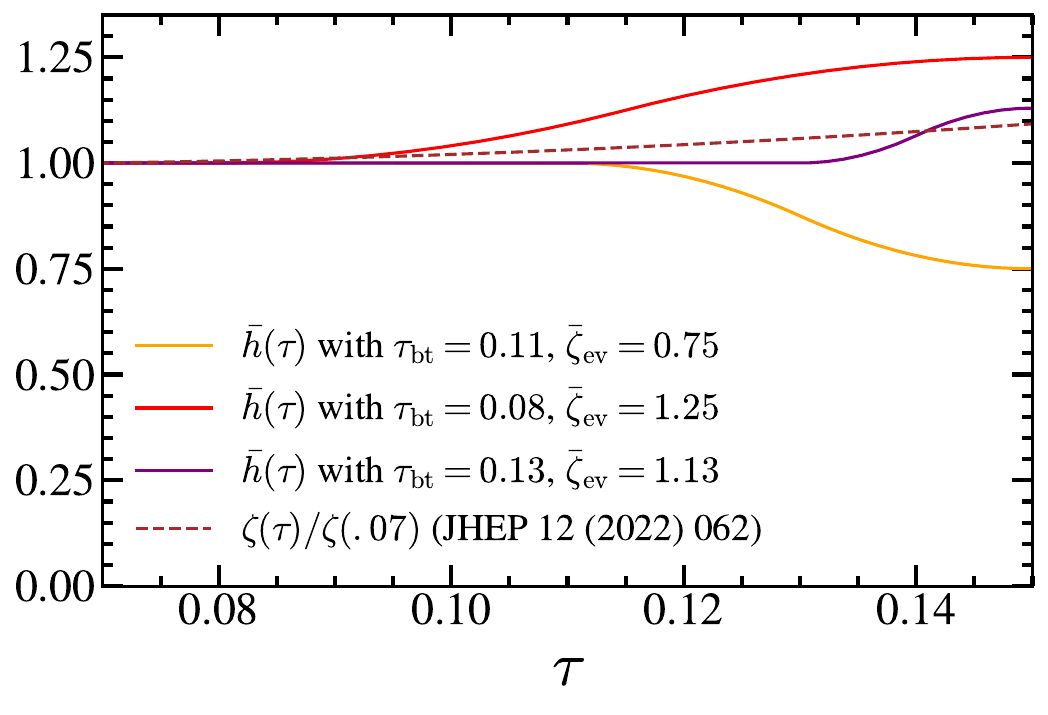}
\caption{\label{fig:zetaBarModel}}
\end{subfigure}
~
\begin{subfigure}[b]{0.477\textwidth}
\includegraphics[width=\textwidth]{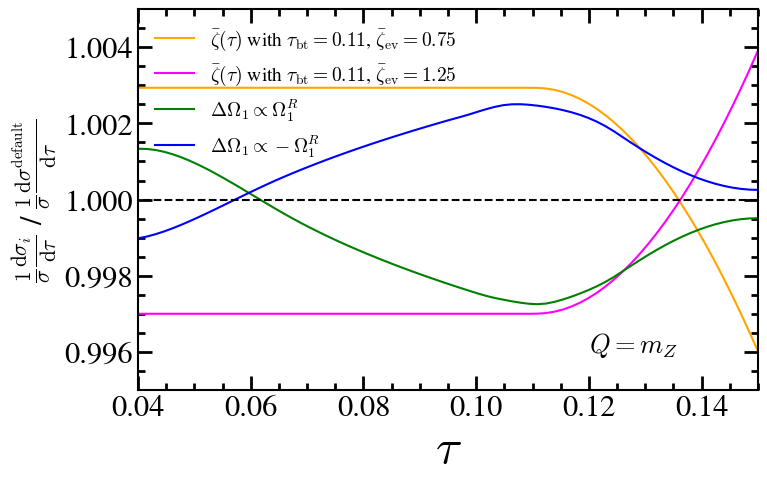}
\caption{\label{fig:uncertaintiesRatio}}
\end{subfigure}

\begin{subfigure}[b]{0.495\textwidth}
\includegraphics[width=\textwidth]{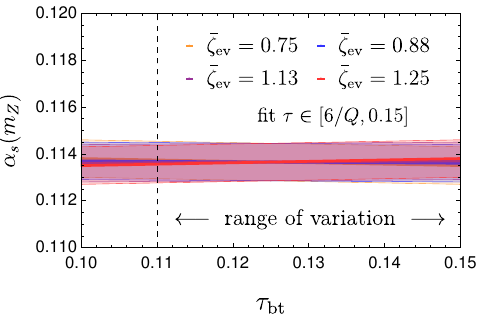}
\caption{\label{fig:alphasTauBt}}
\end{subfigure}
~
\begin{subfigure}[b]{0.482\textwidth}
\includegraphics[width=\textwidth]{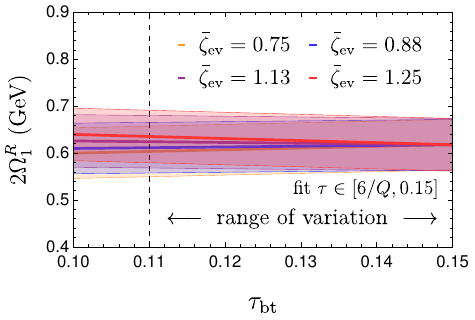}
\caption{\label{fig:omega1TauBt}}
\end{subfigure}
\caption{\label{fig:paramsTauBt}
Panel (a) shows deviations from the dijet treatment of power corrections versus $\tau$ from different model functions (red, purple, and yellow curves). The dashed line denotes the model of Ref.~\cite{Nason:2023asn} normalized by its value at $\tau=0.07$. Panel (b) shows the impact on the differential cross section of the two largest dijet deviation models (yellow and magenta curves) and the model used to estimate uncertainties from the treatment of power corrections in the dijet nonsingular distribution (blue and green curves). The lower two panels (c) and (d) depict the results for $\alpha_s(m_Z)$ and $\Omega_1^R$ when varying $\tau_{\rm bt}$ and $\bar \zeta_{\rm ev}$ in Eq.~(\ref{eq:modelFuncDef}). The uncertainty bands contain only experimental uncertainties for the fit range $(6\,{\rm GeV})/Q \le \tau \le 0.15$.}
\end{figure}

For the final $\alpha_s$ determination discussed in Sec.~\ref{sec:inputUpdate} we therefore apply the following changes compared to the 2010 analysis of Ref.~\cite{Abbate:2010xh}:
\begin{itemize}
\item[(1)] We reduce the default fit range from $[(6\,{\rm GeV})/Q,0.33]$ used in Ref.~\cite{Abbate:2010xh} to the range
$[(6\,{\rm GeV})/Q,0.15]$. For $\tau < 0.15$ the resummation of logarithms based on the dijet factorization is clearly valid. On the other hand, the impact of potential non-perturbative corrections from three-jet configuration is expected to be small in this new fit region, so that they can be accounted for as an additional source of uncertainty.

\item[(2)] We treat the deviations from the dijet treatment of the non-perturbative corrections due to three-jet configurations by making the reference shape function $F_\tau(R_0,\mu_0,k)$ of Eq.~(\ref{eq:shapeFun}) $\tau$-dependent through the rescalings $\lambda\to\lambda \bar h(\tau)$ and $\Delta_0\to\Delta_0 \bar h(\tau)$, which correspond to the replacement $\Omega_1\to \Omega_1 \bar h(\tau)$ for the non-perturbative correction in the OPE region. For the setup in which the renormalon is subtracted, identical rescalings are applied to $\delta(R,\mu_s)$ and $\bar \Delta(R,\mu_s)$.
This implements the general formula of Eq.~(\ref{eq:cumiulantdiffferencev2}) with a \mbox{$\tau$-dependence} on the RHS and follows the suggestion of Refs.~\cite{Caola:2022vea,Nason:2023asn} to account for deviations of the dijet treatment of hadronization corrections. However, we employ a different function $\bar h(\tau)$ that is consistent with the presence of a dijet regime and has the following form:
\begin{eqnarray}
\label{eq:modelFuncDef}
\bar h (\tau) = \left\{\begin{array}{lrcl}
1 &0 &\le& \tau < \tau_{\rm bt}
\\
\zeta(1 ,0,0,\bar\zeta_{\rm ev},\tau_{\rm bt},\tau_{\rm et},\tau)\qquad& \tau_{\rm bt} &\le& \tau < \tau_{\rm et}
\\
\bar \zeta_{\rm ev} & \tau_{\rm et} &\le& \tau < 0.5
\end{array}
\right.
\,.
\end{eqnarray}
This function equals unity for $0\le\tau<\tau_{\rm bt}$ in agreement with the dijet factorization. We vary $\tau_{\rm bt}$ in the interval $0.11\le \tau_{\rm bt}\le \tau_{\rm et}=0.225$ in our final analysis following our previous arguments. For $\tau > \tau_{\rm et}$ the function is again constant, attaining the value $\bar \zeta_{\rm ev}$. The parameter $\bar \zeta_{\rm ev}$ will be varied up and down around the value $1$ to account for the ignorance on the size and the sign of potential non-perturbative contributions from the three-jet configurations.
The value of $\tau_{\rm et}=0.225$ is outside our fit range, and therefore not varied for this analysis.
The function $\zeta$
applied in the region $\tau_{\rm bt}<\tau < \tau_{\rm et}=0.225$ is the double-quadratic already used in the construction of the profiles (see Sec.~\ref{subsec:profiles}), and smoothly interpolates between the lower and upper flat regions. To account for the uncertainties related to the unknown three-jet non-perturbative effects we randomly vary $\bar \zeta_{\rm ev}$ in the range $[0.75, 1.25]$ along with the profile function parameters, as described in Sec.~\ref{subsec:profiles}. The variation interval of $\bar \zeta_{\rm ev}$ is based on the assumption that at $\tau=0.15$ the maximal deviation from the dijet treatment of the non-perturbative corrections is $25\%$. Given that at $\tau=0.15$ the singular contributions are still larger than the non-singular contribution by more than a factor of $5$, we believe that this variation is conservative. The shape of $\bar h (\tau)$ for various values for $\bar \zeta_{\rm ev}$ and $\tau_{\rm bt}$ is shown in Fig.~\ref{fig:zetaBarModel}. For the cases $(\tau_{\rm bt},\bar\zeta_{\rm ev})=(0.11,0.75)$ (orange line) and $(0.11,1.25)$ (red line) we also show the ratio of the modified distribution with respect to the default distribution in Fig.~\ref{fig:uncertaintiesRatio}. We see that the overall relative size of the three-jet non-perturbative effects is at the level of three permille. Note that the additional three-jet power correction impacts the distribution for any $\tau$ through the normalization to the integrated cross section, see Sec.~\ref{sec:normchoice}. This is the reason why the ratio deviates from unity for $\tau<0.11$ even though $\bar h(\tau)=1$ in that range.

\item[(3)] We model deviations from our default treatment of the dijet non-singular hadronization corrections by introducing a non-singular term with a different leading power correction parameter which we add to our default non-singular factorization formula, see Eq.~(\ref{eq:Om1nonSing}). This additional term has the form
\begin{align}
\label{eq:factorizationFormula2}
\Delta \bigg[ \frac{\text{d}\sigma_{\rm ns}}{\text{d}\tau} \bigg]^{\rm NP} &=
\frac{\text{d} \hat{\sigma}_{\rm ns}}{\text{d}\tau} \biggl[\tau - \frac{2\Delta \Omega_1(\tau)}{Q}\biggr] - \frac{\text{d} \hat{\sigma}_{\rm ns}}{\text{d}\tau} (\tau)\,,
\end{align}
with
\begin{align}
\Delta \Omega_1(\tau) &= \Omega_1^R \,P(\tau)\,.
\end{align}
The overall scaling of the additional non-perturbative correction is set by $\Omega_1^R$ and the function $P(\tau)$ parametrizes the deviations. It is reasonable to assume that $P(\tau)$ is a function with size of order unity that should be rather flat in the tail region, where we carry out the $\alpha_s$ fits, and smoothly decreases in the region where the three-jet power correction increases. We also note that the ansatz of Eq.~(\ref{eq:factorizationFormula2}) is appropriate only in the tail region where the OPE can be applied, see Sec.~\ref{subsec:nonpertCorr}. If we apply it with a flat and finite $P(\tau)$ in the peak region, where the correct implementation would be in terms of a shape function, the logarithmic behavior of the non-singular cross section $\text{d}\sigma_{\rm ns}/\text{d}\tau$ for $\tau\to 0$ will lead to an unphysical singular behavior in Eq.~(\ref{eq:factorizationFormula2}). To avoid this, $P(\tau)$ needs to vanish for $\tau\to 0$. This choice is also consistent with the fact that the additional non-singular non-perturbative corrections should have an even smaller impact in the peak region.
The functional form of $P(\tau)$ we adopt is thus given by
\begin{eqnarray}
\label{eq:Ptau}
P (\tau) = \left\{\begin{array}{lrcl}
0& 0 &\le& \tau < 0.01
\\
\zeta(0,0,0,0.25,0.1,6/Q,\tau)&0.01 &\le& \tau < 6/Q
\\
\bar \zeta_{\rm ns} & 6/Q &\le& \tau < 0.1
\\
\zeta(0.25,0,0,0,0.1,0.15,\tau)\qquad& 0.1 &\le& \tau < 0.15
\\
0 & 0.15& \le &\tau < 0.5
\end{array}
\right.\,,
\end{eqnarray}
where we note that the particular decreasing form of $P(\tau)$ for $\tau<6/Q$, which is outside of our fit interval, can also be interpreted as a realization of a shape function for the peak region in the context of the ansatz in Eq.~(\ref{eq:factorizationFormula2}). In our fits
we vary $\bar \zeta_{\rm ns}$ conservatively between $\pm 1$. The relative size of these additional non-singular hadronization corrections relative to the default distribution is shown in Fig.~\ref{fig:uncertaintiesRatio} for $\bar \zeta_{\rm ns}=\pm 1$ (blue and green curves). We see that the overall magnitude of the effects caused by these additional dijet non-singular non-perturbative corrections are also small, below three per mille, and thus comparable to the effect of three-jet power corrections. However, both effects differ quite a lot concerning their $\tau$ dependence. While the three-jet power correction yields varying shape modifications only outside the dijet regime, the additional dijet non-singular hadronization effects show significant shape variations in the dijet regime and decrease for larger $\tau$ values.

\end{itemize}

Before we start our final analysis in the next section, let us take a brief look at the impact of the variations caused by the three-jet power correction model and the additional dijet non-singular hadronization correction. To that end, we carry out toy fits to the experimental data in the interval $[(6\,{\rm GeV})/Q,0.15]$, showing again only the experimental uncertainties to better see the impact of the effects. In Fig.~\ref{fig:alphasTauBt} we show the results for $\alpha_s(m_Z)$ using $\bar \zeta_{\rm ev}=0.75, 0.88, 1.13, 1.25$, as a function of $\tau_{\rm bt}$ covering values above $0.11$ (which is the choice we adopt in our final analysis) and below. The central colored solid lines are the best fit and the corresponding colored bands represent the experimental uncertainties. We see that $\alpha_s(m_z)$ remains stable around $0.114$ and that the most important effect is the slightly larger uncertainty caused by the variations of $\tau_{\rm bt}$ and $\bar\zeta_{\rm ev}$. We see that for $\alpha_s$ the effects from the three-jet power corrections vanish for $\tau_{\rm bt}\approx 0.126$ and lead to mild and continuously increasing uncertainties for smaller and larger $\tau$ values. For $\tau_{\rm bt}$ outside the fit range, the uncertainty bands freeze and does not increase further. For $\tau_{\rm bt}\ge 0.11$ we obtain an additional uncertainty of $\delta\alpha_s(m_Z)=\pm 0.0002$. The outcome for $\Omega_1^R$, which is illustrated in Fig.~\ref{fig:omega1TauBt}, remains stable as well with a value of around $0.3$\,GeV, but the additional uncertainty, which vanishes for $\tau_{\rm bt}>0.15$, continuously increases for decreasing $\tau_{\rm bt}$. The uncertainty band again freezes for $\tau_{\rm bt}< 6/Q$. For $\tau_{\rm bt}\ge 0.11$ we obtain an additional uncertainty of $\delta\Omega_1^R=\pm 0.01$\,GeV. The impact of the additional dijet non-singular hadronization corrections on the fit results for $\alpha_s(m_Z)$ and $\Omega_1^R$ depends linearly on $\bar\zeta_{\rm ns}$. For $\bar\zeta_{\rm ns}=\pm 1$ the additional uncertainties amount to $\delta\alpha_s(m_Z)=\pm 0.00015$ and $\delta\Omega_1^R=\pm 0.004$\,GeV.

Overall, the effects from the additional three-jet and dijet non-singular hadronization corrections are both an order of magnitude smaller than the perturbative uncertainties from the variations of the profile functions and the experimental uncertainties. Nevertheless, we include them in our final error bucket for completeness.

\section{Dijet $\alpha_s$ Extraction \label{sec:inputUpdate}}

In this section we present the results of our final strong coupling determination using our N$^3$LL$^{\prime}$ $+\mathcal{O}(\alpha_s^3)$ resummed thrust distribution in the R-gap scheme, which represents the main outcome of the phenomenological analysis in this article. All details on the implementation of our theory prediction have been given in Sec.~\ref{sec:thrustDistr}, which discusses how we include and go beyond the formalism developed in Ref.~\cite{Abbate:2010xh}.
We remind the reader that we
will primarily quote results without including finite bottom quark mass and QED corrections, in order to facilitate direct comparison with other analyses, even though such effects can be included (as was done in Ref.~\cite{Abbate:2010xh}).
Our theory predictions are based on the most comprehensive and rigorous combination of all known components including fixed-order, resummation, and hadronization effects, including both the leading-power dijet factorization theorem and beyond.

A key aspect of our analysis is accounting
for uncertainties associated to the potential impact of three-jet non-perturbative power corrections for which, currently, no first-principle QCD treatment exists. This is done
in a conservative manner by reducing the thrust fit region from the $[(6\,{\rm GeV})/Q,0.33]$ used in Ref.~\cite{Abbate:2010xh}
to $[(6\,{\rm GeV})/Q,0.15]$, which diminishes the potential contributions from three-jet hadronization effects significantly, and by considering a model for these three-jet power corrections which allows us to assess this additional sources of uncertainty (Sec.~\ref{subsec:3jetuncertainties}). Furthermore, we account for the uncertainties coming from modelling subleading dijet power corrections (Sec.~\ref{subsec:3jetuncertainties}) and from subleading shape function OPE ($\Omega_2$) corrections (Sec.~\ref{subsubsec:reviewGap}).
These uncertainties are assessed without performing fits for any additional hadronic parameters.
Overall, these three sources of non-perturbative effects yield uncertainties that are much smaller than our experimental and perturbative uncertainties.\footnote{On the perturbative side, we do not incorporate resummation of leading-logarithms on the next-to-leading partonic corrections~\cite{Moult:2018jjd,Moult:2019uhz,Beneke:2022obx}. We believe these should be small since the dijet cross section overly dominates in the fit region. The corresponding uncertainties should be covered by the variations of the renormalization scales.} A discussion on all uncertainties we include is the main content of Sec.~\ref{subsec:fitResults}.

We emphasize that our combined default theoretical N$^3$LL$^{\prime}$ $+\mathcal{O}(\alpha_s^3)$ resummed thrust distribution
(without the effects of three-jet and subleading dijet power corrections),
provides an excellent description of the experimental data, not only for the thrust interval used for the $\alpha_s$ extraction, but also for all $\tau$ and $Q$ values. This is demonstrated in Sec.~\ref{subsec:comparison}.

\subsection{Results from Fits}
\label{subsec:fitResults}

\begin{figure}[t!]
\centering
\begin{subfigure}[b]{0.485\textwidth}
\includegraphics[width=\textwidth]{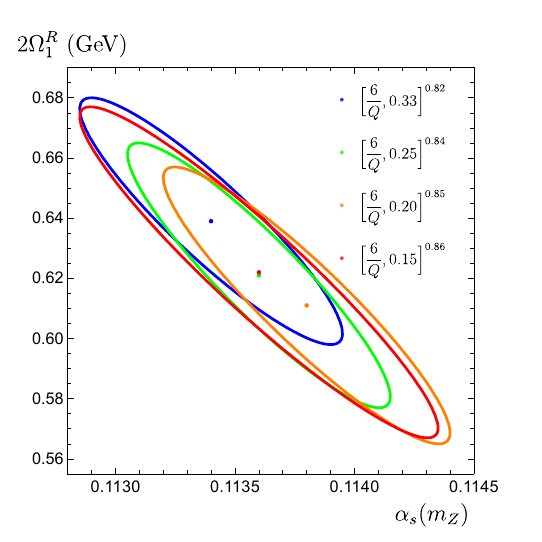}
\caption{\label{fig:expErrEllipse}}
\end{subfigure}
~
\begin{subfigure}[b]{0.485\textwidth}
\includegraphics[width=\textwidth]{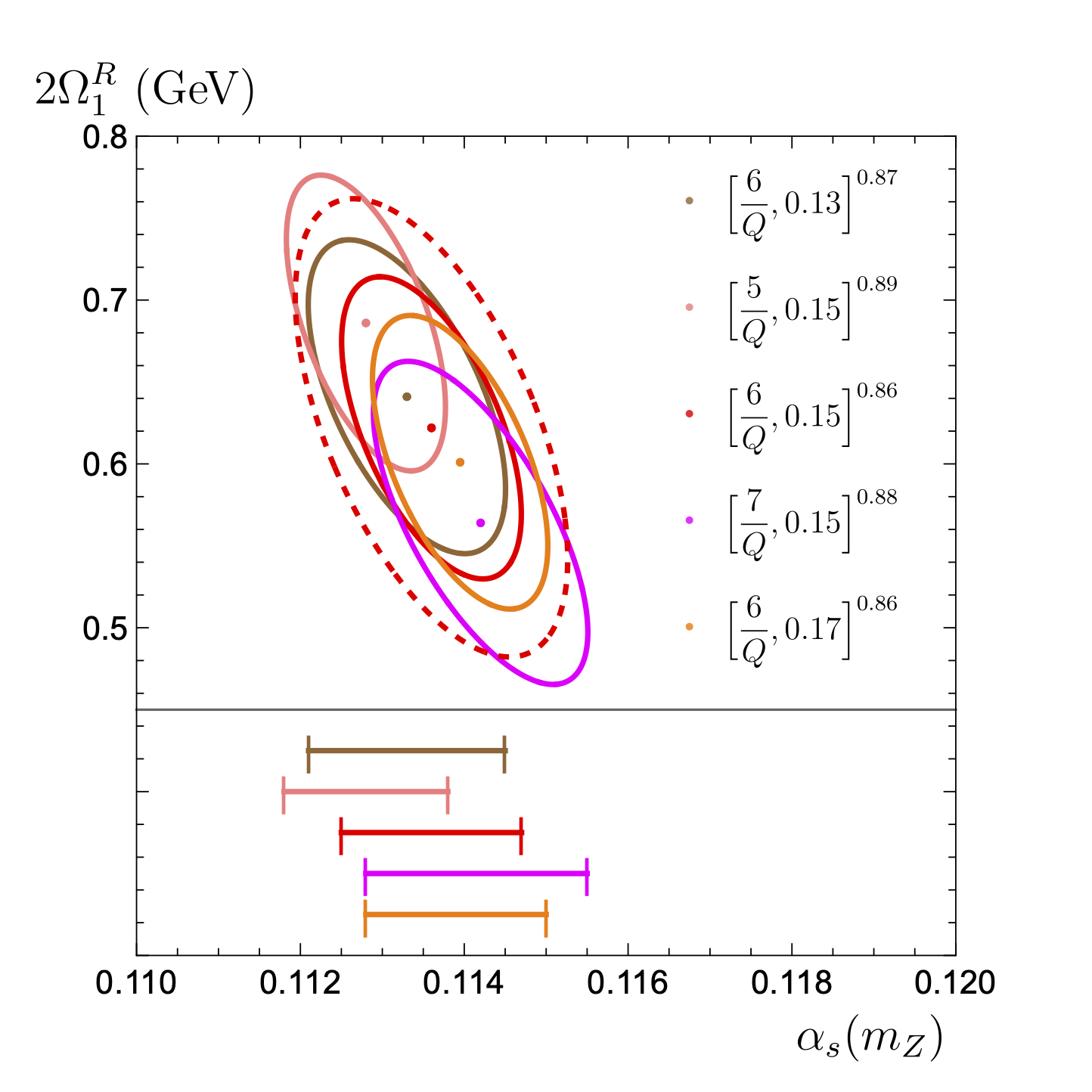}
\caption{\label{fig:dijetVariation}}
\end{subfigure}
\caption{\label{fig:expDijetvariation}
Panel (a) shows results for our default fit to $\alpha_s$ and $\Omega_1^R$ using different datasets $\tau \in [\tau_{\rm min},\tau_{\rm max}]^{\chi^2/{\rm dof}}$
with larger choices of the upper limit. Here the error ellipses only contain experimental uncertainties.
Panel (b) shows variations of the fit range about our default choice of $\tau \in [6\,{\rm GeV}/Q,0.15]$. Here the error ellipses contain both experimental and perturbative uncertainties.
The solid ellipses in both (a) and (b) correspond to $39\%$ confidence level (1-$\sigma$ in each of the parameters). In panel (b) we show these projections by error bars on the $x$-axis, and the dashed error ellipse corresponds to 68\% confidence level in two dimensions.}
\end{figure}

We start by briefly discussing the impact of changing the fit interval
from $[(6\,{\rm GeV})/Q,0.33]$ used in Ref.~\cite{Abbate:2010xh} to the new range
$[(6\,{\rm GeV})/Q,0.15]$. As already elaborated on in Sec.~\ref{subsec:stability}, the stability of our theory description in the dijet regime implies that the fit results in both cases are consistent. In Fig.~\ref{fig:expErrEllipse} the fit results for $\alpha_s(m_Z)$ are displayed for the upper bounds $\tau=0.15$ (red), $0.20$ (orange), $0.25$ (green) and $0.33$ (blue). The error ellipses (see Sec.~\ref{sec:expData}) only contain experimental uncertainties (i.e.\ renormalization-scale, three-jet, and subleading dijet power-correction uncertainties are not displayed) to better illustrate their dependence on the upper bounds. The ellipses, which also illustrate the correlation between $\alpha_s(m_Z)$ and $\Omega_1^R$, correspond to $39\%$ confidence level (CL), i.e.\ they are 1-$\sigma$ when projected onto a single parameter. The superscripts on the fit interval represent the $\chi^2$/dof value of the best fit. We see that the central values with $\alpha_s(m_Z)$ between $0.1134$ and $0.1138$ and the reduced $\chi^2$ with values between $0.82$ and $0.86$ are very stable, and that the primary change due to reducing the upper bound from $0.33$ to $0.15$ is an increase of the experimental uncertainty by about $30\%$.

\begin{table}[t!]
\centering
\begin{tabular}{c | c | c | c}
& $\delta\alpha_s(m_Z)$ & $\delta \Omega_1^R$
& Included in~\cite{Abbate:2010xh} \\ [0.5ex]
\hline
Experiment & $0.0003$ & $0.010$ & $\checkmark$ \\
$\Omega_1$/$\alpha_s$ & $0.0007$ & $0.026$ & $\checkmark$ \\
\bf{Total Experiment + $\Omega_1$/$\alpha_s$} & $\mathbf{0.0008}$ & $\mathbf{0.028}$ & $\checkmark$ \\
\hline
$\Omega_2$ hadronization & $0.0002$ & $0.013$ & $\checkmark$ \\
3jet hadronization & $0.0002$ & $0.010$ \\
Subleading power dijet & $0.0002$ & $0.004$ \\
{\bf Total subleading hadronization} & $\mathbf{0.0003}$ & $\mathbf{0.017}$ \\
\hline
Perturbative & $\mathbf{0.0008}$ & $\mathbf{0.037}$ & $\checkmark$ \\
\hline
{\bf Total} & $\mathbf{0.0012}$ & $\mathbf{0.049}$ \\
\end{tabular}
\caption{Sources of uncertainties for best-fit results. The first column contains the source of uncertainty. The second column quotes the contribution to the error on $\alpha_s(m_Z)$, whereas the third column quotes the corresponding contribution to the overall uncertainty on $\Omega_1^R$. It is also indicated in the last column which of these uncertainties were included in the 2010 analysis of Ref.~\cite{Abbate:2010xh}.}
\label{tab:uncertaintiesListed}
\end{table}

In the context of fit range stability, we also assess results from variations around our default by considering fits with the ranges $\tau \in [(6\pm 1\,{\rm GeV})/Q,0.15\pm 0.02]$. The outcome is shown in Fig.~\ref{fig:dijetVariation}, where the ellipses now contain both experimental and perturbative uncertainties. We observe that all best fit values for $\alpha_s(m_Z)$, represented by the dots in the center of each ellipse, are again contained within the perturbative uncertainty obtained from our default choice for the fit range (cf.~Sec.~\ref{subsec:stability}). This demonstrates that the theoretical uncertainty associated with varying the fit range provides a fully compatible estimate to our perturbative uncertainty from scale variation in the resummed analysis, just as was seen in Sec.~\ref{subsec:stability}. For this reason, we do not view the variation of the fit range as a new source of uncertainty.

As a last comment on the contributions to the total uncertainties on our best-fit results, we note that both perturbative as well as experimental uncertainties are included in an analogous manner to Ref.~\cite{Abbate:2010xh}, described in detail in Sec.~\ref{sec:expData}.

The final result for the $\alpha_s(m_Z)$-$\Omega_1^R$ fit using the interval $[(6\,{\rm GeV})/Q,0.15]$ and accounting for all uncertainties listed in Table~\ref{tab:uncertaintiesListed} reads
\begin{align}\label{eq:finalFitResults}
\alpha_s(m_Z) &= 0.1136 \pm 0.0012_{\text{tot}}\,,
\\
\Omega_1^R &= 0.311 \pm 0.049_{\text{tot}} ~ {\rm GeV}\,,\nn
\\
\chi^2 / \text{dof} &= 0.86\,.\nn
\end{align}
These results represent the main outcome of the phenomenological analysis in this article.
The minimal $\chi^2/\text{dof}$ value, computed by averaging over the individual $500$ results obtained in the random scan, indicates an excellent agreement between our theoretical description and the experimental data.

The table shows all individual sources of hadronization uncertainties considered. They include the effects of
(dimension-two) $\Omega_2$ effects in the OPE of the shape function, of potential three-jet power corrections as well as non-perturative contributions from subleading power dijet corrections. For $\alpha_s(m_Z)$ each of these uncertainties amounts to $\delta\alpha_s(m_Z)=0.0002$.
For $\Omega_1^R$ the $\Omega_2$ and three-jet power uncertainties amount to
$(\delta\Omega_1^R)_{\Omega_2}=0.013$ and
$(\delta \Omega_1^R)_{3-{\rm jet}}=0.010$, respectively, which are larger that the uncertainty from subleading power dijet effects, which is $(\delta\Omega_1^R)_{\rm dijet\,pc}=0.004$.
All these uncertainties each are comparable in size to the experimental uncertainties which are
$(\delta\alpha_s(m_Z))_{\rm exp}=0.0003$
and
$(\delta \Omega_1^R)_{\rm exp}=0.010$.
The dominant sources of uncertainties are the perturbative error from scale variation which amount to
$(\delta\alpha_s(m_Z))_{\rm pert}=0.0008$ and
$(\delta\Omega_1^R)_{\rm pert}=0.037$.
We remind the reader that QED corrections will lower $\alpha_s(m_Z)$ by $\Delta\alpha_s(m_Z)_{\rm QED}= - 0.0005$, and that finite bottom mass effects are negligible, see Sec.~\ref{subsec:additionalUncertainties}.

\begin{figure}[t!]
\centering
\begin{subfigure}[b]{0.485\textwidth}
\includegraphics[width=\textwidth]{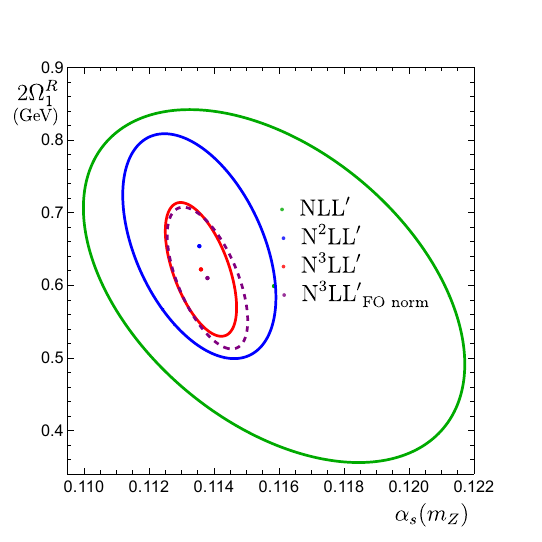}\vspace{-0.3cm}
\caption{\label{fig:oboConvergenceParams}}
\end{subfigure}
~
\begin{subfigure}[b]{0.485\textwidth}
\includegraphics[width=\textwidth]{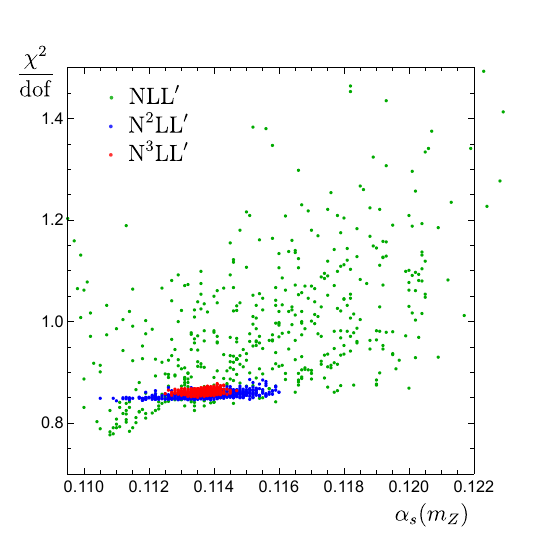}\vspace{-0.3cm}
\caption{\label{fig:chi2ScatterPlot}}
\end{subfigure}
\caption{\label{fig:oboConvergence}
Distribution of best-fit points at different orders for the default fit range \mbox{$\tau \in [(6\,{\rm GeV})/Q,0.15]$}, shown in the $\alpha_s(m_Z)$-$2\Omega_1^R$ plane in panel (a), and in the $\chi^2/{\rm dof}$-$\alpha_s(m_Z)$ plane in panel (b). Ellipses contain both experimental and theoretical uncertainties.}
\end{figure}

At this point it is also instructive to have a closer look at the order-by-order convergence of the fits and the distribution of the individual minimal $\chi^2/\text{dof}$ values for the $500$ profile functions. In Fig.~\ref{fig:oboConvergenceParams} the $39\%$ CL ellipses (accounting for experimental and theoretical uncertainties) are displayed at N$^3$LL$^{\prime}$+${\cal O}(\alpha_s^3)$ (red), N$^2$LL$^{\prime}$+${\cal O}(\alpha_s^2)$ (blue) and NLL$^{\prime}$+${\cal O}(\alpha_s)$ (green). The dashed dark red ellipse arises from using the fixed-order norm for normalizing the N$^3$LL$^{\prime}$+${\cal O}(\alpha_s^3)$ differential distribution. We see that the results exhibit excellent order-by-order convergence. Furthermore, using the fixed-order total cross section for the normalization leads to an almost identical result at the highest order. While the NLL$^{\prime}$+${\cal O}(\alpha_s)$ results for $\alpha_s(m_Z)$ are compatible with the current world average, albeit with huge uncertainties, the fit results for all higher orders are consistently below the world average centering around $0.1135$ in a stable way. In Fig.~\ref{fig:chi2ScatterPlot} we show the distribution of the minimal $\chi^2/\text{dof}$ values from the individual fits for the different profile functions as a function of the best-fit $\alpha_s(m_Z)$ values. As for the error ellipses we clearly see a stabilization of the distribution beyond NLL$^{\prime}$+${\cal O}(\alpha_s)$ and an excellent convergence for the N$^2$LL$^{\prime}$+${\cal O}(\alpha_s^2)$ and N$^3$LL$^{\prime}$+${\cal O}(\alpha_s^3)$ orders.

It is also useful to compare the fit results displayed in Eq.~(\ref{eq:finalFitResults}) and Table~\ref{tab:uncertaintiesListed} to those obtained with the fit approach of the 2010 analysis of Ref.~\cite{Abbate:2010xh}. This way we can assess the effect of the theoretical updates in the N$^3$LL$^{\prime}$+${\cal O}(\alpha_s^3)$ resummed thrust distribution, namely the non-singular distribution, and the ${\cal O}(\alpha_s^3)$ jet and soft function non-logarithmic coefficients $j_3$ and $s_3$. The corresponding results using the fit interval $[(6\,{\rm GeV})/Q,0.33]$, the 2010 profile functions, and fixed-order normalization, while ignoring the three-jet and subleading dijet power corrections read\footnote{In contrast if we use the new 2024 profiles this shifts the central value by $\Delta \alpha_s(m_Z)=-0.0006$, with a perturbative uncertainty of $\pm 0.0010$.}
\begin{align}\label{eq:2010numbers}
\alpha_s(m_Z) &= 0.1140 \pm 0.0008_{\text{pert}}
\pm 0.0006_{\text{exp}+\Omega_1}
[ \pm 0.0002_{\text{had}\Omega_2} ]
\,, \\[5pt]
\Omega_1^R &= 0.332 \pm 0.045_{\text{pert}}
\pm 0.024_{\text{exp}+\alpha_s}
[ \pm 0.013_{\text{had}\Omega_2} ]
\,{\rm GeV}
\,, \nonumber\\[5pt]
\chi^2 / \text{dof} &= 0.85\,.\nonumber
\end{align}
The results compare well to those in Ref.~\cite{Abbate:2010xh}, see their Fig.~14 and the bottom lines of Tables~IV and V. The full breakdown of the uncertainties is displayed in Eq.~(68), where QED and $b$-quark mass effects are included, leading to the downward shift $\Delta_{{\rm QED},m_b}\alpha_s(m_Z)= -0.0004$. The updates concerning the ${\cal O}(\alpha_s^3)$ soft and jet function coefficients $s_3$ and $j_3$, and the ${\cal O}(\alpha_s^3)$ non-singular distribution we have implemented in this analysis have negligible effects on the central value of $\alpha_s(m_Z)$ and yield an increase in $\Omega_1^R$ by about $10$\,MeV. The experimental uncertainties remain unchanged and the perturbative uncertainty of the strong coupling decreases by $0.0001$. The reduced $\chi^2$ of this update $\chi^2/\text{dof}= 0.85$ is slightly better than in Ref.~\cite{Abbate:2010xh}, where $\chi^2/\text{dof}= 0.91$ was found, confirming that the newly available perturbative ingredients translate into a better agreement with data. Our results are also in very good agreement with the recent results quoted in Ref.~\cite{Bell:2023dqs}, which is a strong consistency check for the three analyses.
Comparing the fit results in Eqs.~(\ref{eq:finalFitResults}) and (\ref{eq:2010numbers}) we see that they are perfectly compatible. This leads us to the conclusion that, within the current perturbative and experimental uncertainties, the phenomenological impact of three-jet power corrections
is small for $\tau>0.15$. This view is also confirmed by the comparison to experimental data discussed in Sec.~\ref{subsec:comparison}.

\subsection{Comparison against Experimental Data
\label{subsec:comparison}}

\begin{figure}[t!]
\centering
\begin{subfigure}[b]{0.475\textwidth}
\includegraphics[width=\textwidth]{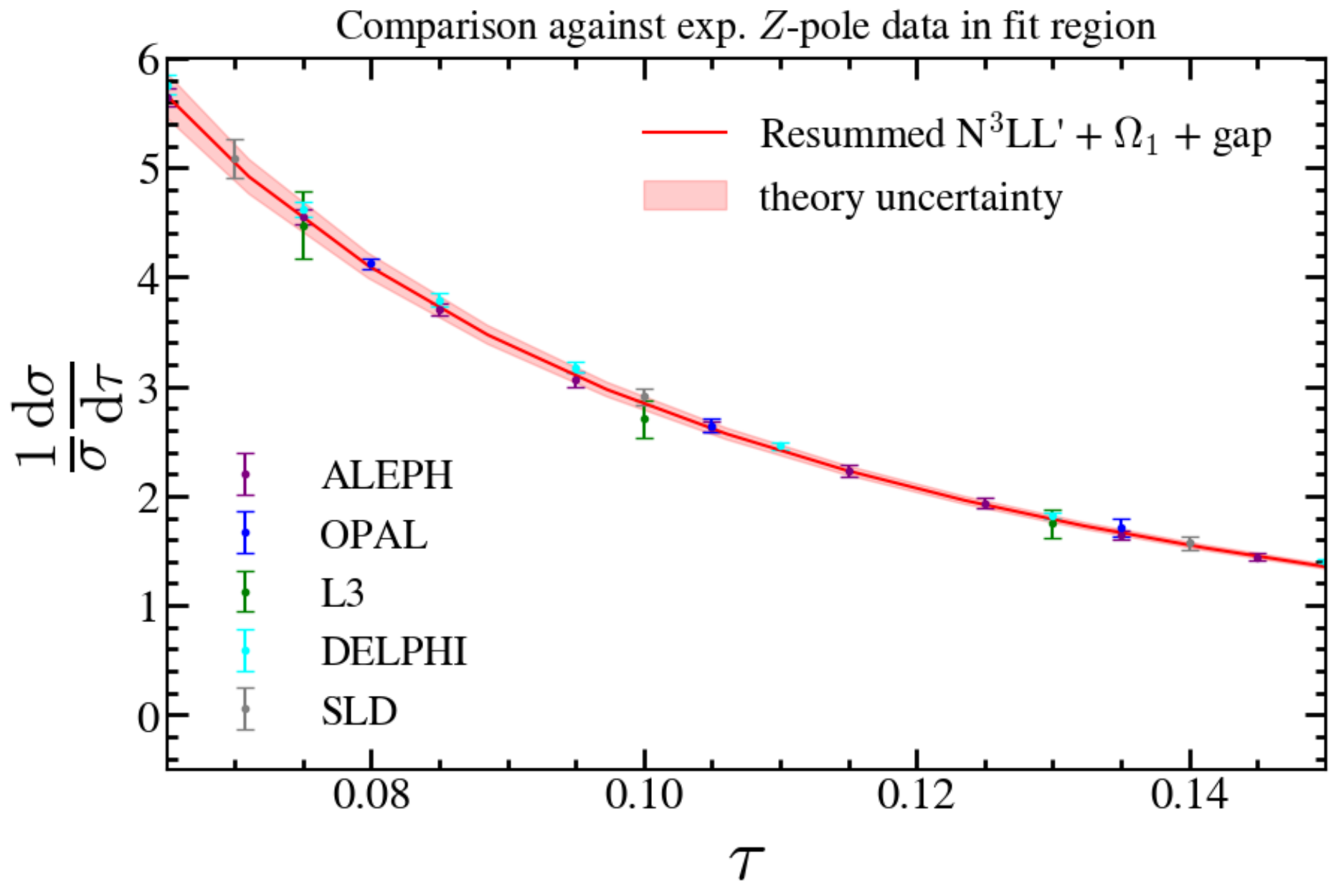}
\caption{\label{fig:fitRegion}}
\end{subfigure}
~
\begin{subfigure}[b]{0.48\textwidth}
\includegraphics[width=\textwidth]{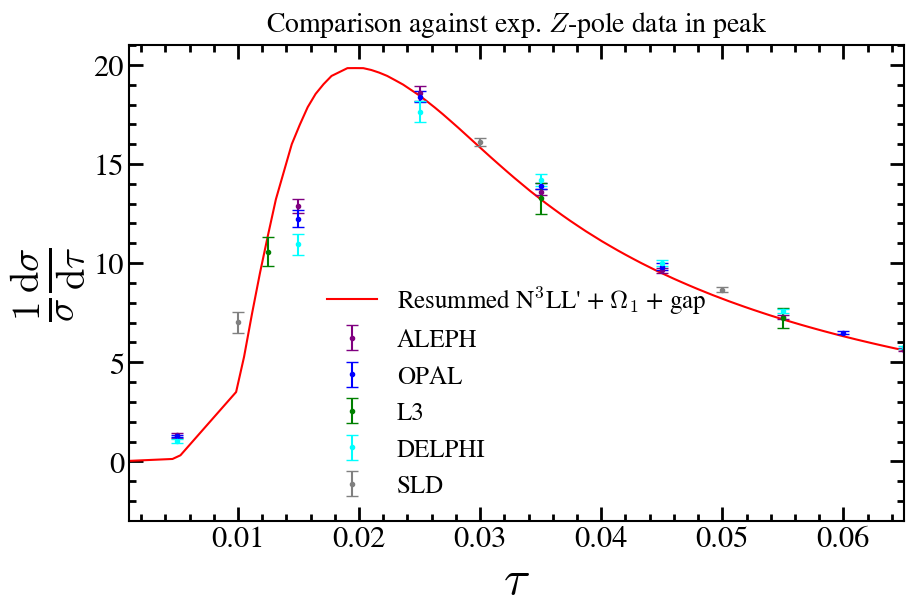}
\caption{\label{fig:peak}}
\end{subfigure}

\begin{subfigure}[b]{0.48\textwidth}
\includegraphics[width=\textwidth]{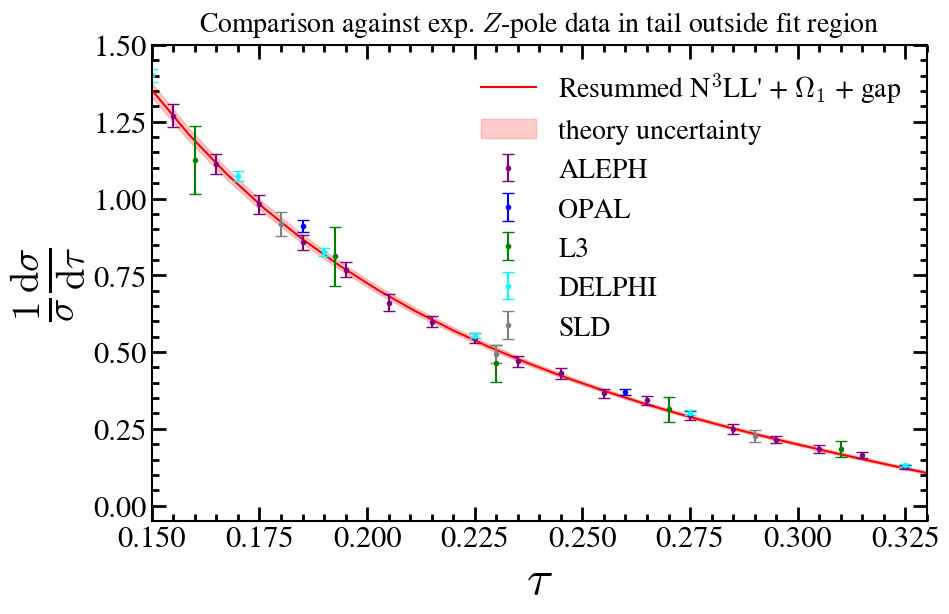}
\caption{\label{fig:tail}}
\end{subfigure}
~
\begin{subfigure}[b]{0.485\textwidth}
\includegraphics[width=\textwidth]{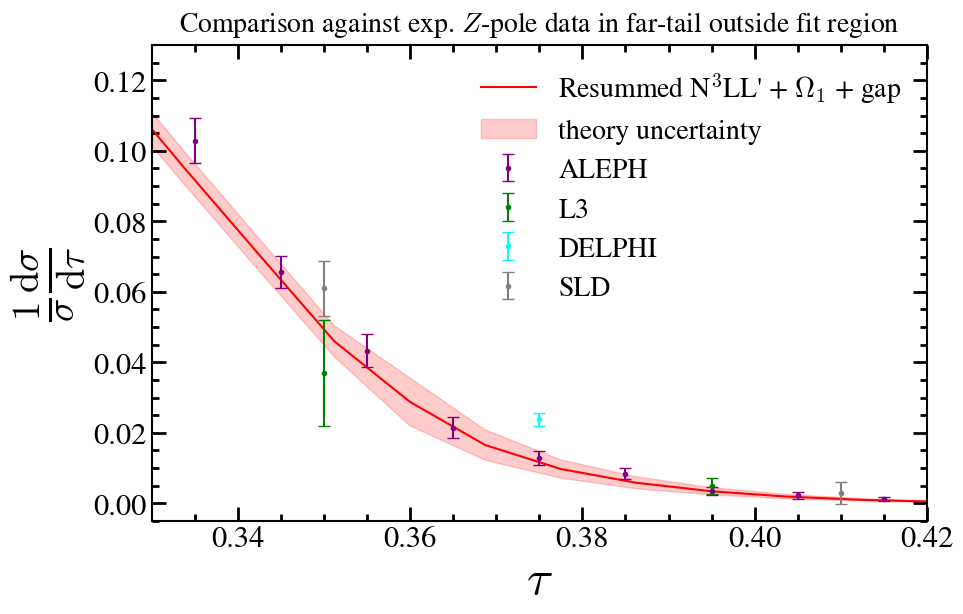}
\caption{\label{fig:farTail}}
\end{subfigure}
\caption{\label{fig:comparison}
Comparison of theory prediction and experimental data at the $Z$-pole in the fit region, panel (a). We also show results outside the fit region in the peak (b), tail (c), and far-tail (d) regions. The theory prediction uses
our default N$^3$LL$^\prime$+${\cal O}(\alpha_s^3)$ results for the cross section. The best fit values for $\alpha_s$ and $\Omega_1^R$ are used.}
\end{figure}

We conclude this section with a comparison of our best theoretical N$^3$LL$^{\prime}$ $+\mathcal{O}(\alpha_s^3)$ thrust distribution and the experimental data. In Fig.~\ref{fig:fitRegion} $Z$-pole ($Q=m_Z$) data and theory are shown in the fit interval $[(6\,{\rm GeV})/Q,0.15]$. The theory curve is our default N$^3$LL$^{\prime}$ thrust prediction (with the default renormalization-scale profile functions, and with the three-jet and subleading dijet power corrections set to zero) using the best-fit results in Eq.~(\ref{eq:finalFitResults}). The displayed red error band arises from the $500$-point random scan over the profile parameters. The analogous comparisons for $Q=44$\,GeV and $Q=189$\,GeV are shown in Figs.~\ref{fig:fitRegionLowQ} and \ref{fig:fitRegionHighQ}, respectively, in the appendix. Overall, we see a very good agreement to the data points within their uncertainties for all thrust values and energies, and in particular with those having the smallest uncertainties.\footnote{In Ref.~\cite{Abbate:2010xh} a very similar comparison was already carried out for the $Z$-pole data.}

The other plots in Figs.~\ref{fig:comparison} show the comparison of the theory prediction to the $Z$-pole data, based on the best-fit results, for thrust values not used in the fit. For example, Fig.~\ref{fig:peak} shows the comparison in the peak region, Fig.~\ref{fig:tail} in the tail region above the fit interval up to $\tau=0.325$, and Fig.~\ref{fig:farTail} in the far-tail and endpoint region for $\tau>0.325$. In Figs.~\ref{fig:comparisonLowQ} and \ref{fig:comparisonHighQ} the analogous plots show the comparison for the other energies.
We again find an excellent agreement with the experimental data, which is comparable to the one visible for the thrust fit intervals. There are some discrepancies visible in the prediction of the peak region shape. However, this is not unexpected since in the peak a more flexible parametrization for the shape function should be implemented, including more parameters than the single one employed in Eq.~(\ref{eq:shapeFun}). This is related to the fact that the shape function OPE of Eq.~(\ref{eq:opeshape}) is not applicable in the peak, where details of the shape function's form become are important.\footnote{Examples of shape function parametrizations suitable for peak fits of event-shapes have been constructed for $B\to X_s\gamma$ in Ref.~\cite{Bernlochner:2020jlt} and for top quark production in Ref.~\cite{Dehnadi:2023msm}. For the dijet tail fit analysis carried out in this article such a sophisticated parametrization is not needed, and we have estimated the uncertainty induced by dropping the next most important parameter $\Omega_2$.}
Interestingly, even in the far-tail region, shown in Figs.~\ref{fig:farTail}, \ref{fig:farTailLowQ} and
\ref{fig:farTailHighQ},
where there is no first-principles argument indicating that the dijet factorization formula should provide a good data description, the data is in fact described very well. This is particularly notable for the very accurate $Z$-pole data.

The important conclusion we can draw from this comparison is that our default N$^3$LL$^{\prime}$ thrust distribution prediction, which is based on the best possible combination of the ingredients related to leading-power dijet factorization, provides an excellent data description for all thrust values. In the context of the discussion on possible sizeable effects caused by three-jet-related non-perturbative power corrections, this implies that for the thrust values where these corrections must be included, the improved theoretical description must yield predictions that are equivalent to our default dijet-based treatment. In other words, all mandatory modifications beyond the dijet-based treatment do for some reason cancel or happen to be very small.

\section{Conclusions \label{sec:conclusions}}

In this paper we have updated the theoretical ingredients used in the 2010 thrust fit of Ref.~\cite{Abbate:2010xh},
and refined this previous determination of $\alpha_s(m_Z)$ by carrying out a fit purely in the dijet region.
While the highest perturbative order is still the same, namely N$^3$LL$^\prime+{\cal O}(\alpha_s^3)$, the various upgrades have a positive impact on the robustness of the results.

This analysis has been
partly motivated by a recent theoretical and phenomenological article~\cite{Nason:2023asn} that has questioned some vital aspects of the theoretical description employed in the 2010 study, most notably the resummation of large logarithms, $\ln\tau$, and the description of non-perturbative corrections in the three-jet region and near the dijet limit.
We have demonstrated that the questions casting doubt on the dijet factorization theorem, and its associated resummation of large logarithms and description of power corrections, are not justified.
In contrast, the need for a modified description of non-perturbative corrections in the three-jet region is in our view justified.

Since the 2010 analysis, four new perturbative ingredients have became known:
the two-loop~\cite{Kelley:2011ng,Monni:2011gb} and three-loop soft functions~\cite{Baranowski:2024vxg}, the three-loop jet function~\cite{Bruser:2018rad,Banerjee:2018ozf} and the 4-loop cusp anomalous dimension~\cite{Boels:2017skl,Henn:2019rmi,Henn:2019swt}. These new ingredients have been included in our computer code, that has been written anew in \texttt{C++}.
Their impact on the final $\alpha_s$ fit is much smaller than the perturbative uncertainty.
The $\mathcal{O}(\alpha_s^2)$ and $\mathcal{O}(\alpha_s^3)$ non-singular distributions have been updated using higher statistics for the former and the more precise outcome of the computer code {\sc CoLoRFulNNLO}~\cite{DelDuca:2016ily} for the latter. Moreover, both parametrizations have been constructed to exactly reproduce the (now known) leading logarithms of the next-to-leading power partonic contribution~\cite{Moult:2018jjd,Moult:2019uhz,Beneke:2022obx}.
Among these updates, the improved three loop nonsingular has the only significant impact on the fit $\alpha_s$ in the smaller fit window, being of similar size as the perturbative uncertainties.
The improvement that has the most prominent impact on our results is the implementation of more flexible renormalization scale profile functions in the different components of our factorization formula that allow using canonical scaling in the three regions (peak, tail and far tail) where theoretical constraints demand a particular $\tau$-scaling for renormalization scales. The profile functions in these three regions are smoothly joined with ``junction'' functions whose specific form mildly affects the cross section. Besides the new functional form of the profile functions, we have shown how to cast their dependence on $\tau$ such that the OPE for the total cross section, i.e.\ the integrated norm, is respected, exhibiting full independence with respect to the leading linear power correction $\Omega_1\propto \Lambda_{\rm QCD}$.
Having implemented this new prescription and including all the updates of our best N$^3$LL$^\prime+{\cal O}(\alpha_s^3)$ prediction we found excellent agreement between the total cross section
obtained from integrating the distribution
and the one obtained in fixed-order perturbation theory.

In the context of these analyses we have also discussed the dependence of our predictions on the choice of the gap subtraction scheme which removes the dominant ${\cal O}(\Lambda_{\rm QCD})$-renormalon from the definition of first non-perturbative shape function moment $\Omega_1$. The gap subtraction plays an important role in achieving the small uncertainties of our best N$^3$LL$^\prime+{\cal O}(\alpha_s^3)$ prediction and the resulting strong coupling fits.
Some reasonable concerns were expressed in Ref.~\cite{Bell:2023dqs} that the gap scheme dependence, which was not examined in the 2010 analysis, may not be covered by the renormalization scale variations within the R-gap scheme.
In this article we have studied a wider class of gap subtraction schemes and checked the results for consistency. We found that the gap subtractions that are constructed in accordance with the summation of logarithms lead to results mutually consistent to each other within the renormalization scale uncertainties at N$^3$LL$^\prime+{\cal O}(\alpha_s^3)$ order. Furthermore, the values of the renormalon-free $\Omega_1$-values in the different gap schemes can be converted to each other in full accordance to the differences of respective perturbative subtraction series.

To confirm that the summation of logarithms arising from dijet factorization is reliable in the thrust fit intervals used in the 2010 analysis, we have critically reexamined the behavior of the perturbative expansions both in the dijet-based renormalization group improved (resummed) and the fixed-order approaches. Up to $\tau=0.3$ we found a perfectly stable convergence for the resummed approach. In contrast, for the fixed-order expansion the higher-order corrections are significantly larger, and the fixed-order series is observed to converge towards the best N$^3$LL$^\prime+{\cal O}(\alpha_s^3)$ resummed prediction. Furthermore, we find that in $\alpha_s$-fits, the resummed prediction together with the updated profile function yields results that are
reasonably
independent of the thrust intervals used for the fit covering ranges with $5/Q$ as the smallest lower and $0.38$ as the largest upper boundary. All fit results for $\alpha_s(m_Z)$ are close to $0.114$. In contrast, the fixed-order expansion at ${\cal O}(\alpha_s^3)$ yields fit results that are depending very strongly on the choice of the fit interval covering $\alpha_s(m_Z)$ values ranging from $0.112$ to $0.120$. Some fit intervals yield strong coupling results perfectly compatible our final result and some yield results compatible with the world average. Overall, we found no evidence that the summation of logarithms based on the dijet factorization were unreliable, and we refute any claims that the opposite were true. In contrast, the resummation of logarithms based on dijet factorization plays an essential role in the ability of our best N$^3$LL$^\prime+{\cal O}(\alpha_s^3)$ prediction to describe experimental data in the thrust tail above the thrust peak, not only in the fit intervals, but for the entire tail region up to $\tau=0.3$.

Addressing the criticism in Ref.~\cite{Nason:2023asn} that the treatment of the leading hadronization corrections arising from dijet factorization were unreliable in the thrust fit intervals used in the 2010 analysis, we have reexamined the physical arguments for the dijet picture.
Their conclusions were based on the three-jet model calculations of Refs.~\cite{Luisoni:2020efy,Caola:2021kzt,Caola:2022vea,Nason:2023asn}, and we discussed the underlying assumptions and interpretation.
In particular we have addressed the limitations of the model
that was based on a small gluon mass and renormalon calculation for $e^+e^-\to q\bar q \gamma$~\cite{Caola:2021kzt,Caola:2022vea,Nason:2023asn}.
These model calculations provide an estimate for the size of power corrections in the three-jet region.
The essential omissions are that the model calculation assumes
that soft non-perturbative modes can always resolve three distinct hard jets for any thrust value and
that the normalization of the three-jet hadronization correction is fixed by the dijet power correction.\footnote{One of the biggest assumptions in the fit of Ref.~\cite{Nason:2023asn} is that three-jet hadronization corrections are universal for different event shapes, which unlike in the dijet case, has not been proven from QCD.}
Thus, these results
(i) cannot be used to justify that the same hadronic parameter describes dijet and three-jet power corrections, since in QCD they involve different hadronic matrix elements,
(ii) are at best a suppressed effect in the dijet dominated region, and
(iii) do not describe the transition region between the dijet regime and the regime where a three-jet description becomes relevant.
To address the fact that different non-perturbative power corrections are associated to dijet and three-jet configurations, we focused on a reduced dijet dominated thrust fit region and parameterized and estimated uncertainties coming from the potential three-jet power corrections.

Our final strong coupling determination, is a simultaneous fit of $\alpha_s(m_Z)$ and $\Omega_1^R$ based on our best N$^3$LL$^\prime+{\cal O}(\alpha_s^3)$ thrust theory description.
To reduce the impact of the three-jet power correction uncertainties in our new $\alpha_s(m_Z)$ fit, we have lowered the upper boundary of the thrust fit intervals from $0.33$ used in the 2010 analysis to $0.15$ for the data from all c.m.\ energies. For the upper bound $0.15$ the terms in the dijet factorization theorem dominate
over the ${\cal O}(\alpha_s^3)$ nonsingular terms (which are also included),
and the effects of potential three-jet power corrections can be accounted for by an additional source of theoretical uncertainty, which we estimated conservatively using a model that yields additional three-jet power corrections.
At the same time we also included an independent uncertainty that arises from the ignorance of subleading power non-perturbative power corrections affecting the nonsingular distribution in the dijet regime. We find that both types of uncertainties are much smaller than the perturbative uncertainties and thus do not play an essential role.
The final result of our analysis is perfectly compatible with the result of the 2010 analysis, albeit with slightly larger experimental uncertainties due to the reduction of the data used in the fit. The final fit results are quoted in Eqs.~(\ref{eq:finalFitResults}) with a breakdown of all individual uncertainties shown in Table~\ref{tab:uncertaintiesListed}.

We also note that using these best fit results and our default N$^3$LL$^\prime+{\cal O}(\alpha_s^3)$ thrust theory prediction we find an excellent description of the experimental data from all c.m.\ energies, not only in the fit ranges, but for all thrust values larger than $0.15$. These results show that the strict leading power dijet-based description of the thrust distribution provides an excellent data description even for thrust values where one may expect sizeable non-perturbative corrections from three-jet or multi-jet final states.

Our final determination of $\alpha_s(m_Z)$ from thrust data is significantly lower than and incompatible with the 2023 PDG world average~\cite{ParticleDataGroup:2024cfk}, $\alpha_s(m_Z)=0.1180\pm 0.0009$. After carefully examining all possible aspects of our theoretical prediction, and having thoroughly estimated all sources of uncertainties from perturbative and non-perturbative origin, we are confident that our result is sound, based on the available data.
We believe it would be worth reexamining the systematic experimental uncertainties in $e^+e^-$ event shapes, for example related to unfolding and extrapolation outside fiducial regions, making use of modern Monte Carlo generators.

Possible next steps for our analysis include pushing the perturbative order and resummation to N$^4$LL, incorporating next-to-leading-power summation of logarithms~\cite{Moult:2018jjd,Moult:2019uhz,Beneke:2022obx}, and having a better understanding of power corrections beyond the dijet approximation.
However, we are confident that including these effects will not lead to a significant change in a strong coupling determination.
It is also possible to include peak data, which invalidates the dijet OPE and requires fitting for the shape function as a whole. Such fits are now technically in reach~\cite{Dehnadi:2023msm} and can be carried out as well at N$^3$LL$^\prime+{\cal O}(\alpha_s^3)$ order for thrust.
One can also consider reassessing uncertainties using the theory nuisance parameter method~\cite{Tackmann:2024kci}, and combined fits with other event shape observables like C-parameter and Heavy Jet Mass, where additional experimental and theoretical correlations in uncertainties are important.
Finally, one could consider other types of event-shapes with different systematics such as energy-energy correlators or jet rates, or apply grooming techniques to reduce the effects of hadronization.

\acknowledgments
The authors would like to thank Paolo Nason, Pier Monni, Guido Bell, Christopher Lee, Alexander Huss, Matthew Schwartz,
and Giulia Zanderighi for useful conversations.
VM and MB are supported by the Spanish MECD grant No.\ PID2022-141910NB-I00, the JCyL grant SA091P24 under program EDU/841/2024, and the EU STRONG-2020 project under Program No.\ H2020-INFRAIA-2018-1, Grant Agreement No.\ 824093. MB is supported by a JCyL scholarship funded by the regional government of Castilla y Le\'on and European Social Fund, 2022 call.
I.S. is supported in part by the U.S. Department of Energy, Office of Nuclear Physics from DE-SC0011090 and the Simons Foundation through the Investigator grant 327942. AHH acknowledges partial support by the FWF Austrian Science Fund under the Project No. P32383-N27 and the Doctoral Program “Particles and Interactions” No. W1252-N27. We thank the Erwin Schr\"odinger Institute where part of this work was completed during the 2023 program on Quantum Field Theory at the Frontiers of the Strong Interaction. IS, AH and GV thank the University of Salamanca for hospitality while parts of this project were completed. MB and IS thank the University of Vienna for hospitality. IS and AHH would like to express thanks to the Mainz Institute for Theoretical Physics (MIAP) of the Cluster of Excellence PRISMA$+$ (Project IC 390831469) for its hospitality and support during the scientic program ``Energy Correlators at the Collider Frontier", July 8-12, 2024.

\appendix

\appendix

\section{Comparison against experiment for different energies}

\begin{figure}[t!]
\centering
\begin{subfigure}[b]{0.485\textwidth}
\includegraphics[width=\textwidth]{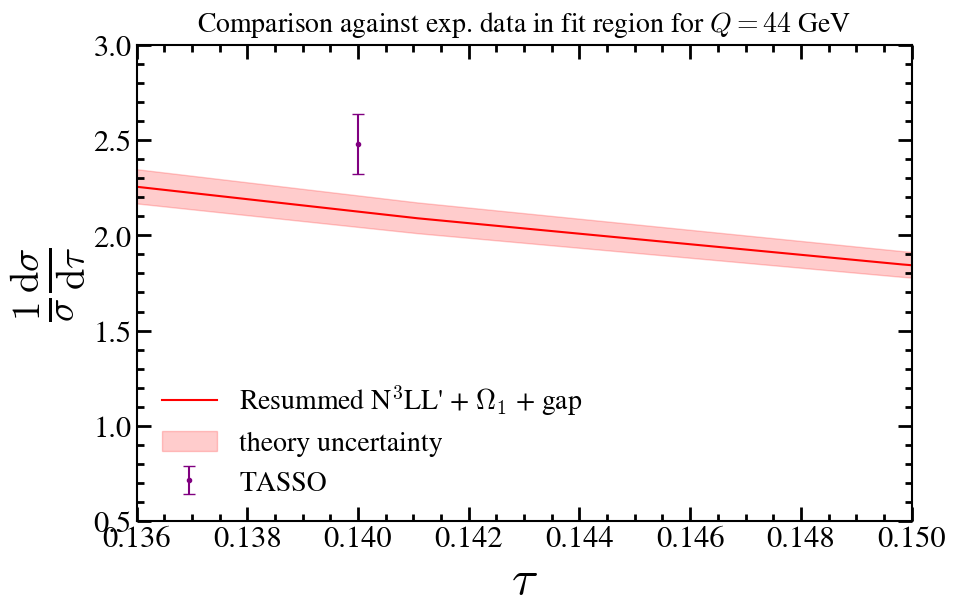}
\caption{\label{fig:fitRegionLowQ}}
\end{subfigure}
~
\begin{subfigure}[b]{0.48\textwidth}
\includegraphics[width=\textwidth]{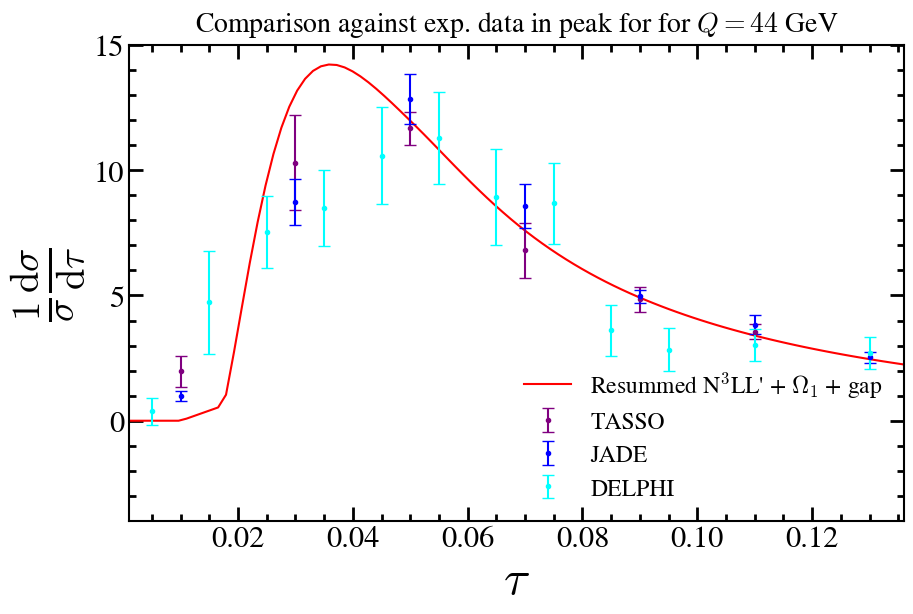}
\caption{\label{fig:peakLowQ}}
\end{subfigure}

\begin{subfigure}[b]{0.48\textwidth}
\includegraphics[width=\textwidth]{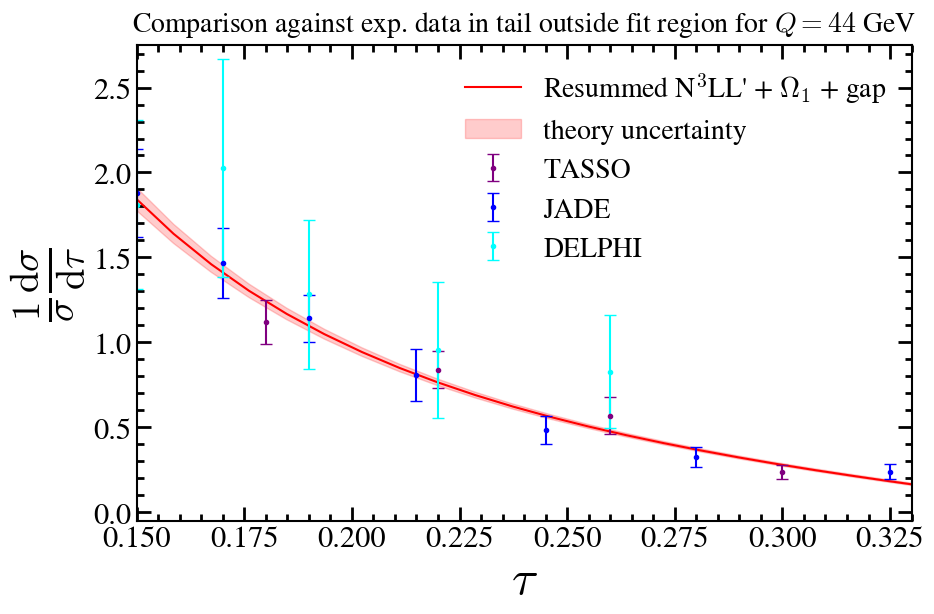}
\caption{\label{fig:tailLowQ}}
\end{subfigure}
~
\begin{subfigure}[b]{0.485\textwidth}
\includegraphics[width=\textwidth]{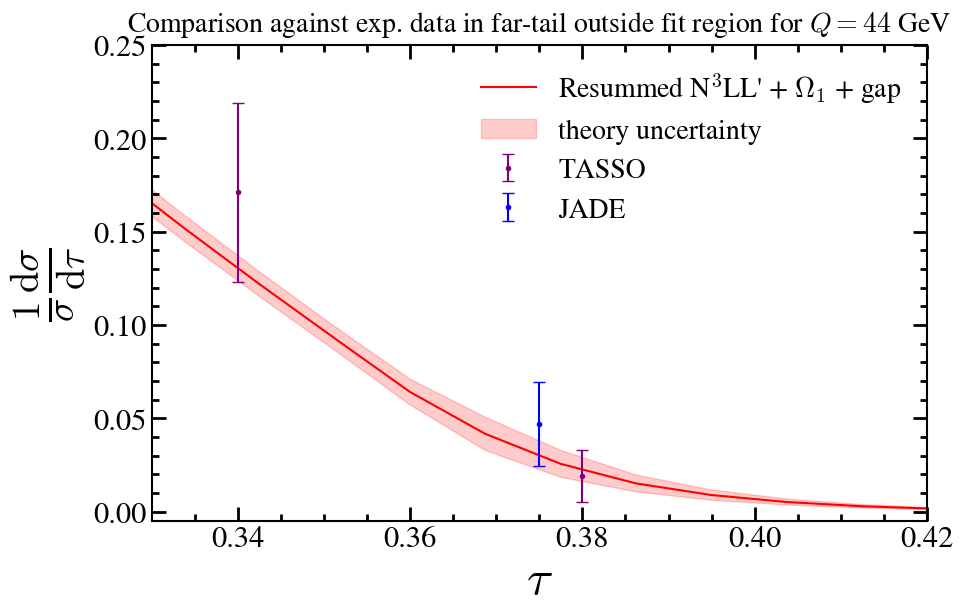}
\caption{\label{fig:farTailLowQ}}
\end{subfigure}
\caption{\label{fig:comparisonLowQ}
Comparison of theory prediction and experimental data for $Q=44$ GeV in the fit region, panel (a). We also show results outside the fit region in the peak (b), tail (c), and far-tail (d) regions. The theory prediction uses
our default N$^3$LL$^\prime$+${\cal O}(\alpha_s^3)$ results for the cross section. The best fit values for $\alpha_s$ and $\Omega_1^R$ are used.}
\end{figure}

\begin{figure}[t!]
\centering
\begin{subfigure}[b]{0.485\textwidth}
\includegraphics[width=\textwidth]{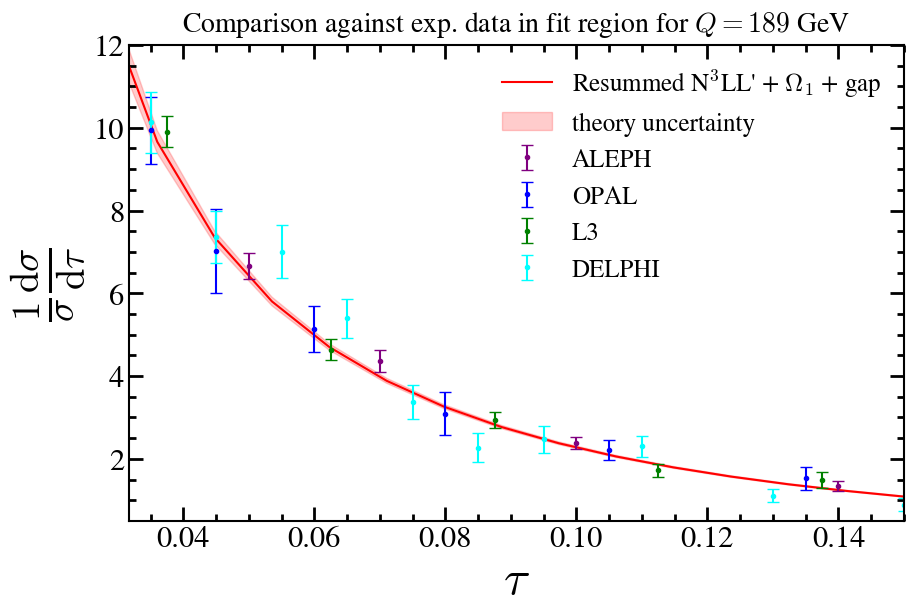}
\caption{\label{fig:fitRegionHighQ}}
\end{subfigure}
~
\begin{subfigure}[b]{0.48\textwidth}
\includegraphics[width=\textwidth]{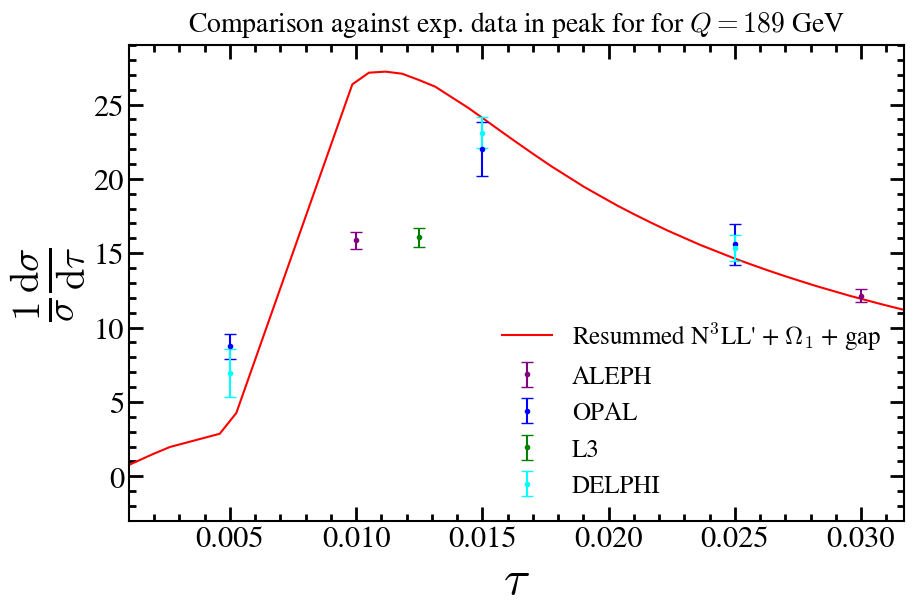}
\caption{\label{fig:peakHighQ}}
\end{subfigure}

\begin{subfigure}[b]{0.48\textwidth}
\includegraphics[width=\textwidth]{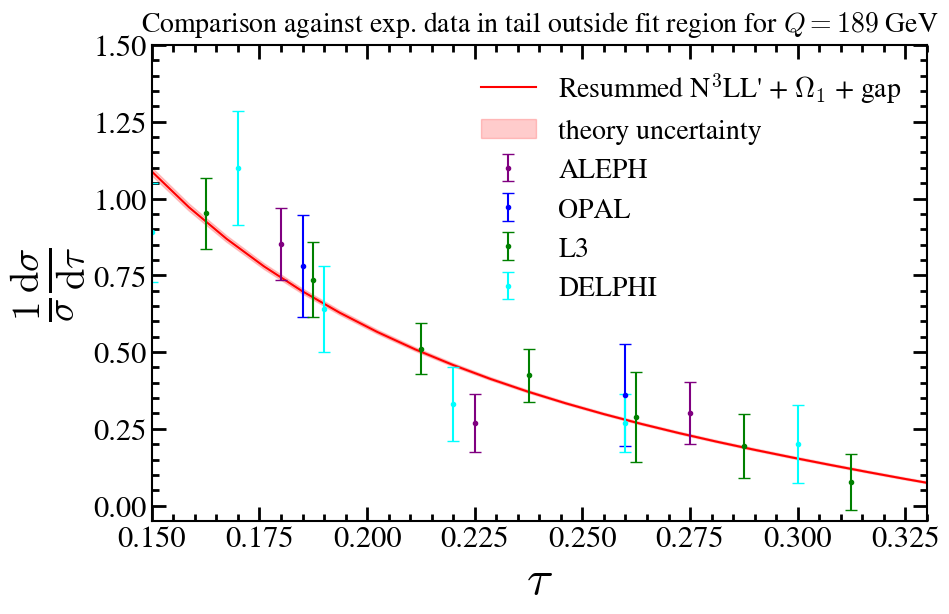}
\caption{\label{fig:tailHighQ}}
\end{subfigure}
~
\begin{subfigure}[b]{0.485\textwidth}
\includegraphics[width=\textwidth]{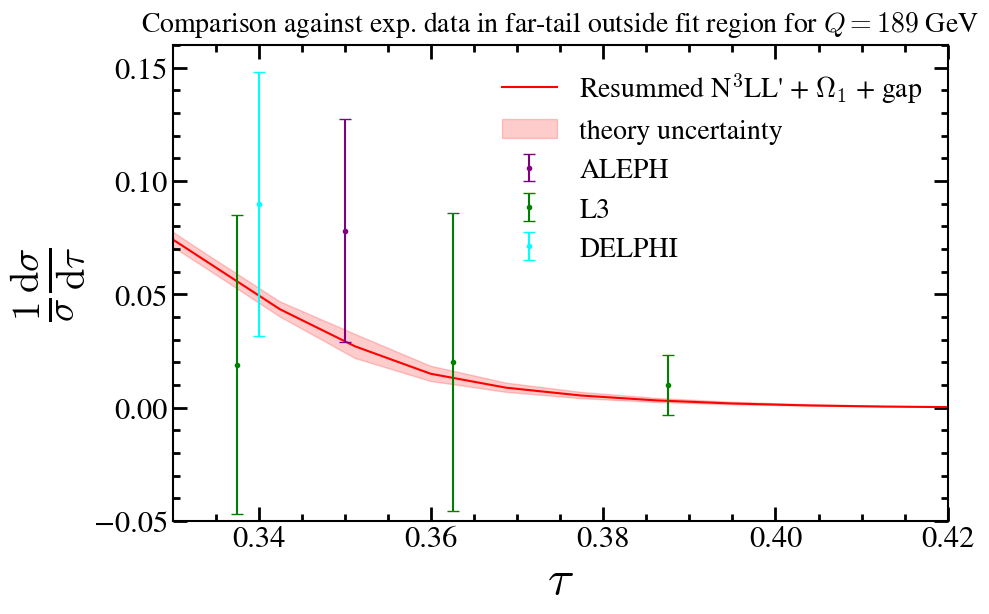}
\caption{\label{fig:farTailHighQ}}
\end{subfigure}
\caption{\label{fig:comparisonHighQ}
Comparison of theory prediction and experimental data for $Q=189$ GeV in the fit region, panel (a). We also show results outside the fit region in the peak (b), tail (c), and far-tail (d) regions. The theory prediction uses
our default N$^3$LL$^\prime$+${\cal O}(\alpha_s^3)$ results for the cross section. The best fit values for $\alpha_s$ and $\Omega_1^R$ are used.}
\end{figure}

In Figs.~\ref{fig:comparisonLowQ} and~\ref{fig:comparisonHighQ} we show, as examples, a comparison of our best theoretical prediction against experimental data for $Q=44$\,GeV and $Q=189$\,GeV. For comparison, the analogous result for $Q=91$\,GeV is given in Fig.~\ref{fig:comparison} in the main text. Once again we partitioned the results into different regions of the thrust spectrum, with one panel showing data used in the fit and three pannels showing data outside the fit region. (Recall that theoretical uncertainties are not shown in the peak panels, \ref{fig:comparisonLowQ}b and~\ref{fig:comparisonHighQ}b.) For these and other values of $Q$ the theoretical description works well outside the fit window. Note that in making comparisons of continuous theory curves to the data points the one must be careful to note that different datasets use bins of different sizes.

\newpage

\addcontentsline{toc}{section}{References}
\bibliographystyle{jhep}
\bibliography{refs}

\providecommand{\href}[2]{#2}\begingroup\raggedright\begin{thebibliography}{100}

\bibitem{TASSO:1990cdg}
{\scshape TASSO} collaboration, W.~Braunschweig et~al., \emph{{Global Jet
  Properties at $14$\,GeV to $44$\,GeV Center-of-mass Energy in $e^+ e^-$
  Annihilation}}, \href{https://doi.org/10.1007/BF01552339}{\emph{Z. Phys. C}
  {\bfseries 47} (1990) 187}.

\bibitem{MovillaFernandez:1997fr}
{\scshape JADE} collaboration, P.~A. Movilla~Fernandez, O.~Biebel, S.~Bethke,
  S.~Kluth and P.~Pfeifenschneider, \emph{{A Study of event shapes and
  determinations of alpha-s using data of $e^+ e^-$ annihilations at $s^{1/2} =
  22$\,GeV to $44$\,GeV}},
  \href{https://doi.org/10.1007/s100520050096}{\emph{Eur. Phys. J. C}
  {\bfseries 1} (1998) 461}
  [\href{https://arxiv.org/abs/hep-ex/9708034}{{\ttfamily hep-ex/9708034}}].

\bibitem{AMY:1989feg}
{\scshape AMY} collaboration, Y.~K. Li et~al., \emph{{Multi - hadron event
  properties in $e^+e^-$ annihilation at $\sqrt{s} = 52$\,GeV to $57$\,GeV}},
  \href{https://doi.org/10.1103/PhysRevD.41.2675}{\emph{Phys. Rev. D}
  {\bfseries 41} (1990) 2675}.

\bibitem{SLD:1994idb}
{\scshape SLD} collaboration, K.~Abe et~al., \emph{{Measurement of
  $\alpha_s(M_Z^2)$ from hadronic event observables at the $Z_0$ resonance}},
  \href{https://doi.org/10.1103/PhysRevD.51.962}{\emph{Phys. Rev. D} {\bfseries
  51} (1995) 962} [\href{https://arxiv.org/abs/hep-ex/9501003}{{\ttfamily
  hep-ex/9501003}}].

\bibitem{L3:1992nwf}
{\scshape L3} collaboration, B.~Adeva et~al., \emph{{Studies of hadronic event
  structure and comparisons with QCD models at the $Z_0$ resonance}},
  \href{https://doi.org/10.1007/BF01558288}{\emph{Z. Phys. C} {\bfseries 55}
  (1992) 39}.

\bibitem{L3:2004cdh}
{\scshape L3} collaboration, P.~Achard et~al., \emph{{Studies of hadronic event
  structure in $e^{+} e^{-}$ annihilation from $30$\,GeV to $209$\,GeV with the
  L3 detector}},
  \href{https://doi.org/10.1016/j.physrep.2004.07.002}{\emph{Phys. Rept.}
  {\bfseries 399} (2004) 71}
  [\href{https://arxiv.org/abs/hep-ex/0406049}{{\ttfamily hep-ex/0406049}}].

\bibitem{DELPHI:2003yqh}
{\scshape DELPHI} collaboration, J.~Abdallah et~al., \emph{{A Study of the
  energy evolution of event shape distributions and their means with the DELPHI
  detector at LEP}},
  \href{https://doi.org/10.1140/epjc/s2003-01198-0}{\emph{Eur. Phys. J. C}
  {\bfseries 29} (2003) 285}
  [\href{https://arxiv.org/abs/hep-ex/0307048}{{\ttfamily hep-ex/0307048}}].

\bibitem{DELPHI:2000uri}
{\scshape DELPHI} collaboration, P.~Abreu et~al., \emph{{Consistent
  measurements of $\alpha_s$ from precise oriented event shape distributions}},
  \href{https://doi.org/10.1007/s100520000354}{\emph{Eur. Phys. J. C}
  {\bfseries 14} (2000) 557}
  [\href{https://arxiv.org/abs/hep-ex/0002026}{{\ttfamily hep-ex/0002026}}].

\bibitem{DELPHI:1999vbd}
{\scshape DELPHI} collaboration, P.~Abreu et~al., \emph{{Energy dependence of
  event shapes and of $\alpha_s$ at LEP-2}},
  \href{https://doi.org/10.1016/S0370-2693(99)00472-4}{\emph{Phys. Lett. B}
  {\bfseries 456} (1999) 322}.

\bibitem{OPAL:2004wof}
{\scshape OPAL} collaboration, G.~Abbiendi et~al., \emph{{Measurement of event
  shape distributions and moments in $e^+ e^- \to$ hadrons at $91$\,GeV --
  $209$\,GeV and a determination of $\alpha_s$}},
  \href{https://doi.org/10.1140/epjc/s2005-02120-6}{\emph{Eur. Phys. J. C}
  {\bfseries 40} (2005) 287}
  [\href{https://arxiv.org/abs/hep-ex/0503051}{{\ttfamily hep-ex/0503051}}].

\bibitem{OPAL:1997asf}
{\scshape OPAL} collaboration, K.~Ackerstaff et~al., \emph{{QCD studies with
  $e^+ e^-$- annihilation data at $161$\,GeV}},
  \href{https://doi.org/10.1007/s002880050462}{\emph{Z. Phys. C} {\bfseries 75}
  (1997) 193}.

\bibitem{OPAL:1999ldr}
{\scshape OPAL} collaboration, G.~Abbiendi et~al., \emph{{QCD studies with $e^+
  e^-$ annihilation data at $172$\,GeV -- $189$\,GeV}},
  \href{https://doi.org/10.1007/s100520050015}{\emph{Eur. Phys. J. C}
  {\bfseries 16} (2000) 185}
  [\href{https://arxiv.org/abs/hep-ex/0002012}{{\ttfamily hep-ex/0002012}}].

\bibitem{ALEPH:2003obs}
{\scshape ALEPH} collaboration, A.~Heister et~al., \emph{{Studies of QCD at
  $e^+ e^-$ centre-of-mass energies between $91$\,GeV and $209$\,GeV}},
  \href{https://doi.org/10.1140/epjc/s2004-01891-4}{\emph{Eur. Phys. J. C}
  {\bfseries 35} (2004) 457}.

\bibitem{dEnterria:2022hzv}
D.~d'Enterria et~al., \emph{{The strong coupling constant: state of the art and
  the decade ahead}}, \href{https://doi.org/10.1088/1361-6471/ad1a78}{\emph{J.
  Phys. G} {\bfseries 51} (2024) 090501}
  [\href{https://arxiv.org/abs/2203.08271}{{\ttfamily 2203.08271}}].

\bibitem{Kluth:2006bw}
S.~Kluth, \emph{{Tests of Quantum Chromo Dynamics at $e^+ e^-$ Colliders}},
  \href{https://doi.org/10.1088/0034-4885/69/6/R04}{\emph{Rept. Prog. Phys.}
  {\bfseries 69} (2006) 1771}
  [\href{https://arxiv.org/abs/hep-ex/0603011}{{\ttfamily hep-ex/0603011}}].

\bibitem{MovillaFernandez:2001ed}
P.~A. Movilla~Fernandez, S.~Bethke, O.~Biebel and S.~Kluth, \emph{{Tests of
  power corrections for event shapes in $e^+ e^-$ annihilation}},
  \href{https://doi.org/10.1007/s100520100750}{\emph{Eur. Phys. J. C}
  {\bfseries 22} (2001) 1}
  [\href{https://arxiv.org/abs/hep-ex/0105059}{{\ttfamily hep-ex/0105059}}].

\bibitem{Farhi:1977sg}
E.~Farhi, \emph{{A QCD Test for Jets}},
  \href{https://doi.org/10.1103/PhysRevLett.39.1587}{\emph{Phys. Rev. Lett.}
  {\bfseries 39} (1977) 1587}.

\bibitem{Parisi:1978eg}
G.~Parisi, \emph{{Super Inclusive Cross-Sections}},
  \href{https://doi.org/10.1016/0370-2693(78)90061-8}{\emph{Phys. Lett. B}
  {\bfseries 74} (1978) 65}.

\bibitem{Donoghue:1979vi}
J.~F. Donoghue, F.~E. Low and S.-Y. Pi, \emph{{Tensor Analysis of Hadronic Jets
  in Quantum Chromodynamics}},
  \href{https://doi.org/10.1103/PhysRevD.20.2759}{\emph{Phys. Rev. D}
  {\bfseries 20} (1979) 2759}.

\bibitem{Clavelli:1979md}
L.~Clavelli, \emph{{Jet Invariant Mass in Quantum Chromodynamics}},
  \href{https://doi.org/10.1016/0370-2693(79)90789-5}{\emph{Phys.Lett.}
  {\bfseries B85} (1979) 111}.

\bibitem{Chandramohan:1980ry}
T.~Chandramohan and L.~Clavelli, \emph{{Consequences of Second Order QCD for
  Jet Structure in $e^+ e^-$ annihilation}},
  \href{https://doi.org/10.1016/0550-3213(81)90224-8}{\emph{Nucl.Phys.}
  {\bfseries B184} (1981) 365}.

\bibitem{Catani:1996jh}
S.~Catani and M.~H. Seymour, \emph{{The Dipole formalism for the calculation of
  QCD jet cross-sections at next-to-leading order}},
  \href{https://doi.org/10.1016/0370-2693(96)00425-X}{\emph{Phys. Lett. B}
  {\bfseries 378} (1996) 287}
  [\href{https://arxiv.org/abs/hep-ph/9602277}{{\ttfamily hep-ph/9602277}}].

\bibitem{Catani:1996vz}
S.~Catani and M.~H. Seymour, \emph{{A General algorithm for calculating jet
  cross-sections in NLO QCD}},
  \href{https://doi.org/10.1016/S0550-3213(96)00589-5}{\emph{Nucl. Phys. B}
  {\bfseries 485} (1997) 291}
  [\href{https://arxiv.org/abs/hep-ph/9605323}{{\ttfamily hep-ph/9605323}}].

\bibitem{Gehrmann-DeRidder:2007nzq}
A.~Gehrmann-De~Ridder, T.~Gehrmann, E.~W.~N. Glover and G.~Heinrich,
  \emph{{Second-order QCD corrections to the thrust distribution}},
  \href{https://doi.org/10.1103/PhysRevLett.99.132002}{\emph{Phys. Rev. Lett.}
  {\bfseries 99} (2007) 132002}
  [\href{https://arxiv.org/abs/0707.1285}{{\ttfamily 0707.1285}}].

\bibitem{Gehrmann-DeRidder:2009fgd}
A.~Gehrmann-De~Ridder, T.~Gehrmann, E.~W.~N. Glover and G.~Heinrich,
  \emph{{NNLO moments of event shapes in $e^+e^-$ annihilation}},
  \href{https://doi.org/10.1088/1126-6708/2009/05/106}{\emph{JHEP} {\bfseries
  05} (2009) 106} [\href{https://arxiv.org/abs/0903.4658}{{\ttfamily
  0903.4658}}].

\bibitem{Weinzierl:2008iv}
S.~Weinzierl, \emph{{NNLO corrections to 3-jet observables in electron-positron
  annihilation}},
  \href{https://doi.org/10.1103/PhysRevLett.101.162001}{\emph{Phys. Rev. Lett.}
  {\bfseries 101} (2008) 162001}
  [\href{https://arxiv.org/abs/0807.3241}{{\ttfamily 0807.3241}}].

\bibitem{Weinzierl:2009ms}
S.~Weinzierl, \emph{{Event shapes and jet rates in electron-positron
  annihilation at NNLO}},
  \href{https://doi.org/10.1088/1126-6708/2009/06/041}{\emph{JHEP} {\bfseries
  06} (2009) 041} [\href{https://arxiv.org/abs/0904.1077}{{\ttfamily
  0904.1077}}].

\bibitem{DelDuca:2016csb}
V.~Del~Duca, C.~Duhr, A.~Kardos, G.~Somogyi and Z.~Tr\'ocs\'anyi,
  \emph{{Three-Jet Production in Electron-Positron Collisions at
  Next-to-Next-to-Leading Order Accuracy}},
  \href{https://doi.org/10.1103/PhysRevLett.117.152004}{\emph{Phys. Rev. Lett.}
  {\bfseries 117} (2016) 152004}
  [\href{https://arxiv.org/abs/1603.08927}{{\ttfamily 1603.08927}}].

\bibitem{DelDuca:2016ily}
V.~Del~Duca, C.~Duhr, A.~Kardos, G.~Somogyi, Z.~Sz\H{o}r, Z.~Tr\'ocs\'anyi
  et~al., \emph{{Jet production in the CoLoRFulNNLO method: event shapes in
  electron-positron collisions}},
  \href{https://doi.org/10.1103/PhysRevD.94.074019}{\emph{Phys. Rev. D}
  {\bfseries 94} (2016) 074019}
  [\href{https://arxiv.org/abs/1606.03453}{{\ttfamily 1606.03453}}].

\bibitem{Catani:1992ua}
S.~Catani, L.~Trentadue, G.~Turnock and B.~R. Webber, \emph{{Resummation of
  large logarithms in $e^+ e^-$ event shape distributions}},
  \href{https://doi.org/10.1016/0550-3213(93)90271-P}{\emph{Nucl. Phys. B}
  {\bfseries 407} (1993) 3}.

\bibitem{Bauer:2000ew}
C.~W. Bauer, S.~Fleming and M.~E. Luke, \emph{{Summing Sudakov logarithms in $B
  \to X_s \gamma$ in effective field theory}},
  \href{https://doi.org/10.1103/PhysRevD.63.014006}{\emph{Phys. Rev.}
  {\bfseries D63} (2000) 014006}
  [\href{https://arxiv.org/abs/hep-ph/0005275}{{\ttfamily hep-ph/0005275}}].

\bibitem{Bauer:2000yr}
C.~W. Bauer, S.~Fleming, D.~Pirjol and I.~W. Stewart, \emph{{An Effective field
  theory for collinear and soft gluons: Heavy to light decays}},
  \href{https://doi.org/10.1103/PhysRevD.63.114020}{\emph{Phys. Rev.}
  {\bfseries D63} (2001) 114020}
  [\href{https://arxiv.org/abs/hep-ph/0011336}{{\ttfamily hep-ph/0011336}}].

\bibitem{Bauer:2001ct}
C.~W. Bauer and I.~W. Stewart, \emph{{Invariant operators in collinear
  effective theory}},
  \href{https://doi.org/10.1016/S0370-2693(01)00902-9}{\emph{Phys. Lett.}
  {\bfseries B516} (2001) 134}
  [\href{https://arxiv.org/abs/hep-ph/0107001}{{\ttfamily hep-ph/0107001}}].

\bibitem{Bauer:2001yt}
C.~W. Bauer, D.~Pirjol and I.~W. Stewart, \emph{{Soft collinear factorization
  in effective field theory}},
  \href{https://doi.org/10.1103/PhysRevD.65.054022}{\emph{Phys. Rev. D}
  {\bfseries 65} (2002) 054022}
  [\href{https://arxiv.org/abs/hep-ph/0109045}{{\ttfamily hep-ph/0109045}}].

\bibitem{Bauer:2002nz}
C.~W. Bauer, S.~Fleming, D.~Pirjol, I.~Z. Rothstein and I.~W. Stewart,
  \emph{{Hard scattering factorization from effective field theory}},
  \href{https://doi.org/10.1103/PhysRevD.66.014017}{\emph{Phys. Rev.}
  {\bfseries D66} (2002) 014017}
  [\href{https://arxiv.org/abs/hep-ph/0202088}{{\ttfamily hep-ph/0202088}}].

\bibitem{Schwartz:2007ib}
M.~D. Schwartz, \emph{{Resummation and NLO matching of event shapes with
  effective field theory}},
  \href{https://doi.org/10.1103/PhysRevD.77.014026}{\emph{Phys. Rev. D}
  {\bfseries 77} (2008) 014026}
  [\href{https://arxiv.org/abs/0709.2709}{{\ttfamily 0709.2709}}].

\bibitem{Becher:2008cf}
T.~Becher and M.~D. Schwartz, \emph{{A precise determination of $\alpha_s$ from
  LEP thrust data using effective field theory}},
  \href{https://doi.org/10.1088/1126-6708/2008/07/034}{\emph{JHEP} {\bfseries
  07} (2008) 034} [\href{https://arxiv.org/abs/0803.0342}{{\ttfamily
  0803.0342}}].

\bibitem{Chien:2010kc}
Y.-T. Chien and M.~D. Schwartz, \emph{{Resummation of heavy jet mass and
  comparison to LEP data}},
  \href{https://doi.org/10.1007/JHEP08(2010)058}{\emph{JHEP} {\bfseries 08}
  (2010) 058} [\href{https://arxiv.org/abs/1005.1644}{{\ttfamily 1005.1644}}].

\bibitem{Hoang:2014wka}
A.~H. Hoang, D.~W. Kolodrubetz, V.~Mateu and I.~W. Stewart,
  \emph{{$C$-parameter distribution at N$^3$LL' including power corrections}},
  \href{https://doi.org/10.1103/PhysRevD.91.094017}{\emph{Phys. Rev. D}
  {\bfseries 91} (2015) 094017}
  [\href{https://arxiv.org/abs/1411.6633}{{\ttfamily 1411.6633}}].

\bibitem{Bell:2018gce}
G.~Bell, A.~Hornig, C.~Lee and J.~Talbert, \emph{{$e^+ e^-$ angularity
  distributions at NNLL$^\prime$ accuracy}},
  \href{https://doi.org/10.1007/JHEP01(2019)147}{\emph{JHEP} {\bfseries 01}
  (2019) 147} [\href{https://arxiv.org/abs/1808.07867}{{\ttfamily
  1808.07867}}].

\bibitem{Banfi:2001bz}
A.~Banfi, G.~P. Salam and G.~Zanderighi, \emph{{Semi-numerical resummation of
  event shapes}},
  \href{https://doi.org/10.1088/1126-6708/2002/01/018}{\emph{JHEP} {\bfseries
  01} (2002) 018} [\href{https://arxiv.org/abs/hep-ph/0112156}{{\ttfamily
  hep-ph/0112156}}].

\bibitem{Banfi:2014sua}
A.~Banfi, H.~McAslan, P.~F. Monni and G.~Zanderighi, \emph{{A general method
  for the resummation of event-shape distributions in $e^{+} e^{−}$
  annihilation}}, \href{https://doi.org/10.1007/JHEP05(2015)102}{\emph{JHEP}
  {\bfseries 05} (2015) 102} [\href{https://arxiv.org/abs/1412.2126}{{\ttfamily
  1412.2126}}].

\bibitem{Webber:1994cp}
B.~R. Webber, \emph{{Estimation of power corrections to hadronic event
  shapes}}, \href{https://doi.org/10.1016/0370-2693(94)91147-9}{\emph{Phys.
  Lett. B} {\bfseries 339} (1994) 148}
  [\href{https://arxiv.org/abs/hep-ph/9408222}{{\ttfamily hep-ph/9408222}}].

\bibitem{Dokshitzer:1995zt}
Y.~L. Dokshitzer and B.~R. Webber, \emph{{Calculation of power corrections to
  hadronic event shapes}},
  \href{https://doi.org/10.1016/0370-2693(95)00548-Y}{\emph{Phys. Lett.}
  {\bfseries B352} (1995) 451}
  [\href{https://arxiv.org/abs/hep-ph/9504219}{{\ttfamily hep-ph/9504219}}].

\bibitem{Akhoury:1995sp}
R.~Akhoury and V.~I. Zakharov, \emph{{On the universality of the leading, 1/Q
  power corrections in QCD}},
  \href{https://doi.org/10.1016/0370-2693(95)00866-J}{\emph{Phys. Lett. B}
  {\bfseries 357} (1995) 646}
  [\href{https://arxiv.org/abs/hep-ph/9504248}{{\ttfamily hep-ph/9504248}}].

\bibitem{Korchemsky:1994is}
G.~P. Korchemsky and G.~F. Sterman, \emph{{Nonperturbative corrections in
  resummed cross-sections}},
  \href{https://doi.org/10.1016/0550-3213(94)00006-Z}{\emph{Nucl. Phys. B}
  {\bfseries 437} (1995) 415}
  [\href{https://arxiv.org/abs/hep-ph/9411211}{{\ttfamily hep-ph/9411211}}].

\bibitem{Korchemsky:1999kt}
G.~P. Korchemsky and G.~F. Sterman, \emph{{Power corrections to event shapes
  and factorization}},
  \href{https://doi.org/10.1016/S0550-3213(99)00308-9}{\emph{Nucl. Phys. B}
  {\bfseries 555} (1999) 335}
  [\href{https://arxiv.org/abs/hep-ph/9902341}{{\ttfamily hep-ph/9902341}}].

\bibitem{Lee:2006nr}
C.~Lee and G.~F. Sterman, \emph{{Momentum Flow Correlations from Event Shapes:
  Factorized Soft Gluons and Soft-Collinear Effective Theory}},
  \href{https://doi.org/10.1103/PhysRevD.75.014022}{\emph{Phys. Rev. D}
  {\bfseries 75} (2007) 014022}
  [\href{https://arxiv.org/abs/hep-ph/0611061}{{\ttfamily hep-ph/0611061}}].

\bibitem{Hoang:2007vb}
A.~H. Hoang and I.~W. Stewart, \emph{{Designing gapped soft functions for jet
  production}},
  \href{https://doi.org/10.1016/j.physletb.2008.01.040}{\emph{Phys. Lett.}
  {\bfseries B660} (2008) 483}
  [\href{https://arxiv.org/abs/0709.3519}{{\ttfamily 0709.3519}}].

\bibitem{Abbate:2010xh}
R.~Abbate, M.~Fickinger, A.~H. Hoang, V.~Mateu and I.~W. Stewart, \emph{{Thrust
  at N$^3$LL with Power Corrections and a Precision Global Fit for
  $\alpha_s(m_Z)$}},
  \href{https://doi.org/10.1103/PhysRevD.83.074021}{\emph{Phys. Rev.}
  {\bfseries D83} (2011) 074021}
  [\href{https://arxiv.org/abs/1006.3080}{{\ttfamily 1006.3080}}].

\bibitem{Nason:1995np}
P.~Nason and M.~H. Seymour, \emph{{Infrared renormalons and power suppressed
  effects in $e^+\, e^-$ jet events}},
  \href{https://doi.org/10.1016/0550-3213(95)00461-Z}{\emph{Nucl. Phys. B}
  {\bfseries 454} (1995) 291}
  [\href{https://arxiv.org/abs/hep-ph/9506317}{{\ttfamily hep-ph/9506317}}].

\bibitem{Gardi:2001ny}
E.~Gardi and J.~Rathsman, \emph{{Renormalon resummation and exponentiation of
  soft and collinear gluon radiation in the thrust distribution}},
  \href{https://doi.org/10.1016/S0550-3213(01)00284-X}{\emph{Nucl. Phys. B}
  {\bfseries 609} (2001) 123}
  [\href{https://arxiv.org/abs/hep-ph/0103217}{{\ttfamily hep-ph/0103217}}].

\bibitem{Dokshitzer:1995qm}
Y.~L. Dokshitzer, G.~Marchesini and B.~R. Webber, \emph{{Dispersive approach to
  power behaved contributions in QCD hard processes}},
  \href{https://doi.org/10.1016/0550-3213(96)00155-1}{\emph{Nucl. Phys. B}
  {\bfseries 469} (1996) 93}
  [\href{https://arxiv.org/abs/hep-ph/9512336}{{\ttfamily hep-ph/9512336}}].

\bibitem{Dokshitzer:1997ew}
Y.~L. Dokshitzer and B.~R. Webber, \emph{{Power corrections to event shape
  distributions}},
  \href{https://doi.org/10.1016/S0370-2693(97)00573-X}{\emph{Phys. Lett. B}
  {\bfseries 404} (1997) 321}
  [\href{https://arxiv.org/abs/hep-ph/9704298}{{\ttfamily hep-ph/9704298}}].

\bibitem{Davison:2009wzs}
R.~A. Davison and B.~R. Webber, \emph{{Non-Perturbative Contribution to the
  Thrust Distribution in $e^+ e^-$ Annihilation}},
  \href{https://doi.org/10.1140/epjc/s10052-008-0836-7}{\emph{Eur. Phys. J. C}
  {\bfseries 59} (2009) 13} [\href{https://arxiv.org/abs/0809.3326}{{\ttfamily
  0809.3326}}].

\bibitem{Hoang:2009yr}
A.~H. Hoang, A.~Jain, I.~Scimemi and I.~W. Stewart, \emph{{R-evolution:
  Improving perturbative QCD}},
  \href{https://doi.org/10.1103/PhysRevD.82.011501}{\emph{Phys. Rev. D}
  {\bfseries 82} (2010) 011501}
  [\href{https://arxiv.org/abs/0908.3189}{{\ttfamily 0908.3189}}].

\bibitem{Hoang:2008yj}
A.~H. Hoang, A.~Jain, I.~Scimemi and I.~W. Stewart, \emph{{Infrared
  Renormalization Group Flow for Heavy Quark Masses}},
  \href{https://doi.org/10.1103/PhysRevLett.101.151602}{\emph{Phys. Rev. Lett.}
  {\bfseries 101} (2008) 151602}
  [\href{https://arxiv.org/abs/0803.4214}{{\ttfamily 0803.4214}}].

\bibitem{Hoang:2017suc}
A.~H. Hoang, A.~Jain, C.~Lepenik, V.~Mateu, M.~Preisser, I.~Scimemi et~al.,
  \emph{{The MSR mass and the $
  \mathcal{O}\left({\Lambda}_{\mathrm{QCD}}\right) $ renormalon sum rule}},
  \href{https://doi.org/10.1007/JHEP04(2018)003}{\emph{JHEP} {\bfseries 04}
  (2018) 003} [\href{https://arxiv.org/abs/1704.01580}{{\ttfamily
  1704.01580}}].

\bibitem{Abbate:2012jh}
R.~Abbate, M.~Fickinger, A.~H. Hoang, V.~Mateu and I.~W. Stewart,
  \emph{{Precision Thrust Cumulant Moments at $N^3$LL}},
  \href{https://doi.org/10.1103/PhysRevD.86.094002}{\emph{Phys. Rev. D}
  {\bfseries 86} (2012) 094002}
  [\href{https://arxiv.org/abs/1204.5746}{{\ttfamily 1204.5746}}].

\bibitem{Hoang:2015hka}
A.~H. Hoang, D.~W. Kolodrubetz, V.~Mateu and I.~W. Stewart, \emph{{Precise
  determination of $\alpha_s$ from the $C$-parameter distribution}},
  \href{https://doi.org/10.1103/PhysRevD.91.094018}{\emph{Phys. Rev. D}
  {\bfseries 91} (2015) 094018}
  [\href{https://arxiv.org/abs/1501.04111}{{\ttfamily 1501.04111}}].

\bibitem{Luisoni:2020efy}
G.~Luisoni, P.~F. Monni and G.~P. Salam, \emph{{$C$-parameter hadronisation in
  the symmetric 3-jet limit and impact on $\alpha_s$ fits}},
  \href{https://doi.org/10.1140/epjc/s10052-021-08941-z}{\emph{Eur. Phys. J. C}
  {\bfseries 81} (2021) 158}
  [\href{https://arxiv.org/abs/2012.00622}{{\ttfamily 2012.00622}}].

\bibitem{Caola:2021kzt}
F.~Caola, S.~Ferrario~Ravasio, G.~Limatola, K.~Melnikov and P.~Nason, \emph{{On
  linear power corrections in certain collider observables}},
  \href{https://doi.org/10.1007/JHEP01(2022)093}{\emph{JHEP} {\bfseries 01}
  (2022) 093} [\href{https://arxiv.org/abs/2108.08897}{{\ttfamily
  2108.08897}}].

\bibitem{Caola:2022vea}
F.~Caola, S.~Ferrario~Ravasio, G.~Limatola, K.~Melnikov, P.~Nason and M.~A.
  Ozcelik, \emph{{Linear power corrections to e$^{+}$e$^{-}$ shape variables in
  the three-jet region}},
  \href{https://doi.org/10.1007/JHEP12(2022)062}{\emph{JHEP} {\bfseries 12}
  (2022) 062} [\href{https://arxiv.org/abs/2204.02247}{{\ttfamily
  2204.02247}}].

\bibitem{Nason:2023asn}
P.~Nason and G.~Zanderighi, \emph{{Fits of \ensuremath{\alpha}$_{s}$ using
  power corrections in the three-jet region}},
  \href{https://doi.org/10.1007/JHEP06(2023)058}{\emph{JHEP} {\bfseries 06}
  (2023) 058} [\href{https://arxiv.org/abs/2301.03607}{{\ttfamily
  2301.03607}}].

\bibitem{Bell:2023dqs}
G.~Bell, C.~Lee, Y.~Makris, J.~Talbert and B.~Yan, \emph{{Effects of renormalon
  scheme and perturbative scale choices on determinations of the strong
  coupling from $e^+e^-$ event shapes}},
  \href{https://doi.org/10.1103/PhysRevD.109.094008}{\emph{Phys. Rev. D}
  {\bfseries 109} (2024) 094008}
  [\href{https://arxiv.org/abs/2311.03990}{{\ttfamily 2311.03990}}].

\bibitem{Fleming:2007qr}
S.~Fleming, A.~H. Hoang, S.~Mantry and I.~W. Stewart, \emph{{Jets from massive
  unstable particles: Top-mass determination}},
  \href{https://doi.org/10.1103/PhysRevD.77.074010}{\emph{Phys. Rev. D}
  {\bfseries 77} (2008) 074010}
  [\href{https://arxiv.org/abs/hep-ph/0703207}{{\ttfamily hep-ph/0703207}}].

\bibitem{vonManteuffel:2020vjv}
A.~von Manteuffel, E.~Panzer and R.~M. Schabinger, \emph{{Cusp and collinear
  anomalous dimensions in four-loop QCD from form factors}},
  \href{https://doi.org/10.1103/PhysRevLett.124.162001}{\emph{Phys. Rev. Lett.}
  {\bfseries 124} (2020) 162001}
  [\href{https://arxiv.org/abs/2002.04617}{{\ttfamily 2002.04617}}].

\bibitem{Henn:2019swt}
J.~M. Henn, G.~P. Korchemsky and B.~Mistlberger, \emph{{The full four-loop cusp
  anomalous dimension in $\mathcal{N}=4$ super Yang-Mills and QCD}},
  \href{https://doi.org/10.1007/JHEP04(2020)018}{\emph{JHEP} {\bfseries 04}
  (2020) 018} [\href{https://arxiv.org/abs/1911.10174}{{\ttfamily
  1911.10174}}].

\bibitem{Henn:2019rmi}
J.~M. Henn, T.~Peraro, M.~Stahlhofen and P.~Wasser, \emph{{Matter dependence of
  the four-loop cusp anomalous dimension}},
  \href{https://doi.org/10.1103/PhysRevLett.122.201602}{\emph{Phys. Rev. Lett.}
  {\bfseries 122} (2019) 201602}
  [\href{https://arxiv.org/abs/1901.03693}{{\ttfamily 1901.03693}}].

\bibitem{Bruser:2019auj}
R.~Br\"user, A.~Grozin, J.~M. Henn and M.~Stahlhofen, \emph{{Matter dependence
  of the four-loop QCD cusp anomalous dimension: from small angles to all
  angles}}, \href{https://doi.org/10.1007/JHEP05(2019)186}{\emph{JHEP}
  {\bfseries 05} (2019) 186}
  [\href{https://arxiv.org/abs/1902.05076}{{\ttfamily 1902.05076}}].

\bibitem{Moch:2018wjh}
S.~Moch, B.~Ruijl, T.~Ueda, J.~A.~M. Vermaseren and A.~Vogt, \emph{{On quartic
  colour factors in splitting functions and the gluon cusp anomalous
  dimension}},
  \href{https://doi.org/10.1016/j.physletb.2018.06.017}{\emph{Phys. Lett. B}
  {\bfseries 782} (2018) 627}
  [\href{https://arxiv.org/abs/1805.09638}{{\ttfamily 1805.09638}}].

\bibitem{Moch:2017uml}
S.~Moch, B.~Ruijl, T.~Ueda, J.~A.~M. Vermaseren and A.~Vogt, \emph{{Four-Loop
  Non-Singlet Splitting Functions in the Planar Limit and Beyond}},
  \href{https://doi.org/10.1007/JHEP10(2017)041}{\emph{JHEP} {\bfseries 10}
  (2017) 041} [\href{https://arxiv.org/abs/1707.08315}{{\ttfamily
  1707.08315}}].

\bibitem{Bruser:2018rad}
R.~Br\"user, Z.~L. Liu and M.~Stahlhofen, \emph{{Three-Loop Quark Jet
  Function}}, \href{https://doi.org/10.1103/PhysRevLett.121.072003}{\emph{Phys.
  Rev. Lett.} {\bfseries 121} (2018) 072003}
  [\href{https://arxiv.org/abs/1804.09722}{{\ttfamily 1804.09722}}].

\bibitem{Banerjee:2018ozf}
P.~Banerjee, P.~K. Dhani and V.~Ravindran, \emph{{Gluon jet function at three
  loops in QCD}}, \href{https://doi.org/10.1103/PhysRevD.98.094016}{\emph{Phys.
  Rev. D} {\bfseries 98} (2018) 094016}
  [\href{https://arxiv.org/abs/1805.02637}{{\ttfamily 1805.02637}}].

\bibitem{Kelley:2011ng}
R.~Kelley, M.~D. Schwartz, R.~M. Schabinger and H.~X. Zhu, \emph{{The two-loop
  hemisphere soft function}},
  \href{https://doi.org/10.1103/PhysRevD.84.045022}{\emph{Phys. Rev. D}
  {\bfseries 84} (2011) 045022}
  [\href{https://arxiv.org/abs/1105.3676}{{\ttfamily 1105.3676}}].

\bibitem{Monni:2011gb}
P.~F. Monni, T.~Gehrmann and G.~Luisoni, \emph{{Two-Loop Soft Corrections and
  Resummation of the Thrust Distribution in the Dijet Region}},
  \href{https://doi.org/10.1007/JHEP08(2011)010}{\emph{JHEP} {\bfseries 08}
  (2011) 010} [\href{https://arxiv.org/abs/1105.4560}{{\ttfamily 1105.4560}}].

\bibitem{Baranowski:2024vxg}
D.~Baranowski, M.~Delto, K.~Melnikov, A.~Pikelner and C.-Y. Wang,
  \emph{{Zero-jettiness soft function to third order in perturbative QCD}},
  \href{https://arxiv.org/abs/2409.11042}{{\ttfamily 2409.11042}}.

\bibitem{Almeida:2014uva}
L.~G. Almeida, S.~D. Ellis, C.~Lee, G.~Sterman, I.~Sung and J.~R. Walsh,
  \emph{{Comparing and counting logs in direct and effective methods of QCD
  resummation}}, \href{https://doi.org/10.1007/JHEP04(2014)174}{\emph{JHEP}
  {\bfseries 04} (2014) 174} [\href{https://arxiv.org/abs/1401.4460}{{\ttfamily
  1401.4460}}].

\bibitem{Mateu:2017hlz}
V.~Mateu and P.~G. Ortega, \emph{{Bottom and Charm Mass determinations from
  global fits to $Q\bar{Q}$ bound states at N$^3$LO}},
  \href{https://doi.org/10.1007/JHEP01(2018)122}{\emph{JHEP} {\bfseries 01}
  (2018) 122} [\href{https://arxiv.org/abs/1711.05755}{{\ttfamily
  1711.05755}}].

\bibitem{Ellis:1980wv}
R.~K. Ellis, D.~A. Ross and A.~E. Terrano, \emph{{The Perturbative Calculation
  of Jet Structure in $e^+ e^-$ Annihilation}},
  \href{https://doi.org/10.1016/0550-3213(81)90165-6}{\emph{Nucl. Phys. B}
  {\bfseries 178} (1981) 421}.

\bibitem{Moult:2016fqy}
I.~Moult, L.~Rothen, I.~W. Stewart, F.~J. Tackmann and H.~X. Zhu,
  \emph{{Subleading Power Corrections for N-Jettiness Subtractions}},
  \href{https://doi.org/10.1103/PhysRevD.95.074023}{\emph{Phys. Rev. D}
  {\bfseries 95} (2017) 074023}
  [\href{https://arxiv.org/abs/1612.00450}{{\ttfamily 1612.00450}}].

\bibitem{Gehrmann-DeRidder:2014hxk}
A.~Gehrmann-De~Ridder, T.~Gehrmann, E.~W.~N. Glover and G.~Heinrich,
  \emph{{EERAD3: Event shapes and jet rates in electron-positron annihilation
  at order $\alpha_s^3$}},
  \href{https://doi.org/10.1016/j.cpc.2014.07.024}{\emph{Comput. Phys. Commun.}
  {\bfseries 185} (2014) 3331}
  [\href{https://arxiv.org/abs/1402.4140}{{\ttfamily 1402.4140}}].

\bibitem{Moult:2018jjd}
I.~Moult, I.~W. Stewart, G.~Vita and H.~X. Zhu, \emph{{First Subleading Power
  Resummation for Event Shapes}},
  \href{https://doi.org/10.1007/JHEP08(2018)013}{\emph{JHEP} {\bfseries 08}
  (2018) 013} [\href{https://arxiv.org/abs/1804.04665}{{\ttfamily
  1804.04665}}].

\bibitem{Moult:2019uhz}
I.~Moult, I.~W. Stewart, G.~Vita and H.~X. Zhu, \emph{{The Soft Quark
  Sudakov}}, \href{https://doi.org/10.1007/JHEP05(2020)089}{\emph{JHEP}
  {\bfseries 05} (2020) 089}
  [\href{https://arxiv.org/abs/1910.14038}{{\ttfamily 1910.14038}}].

\bibitem{Beneke:2022obx}
M.~Beneke, M.~Garny, S.~Jaskiewicz, J.~Strohm, R.~Szafron, L.~Vernazza et~al.,
  \emph{{Next-to-leading power endpoint factorization and resummation for
  off-diagonal \textquotedblleft{}gluon\textquotedblright{} thrust}},
  \href{https://doi.org/10.1007/JHEP07(2022)144}{\emph{JHEP} {\bfseries 07}
  (2022) 144} [\href{https://arxiv.org/abs/2205.04479}{{\ttfamily
  2205.04479}}].

\bibitem{Korchemsky:2000kp}
G.~P. Korchemsky and S.~Tafat, \emph{{On power corrections to the event shape
  distributions in QCD}},
  \href{https://doi.org/10.1088/1126-6708/2000/10/010}{\emph{JHEP} {\bfseries
  10} (2000) 010} [\href{https://arxiv.org/abs/hep-ph/0007005}{{\ttfamily
  hep-ph/0007005}}].

\bibitem{Ligeti:2008ac}
Z.~Ligeti, I.~W. Stewart and F.~J. Tackmann, \emph{{Treating the b quark
  distribution function with reliable uncertainties}},
  \href{https://doi.org/10.1103/PhysRevD.78.114014}{\emph{Phys. Rev. D}
  {\bfseries 78} (2008) 114014}
  [\href{https://arxiv.org/abs/0807.1926}{{\ttfamily 0807.1926}}].

\bibitem{Manohar:1994kq}
A.~V. Manohar and M.~B. Wise, \emph{{Power suppressed corrections to hadronic
  event shapes}},
  \href{https://doi.org/10.1016/0370-2693(94)01504-6}{\emph{Phys. Lett. B}
  {\bfseries 344} (1995) 407}
  [\href{https://arxiv.org/abs/hep-ph/9406392}{{\ttfamily hep-ph/9406392}}].

\bibitem{ParticleDataGroup:2022pth}
{\scshape Particle Data Group} collaboration, R.~L. Workman et~al.,
  \emph{{Review of Particle Physics}},
  \href{https://doi.org/10.1093/ptep/ptac097}{\emph{PTEP} {\bfseries 2022}
  (2022) 083C01}.

\bibitem{Hoang:2008fs}
A.~H. Hoang and S.~Kluth, \emph{{Hemisphere Soft Function at
  $\mathcal{O}(\alpha_s^2)$ for Dijet Production in $e^+ e^-$ Annihilation}},
  \href{https://arxiv.org/abs/0806.3852}{{\ttfamily 0806.3852}}.

\bibitem{Bachu:2020nqn}
B.~Bachu, A.~H. Hoang, V.~Mateu, A.~Pathak and I.~W. Stewart, \emph{{Boosted
  top quarks in the peak region with N$^3$LL resummation}},
  \href{https://doi.org/10.1103/PhysRevD.104.014026}{\emph{Phys. Rev. D}
  {\bfseries 104} (2021) 014026}
  [\href{https://arxiv.org/abs/2012.12304}{{\ttfamily 2012.12304}}].

\bibitem{Dehnadi:2023msm}
B.~Dehnadi, A.~H. Hoang, O.~L. Jin and V.~Mateu, \emph{{Top quark mass
  calibration for Monte Carlo event generators \textemdash{} an update}},
  \href{https://doi.org/10.1007/JHEP12(2023)065}{\emph{JHEP} {\bfseries 12}
  (2023) 065} [\href{https://arxiv.org/abs/2309.00547}{{\ttfamily
  2309.00547}}].

\bibitem{Salam:2001bd}
G.~P. Salam and D.~Wicke, \emph{{Hadron masses and power corrections to event
  shapes}}, \href{https://doi.org/10.1088/1126-6708/2001/05/061}{\emph{JHEP}
  {\bfseries 05} (2001) 061}
  [\href{https://arxiv.org/abs/hep-ph/0102343}{{\ttfamily hep-ph/0102343}}].

\bibitem{Mateu:2012nk}
V.~Mateu, I.~W. Stewart and J.~Thaler, \emph{{Power Corrections to Event Shapes
  with Mass-Dependent Operators}},
  \href{https://doi.org/10.1103/PhysRevD.87.014025}{\emph{Phys. Rev. D}
  {\bfseries 87} (2013) 014025}
  [\href{https://arxiv.org/abs/1209.3781}{{\ttfamily 1209.3781}}].

\bibitem{Denner:2010ia}
A.~Denner, S.~Dittmaier, T.~Gehrmann and C.~Kurz, \emph{{Electroweak
  corrections to hadronic event shapes and jet production in $e^+e^-$
  annihilation}},
  \href{https://doi.org/10.1016/j.nuclphysb.2010.04.009}{\emph{Nucl. Phys. B}
  {\bfseries 836} (2010) 37} [\href{https://arxiv.org/abs/1003.0986}{{\ttfamily
  1003.0986}}].

\bibitem{Lepenik:2019jjk}
C.~Lepenik and V.~Mateu, \emph{{NLO Massive Event-Shape Differential and
  Cumulative Distributions}},
  \href{https://doi.org/10.1007/JHEP03(2020)024}{\emph{JHEP} {\bfseries 03}
  (2020) 024} [\href{https://arxiv.org/abs/1912.08211}{{\ttfamily
  1912.08211}}].

\bibitem{Bris:2020uyb}
A.~Bris, V.~Mateu and M.~Preisser, \emph{{Massive event-shape distributions at
  N$^2$LL}}, \href{https://doi.org/10.1007/JHEP09(2020)132}{\emph{JHEP}
  {\bfseries 09} (2020) 132}
  [\href{https://arxiv.org/abs/2006.06383}{{\ttfamily 2006.06383}}].

\bibitem{gcccompileprocess}
NTU, ``Gcc and make compiling, linking and building c/c++ applications.''
  \url{https://www3.ntu.edu.sg/home/ehchua/programming/cpp/gcc_make.html},
  2013.

\bibitem{gfortran}
\emph{{GFortran, Gnu compiler collection (gcc), Version 8.1.0}}. Copyright (C)
  2018 Free Software Foundation, Inc., 2018.

\bibitem{galassi2018scientific}
M.~e.~a. Galassi, \emph{Gnu scientific library reference manual},  2018.

\bibitem{quadpackpp}
``{Quadpack++}.'' \url{https://github.com/drjerry/quadpackpp}.

\bibitem{Rossum:1995:PRM:869369}
G.~Rossum, \emph{Python reference manual},  tech. rep., Amsterdam, The
  Netherlands, The Netherlands, 1995.

\bibitem{swig}
``{SWIG}.'' \url{http://swig.org/}.

\bibitem{Dissertori:2007xa}
G.~Dissertori, A.~Gehrmann-De~Ridder, T.~Gehrmann, E.~W.~N. Glover, G.~Heinrich
  and H.~Stenzel, \emph{{First determination of the strong coupling constant
  using NNLO predictions for hadronic event shapes in $e^+ e^-$
  annihilations}},
  \href{https://doi.org/10.1088/1126-6708/2008/02/040}{\emph{JHEP} {\bfseries
  02} (2008) 040} [\href{https://arxiv.org/abs/0712.0327}{{\ttfamily
  0712.0327}}].

\bibitem{Becher:2009th}
T.~Becher and M.~D. Schwartz, \emph{{Direct photon production with effective
  field theory}}, \href{https://doi.org/10.1007/JHEP02(2010)040}{\emph{JHEP}
  {\bfseries 02} (2010) 040} [\href{https://arxiv.org/abs/0911.0681}{{\ttfamily
  0911.0681}}].

\bibitem{Gehrmann:2012sc}
T.~Gehrmann, G.~Luisoni and P.~F. Monni, \emph{{Power corrections in the
  dispersive model for a determination of the strong coupling constant from the
  thrust distribution}},
  \href{https://doi.org/10.1140/epjc/s10052-012-2265-x}{\emph{Eur. Phys. J. C}
  {\bfseries 73} (2013) 2265}
  [\href{https://arxiv.org/abs/1210.6945}{{\ttfamily 1210.6945}}].

\bibitem{Gracia:2021nut}
N.~G. Gracia and V.~Mateu, \emph{{Toward massless and massive event shapes in
  the large-\ensuremath{\beta}$_{0}$ limit}},
  \href{https://doi.org/10.1007/JHEP07(2021)229}{\emph{JHEP} {\bfseries 07}
  (2021) 229} [\href{https://arxiv.org/abs/2104.13942}{{\ttfamily
  2104.13942}}].

\bibitem{Hoang:2018zrp}
A.~H. Hoang, S.~Pl\"atzer and D.~Samitz, \emph{{On the Cutoff Dependence of the
  Quark Mass Parameter in Angular Ordered Parton Showers}},
  \href{https://doi.org/10.1007/JHEP10(2018)200}{\emph{JHEP} {\bfseries 10}
  (2018) 200} [\href{https://arxiv.org/abs/1807.06617}{{\ttfamily
  1807.06617}}].

\bibitem{Hoang:2024zwl}
A.~H. Hoang, O.~L. Jin, S.~Pl\"atzer and D.~Samitz, \emph{{Matching
  Hadronization and Perturbative Evolution: The Cluster Model in Light of
  Infrared Shower Cutoff Dependence}},
  \href{https://arxiv.org/abs/2404.09856}{{\ttfamily 2404.09856}}.

\bibitem{Bauer:2011uc}
C.~W. Bauer, F.~J. Tackmann, J.~R. Walsh and S.~Zuberi, \emph{{Factorization
  and Resummation for Dijet Invariant Mass Spectra}},
  \href{https://doi.org/10.1103/PhysRevD.85.074006}{\emph{Phys. Rev. D}
  {\bfseries 85} (2012) 074006}
  [\href{https://arxiv.org/abs/1106.6047}{{\ttfamily 1106.6047}}].

\bibitem{Larkoski:2017jix}
A.~J. Larkoski, I.~Moult and B.~Nachman, \emph{{Jet Substructure at the Large
  Hadron Collider: A Review of Recent Advances in Theory and Machine
  Learning}}, \href{https://doi.org/10.1016/j.physrep.2019.11.001}{\emph{Phys.
  Rept.} {\bfseries 841} (2020) 1}
  [\href{https://arxiv.org/abs/1709.04464}{{\ttfamily 1709.04464}}].

\bibitem{Fleming:2007xt}
S.~Fleming, A.~H. Hoang, S.~Mantry and I.~W. Stewart, \emph{{Top Jets in the
  Peak Region: Factorization Analysis with NLL Resummation}},
  \href{https://doi.org/10.1103/PhysRevD.77.114003}{\emph{Phys. Rev. D}
  {\bfseries 77} (2008) 114003}
  [\href{https://arxiv.org/abs/0711.2079}{{\ttfamily 0711.2079}}].

\bibitem{Stewart:2014nna}
I.~W. Stewart, F.~J. Tackmann and W.~J. Waalewijn, \emph{{Dissecting Soft
  Radiation with Factorization}},
  \href{https://doi.org/10.1103/PhysRevLett.114.092001}{\emph{Phys. Rev. Lett.}
  {\bfseries 114} (2015) 092001}
  [\href{https://arxiv.org/abs/1405.6722}{{\ttfamily 1405.6722}}].

\bibitem{Bhattacharya:2022dtm}
A.~Bhattacharya, M.~D. Schwartz and X.~Zhang, \emph{{Sudakov shoulder
  resummation for thrust and heavy jet mass}},
  \href{https://doi.org/10.1103/PhysRevD.106.074011}{\emph{Phys. Rev. D}
  {\bfseries 106} (2022) 074011}
  [\href{https://arxiv.org/abs/2205.05702}{{\ttfamily 2205.05702}}].

\bibitem{Feige:2017zci}
I.~Feige, D.~W. Kolodrubetz, I.~Moult and I.~W. Stewart, \emph{{A Complete
  Basis of Helicity Operators for Subleading Factorization}},
  \href{https://doi.org/10.1007/JHEP11(2017)142}{\emph{JHEP} {\bfseries 11}
  (2017) 142} [\href{https://arxiv.org/abs/1703.03411}{{\ttfamily
  1703.03411}}].

\bibitem{Moult:2019mog}
I.~Moult, I.~W. Stewart and G.~Vita, \emph{{Subleading Power Factorization with
  Radiative Functions}},
  \href{https://doi.org/10.1007/JHEP11(2019)153}{\emph{JHEP} {\bfseries 11}
  (2019) 153} [\href{https://arxiv.org/abs/1905.07411}{{\ttfamily
  1905.07411}}].

\bibitem{Bernlochner:2020jlt}
{\scshape SIMBA} collaboration, F.~U. Bernlochner, H.~Lacker, Z.~Ligeti, I.~W.
  Stewart, F.~J. Tackmann and K.~Tackmann, \emph{{Precision Global
  Determination of the $B\to X_s\gamma$ Decay Rate}},
  \href{https://doi.org/10.1103/PhysRevLett.127.102001}{\emph{Phys. Rev. Lett.}
  {\bfseries 127} (2021) 102001}
  [\href{https://arxiv.org/abs/2007.04320}{{\ttfamily 2007.04320}}].

\bibitem{Boels:2017skl}
R.~H. Boels, T.~Huber and G.~Yang, \emph{{Four-Loop Nonplanar Cusp Anomalous
  Dimension in N=4 Supersymmetric Yang-Mills Theory}},
  \href{https://doi.org/10.1103/PhysRevLett.119.201601}{\emph{Phys. Rev. Lett.}
  {\bfseries 119} (2017) 201601}
  [\href{https://arxiv.org/abs/1705.03444}{{\ttfamily 1705.03444}}].

\bibitem{ParticleDataGroup:2024cfk}
{\scshape Particle Data Group} collaboration, S.~Navas et~al., \emph{{Review of
  particle physics}},
  \href{https://doi.org/10.1103/PhysRevD.110.030001}{\emph{Phys. Rev. D}
  {\bfseries 110} (2024) 030001}.

\bibitem{Tackmann:2024kci}
F.~J. Tackmann, \emph{{Beyond Scale Variations: Perturbative Theory
  Uncertainties from Nuisance Parameters}},
  \href{https://arxiv.org/abs/2411.18606}{{\ttfamily 2411.18606}}.

\end{thebibliography}\endgroup

\end{document}